\documentclass[acmsmall]{acmart}
\AtBeginDocument{%
  }

\setcopyright{acmlicensed}
\copyrightyear{2026}
\acmYear{2026}
\acmDOI{XXXXXXX.XXXXXXX}

\acmJournal{TRETS}
\acmVolume{1}
\acmNumber{1}
\acmArticle{1}
\acmMonth{5}

\usepackage{graphicx}
\graphicspath{{./pic/}}
\usepackage{tabularx}
\usepackage{amsfonts}

\DeclareRobustCommand{\hercules}{\mbox{{\scshape Hercules}}}
\DeclareRobustCommand{\stannic}{\mbox{{\scshape Stannic}}}

\colorlet{revAcolor}{blue!20}
\colorlet{revBcolor}{green!20}
\colorlet{revCcolor}{magenta!20}

\newcommand{\bem}[1]{{\bf\em #1}}

\newcommand{\ceil}[1]{\lceil {#1} \rceil}
\newcommand{\bangle}[1]{$\langle${#1}$\rangle$}

\newcommand{\eg}{\mbox{{\em e.g.}}}
\newcommand{\ie}{\mbox{{\em i.e.}}}
\newcommand{\cf}{\mbox{{c.f.}}}

\newcommand{\ept}{\epsilon}

\newtheorem{definition}{\bf Definition}

\def\EPT{\hat\epsilon}

\def\latency{Latency}

\def\fairness{Fairness}

\def\lbalance{Load Balancing}

\usepackage{hyperref}
\hypersetup{
    colorlinks=true,
    citecolor=blue,
    linkcolor=blue,
    urlcolor=blue,
}
\usepackage{resizegather}
\usepackage{subcaption}
\usepackage{caption}
\usepackage[export]{adjustbox}
\usepackage{bm}
\usepackage{mathtools}
\usepackage{enumitem}
\usepackage[table]{xcolor}
\usepackage{comment}
\usepackage[sort]{cleveref}
\usepackage{pifont}
\usepackage{soul}
\usepackage{wrapfig}

\begin{document}


\title{$\stannic$: \underline{S}ystolic S\underline{T}och\underline{A}stic O\underline{N}li\underline{N}e Schedul\underline{I}ng Ac\underline{C}elerator}
\author{Adam H. Ross}
\orcid{0009-0000-6123-0377}
\email{aross50@uic.edu}
\affiliation{
  \institution{Electrical and Computer Eng. Dept, University of Illinois Chicago}
  \city{Chicago}
  \state{Illinois}
  \country{USA}
}

\author{Vairavan Palaniappan}
\orcid{0009-0002-1281-125X}
\email{vpala8@uic.edu}
\affiliation{
  \institution{Electrical and Computer Eng. Dept, University of Illinois Chicago}
  \city{Chicago}
  \state{Illinois}
  \country{USA}
}


\author{Debjit Pal}
\orcid{0000-0003-3722-5126}
\email{dpal2@uic.edu}
\affiliation{
  \institution{Electrical and Computer Eng. Dept, University of Illinois Chicago}
  \city{Chicago}
  \state{Illinois}
  \country{USA}
}

\renewcommand{\shortauthors}{Ross et al.}
\begin{abstract}

Efficient workload scheduling is a critical challenge in modern heterogeneous computing environments, particularly in high-performance computing (HPC) systems. Traditional software-based schedulers struggle to efficiently balance workloads due to {\em scheduling overhead}, {\em lack of adaptability to stochastic workloads}, and {\em suboptimal resource utilization}. The scheduling problem further compounds in the context of shared HPC clusters, where {\em job arrivals} and {\em processing times} are inherently stochastic. Prediction of these elements is possible, but it introduces additional overhead. To perform this complex scheduling, we developed two FPGA-assisted hardware accelerator microarchitectures, $\hercules$ and $\stannic$. $\hercules$ adopts a task-centric abstraction of stochastic scheduling, whereas $\stannic$ inherits a schedule-centric abstraction. These hardware-assisted solutions leverage \bem{parallelism}, \bem{pre-calculation}, and \bem{spatial memory} access to significantly accelerate scheduling. We accelerate a non-preemptive stochastic online scheduling algorithm to produce heterogeneity-aware schedules in near real time. 

With $\hercules$, we achieved a speedup of {\em up to} 1060$\times$ over a baseline C/C++ implementation, demonstrating the efficacy of a hardware-assisted acceleration for heterogeneity-aware stochastic scheduling. With Stannic, we further improved efficiency, achieving a 7.5$\times$ reduction in latency per computation iteration and a 14$\times$ increase in the target heterogeneous system size. Experimental results show that the resulting schedules demonstrate efficient machine utilization and low average job latency in stochastic contexts.
\end{abstract}



\received{27 November 2025}
\received[revised]{24 March 2026}

\maketitle

\section{Introduction}\label{sec:intro}

In modern high-performance computing environments, heterogeneous processing elements (PEs) such as 
Central Processing Units (CPUs), Graphics Processing Units (GPUs), Field-Programmable Gate Arrays (FPGAs), Neural Processing Units (NPUs), and other application-specific accelerators are increasingly being deployed to meet the demands of diverse 
computing tasks. These systems promise improved performance and energy efficiency by scheduling tasks 
to the most suitable PEs based on computation patterns and objectives. {\em However, 
scheduling in such heterogeneous PEs remains a fundamental and computationally complex challenge} \cite{MPEFT, HEFT}.

\begin{wrapfigure}[20]{r}{1.78in}
    \centering
    \includegraphics[width=1.78in, 
    ]{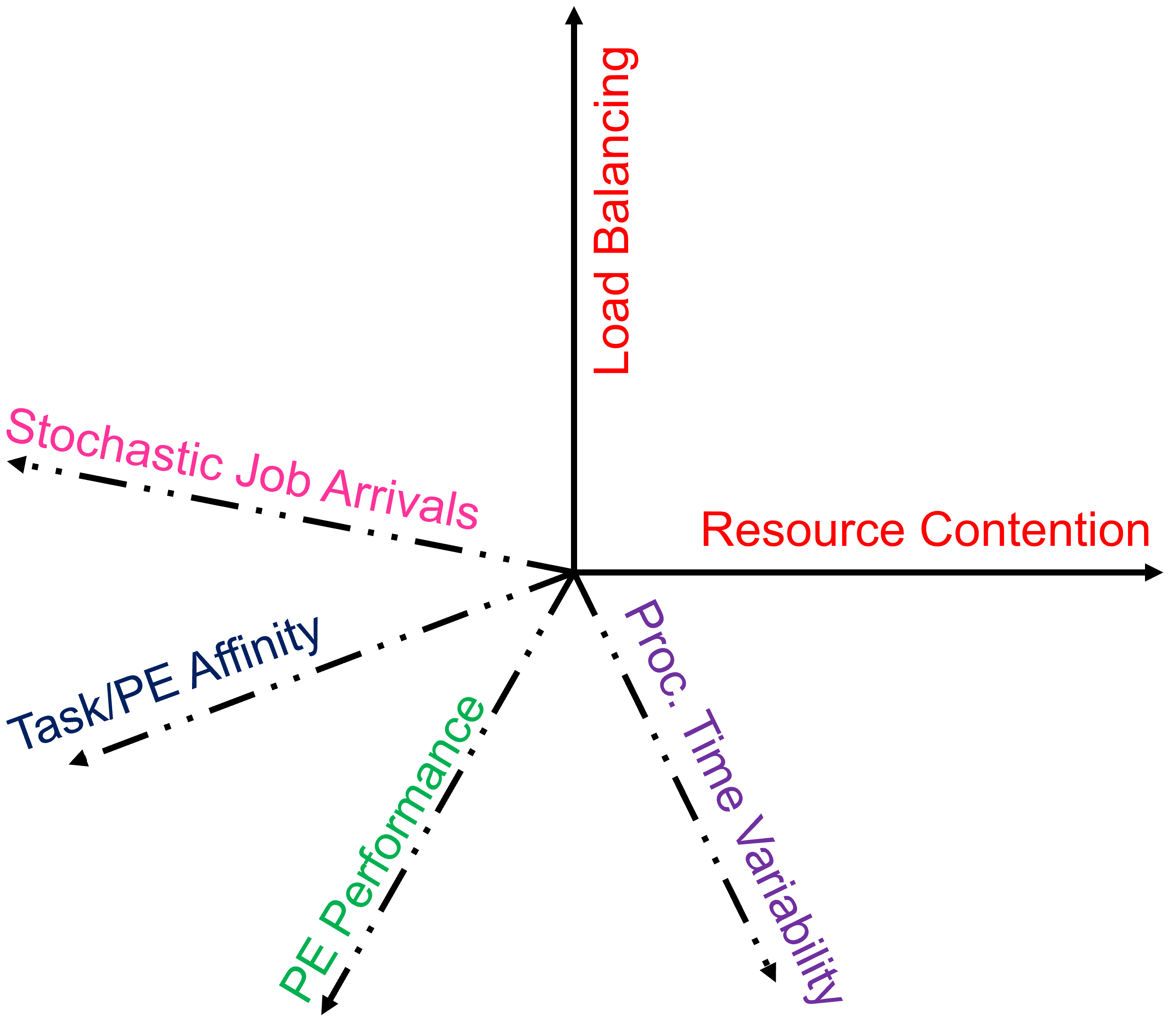}
    \caption{
    {\bf Illustration of the decision space for Stochastic, Online Heterogeneous Scheduling}. Load Balancing and Resource Contention are standard considerations in multi-machine scheduling, with the other axis presenting new considerations in heterogeneous system 
    context.
    }
    \label{fig:descision_space}
\end{wrapfigure}

Unlike homogeneous systems, where scheduling primarily involves load balancing and fair resource allocation, heterogeneous systems introduce a complex multi-dimensional decision space -- scheduling decisions must account for varying processing capabilities, task-PE affinities, execution time variability and unpredictability of tasks' arrival, contention for shared PE resources, and in many cases, strict timing or energy constraints. Consequently, scheduling targeting heterogeneous systems 
becomes a complex {\em search and optimization} in multi-dimensional decision space as shown 
in \Cref{fig:descision_space}. 
Moreover, task characteristics are often not known in advance or may change at runtime, making {\em static or offline scheduling approaches impractical}. 
For example, recent studies on multi-cluster heterogeneous scheduling (e.g., Wang et al. \cite{Wang_multi_cluster}) demonstrate that as the scale of task bursts increases, the combinatorial complexity of offline global batch matching causes scheduling overhead to compound into seconds. This severe computational bottleneck subsequently inflates total execution times, underscoring the critical need for a high-throughput, online scheduling accelerator.
As the system complexity increases, the cost of making optimal or near-optimal scheduling decisions increases rapidly. 
\bem{These challenges demand fast, adaptive, scalable, scheduling technique capable of reasoning under uncertainty, all the while maintaining low computational overhead}. Addressing these goals simultaneously is non-trivial and remains a critical bottleneck in leveraging the full 
potential of heterogeneous computing systems.

Recent scheduling techniques for heterogeneous systems 
aim to improve efficiency and load balancing, but they suffer from high scheduling overhead and inconsistent convergence~\cite{exptaskschjuanfang2020, fdetsadetswan2021, habibpour2024improved}. Other schedulers 
offer adaptability but rely on predictive models and often neglect energy or scalability concerns~\cite{liu2024uncertainty, du2019feature}. More recent scheduling strategies~\cite{schheteroserifyesil2022,naithani2017reliability,raca2022runtime,orr2021optimal} 
provide strong theoretical guarantees but face limitations in dynamic environments, task diversity, and hardware applicability due to their overwhelming computational complexity. Overall, most existing methods are either too complex for real-time use, lack scalability, or fail to adapt under uncertainty.


In our previous work~\cite{HERCULES}, we 
developed a {\em hardware-accelerated scheduler
targeting heterogeneous computing resources to address 
the problem of efficient, low-overhead online scheduling in heterogeneous
systems under unpredictable task arrival
and runtime variability}. We leverage a simple 
and adaptable 
algorithm (Stochastic Online Scheduling Algorithm or SOSA) 
to addresses online scheduling in heterogeneous systems~\cite{jager2023improved} and extend it to make it amenable for hardware implementation. We adopt a {\em task-centric} abstraction of SOSA in our prior work. A critical aspect of this algorithm is that it does not require precise task profiling; 
instead, it relies on 
estimates of processing time, making it practical for dynamic environments where such precise task information is often unavailable and/or unreliable but can be estimated with high confidence based on prior task history. Using this method in \hercules, we were able to demonstrate efficient scheduling, while achieving up to 1093$\times$ speedup over software implementations of the same scheduling methodology. 

In this work, 
we identify and address the shortcomings of our prior work 
and develop $\stannic$ to further optimize scheduling iteration speed, hardware resource utilization, and scalability through routing simplification. 
Crucially, we use the same SOSA algorithm~\cite{jager2023improved} as the foundation for both hardware accelerator architectures, however, we adopt a schedule-centric abstraction to implement $\stannic$. 
The architectural difference of $\stannic$ as compared to our prior work is primarily due to the 
different abstractions/perspectives of the SOSA algorithmic flow.

Specifically, we note the timing and routing overheads severely restricted the performance of the 
the pipeline architecture used in~\cite{HERCULES}, 
and use this insight 
as motivation to devise a new systolic architecture for \stannic\ that leverages the spatial locality of priority-ordered jobs. 
\stannic\ outperforms our prior work 
with an average of 7.5$\times$ less cycles per scheduling iteration, all while requiring fewer hardware components and routing, allowing \stannic\ to scale to much larger system configurations than its predecessor, being capable of tracking up to 140 machines compared to \hercules's maximum of 10. 

In this paper, we make the following additional technical contributions over \cite{HERCULES} to advance the field of stochastic online heterogeneous scheduling.
\begin{enumerate}
    \item We systematically analyze our prior $\mu$architecture, $\hercules$, 
    to localize 
    various performance bottlenecks, and augment them 
    to develop a new 
    $\mu$-architecture, $\stannic$, to further optimize online stochastic scheduling performance. 

    \item We present an alternative, 
    schedule-centric abstraction 
    of the SOSA 
    algorithmic flow to streamline memory access to significantly enhance computation performance. 

    \item We present an extensive set of experimental results demonstrating the effectiveness, efficiency, and adaptability of the SOSA 
    in scheduling tasks with diverse characteristics on a set of heterogeneous computing resources in near real-time with a power envelope of 21 Watts, achieving over 1968$\times$ speedup over a C/C++ baseline software implementation. 
    \item We present a set of quantitative comparison metrics 
    for our two architectures, demonstrating performance improvements across design iterations. 
\end{enumerate}

The paper is organized as follows.~\Cref{sec:prelim_and_background} details preliminaries and necessary background knowledge and 
~\Cref{sec:sos_overview} discusses the scheduling algorithm and its discretization. 
\Cref{sec:asic_design} explains the $\hercules$ microarchitecture. ($\mu$architecture) 
\Cref{sec:bottlenecks} analyzes $\hercules$ $\mu$architecture and localizes 
various 
performance bottlenecks. \Cref{sec:Systolic_SOSA} then presents the $\stannic$ $\mu$architecture addressing the 
identified bottlenecks. 
\Cref{sec:exp_setup,sec:exp_results} discuss 
the experimental setup and results, respectively, which were performed to quantitatively compare and contrast the performance of the 
schedulers. ~\Cref{sec:rel_work} surveys the related work followed by our conclusion in~\Cref{sec:Conclusion}. 

\section{Preliminaries and Background}\label{sec:prelim_and_background}


In this section, first we present key conventions and definitions along with a representative example that we 
use throughout our work. 
Later in this section, we provide a high-level overview of the 
stochastic online 
scheduling algorithm~\cite{jager2023improved}. Finally, 
we present necessary 
background 
on systolic architecture in our context.

\noindent {\bf Conventions}: A computer program can be viewed as a sequence of instructions. In our formalization, we leave the definition of computer program implicit, but we will treat it as a pair $\langle I, t \rangle$ where $t \in \mathbb{Z}^+$. Informally, $I$ represents the set of instructions and $t$ represents the number of cycles required to execute $I$. Given a program $P = \langle I, t \rangle$, we will refer to $t$ as execution time of $P$, denoted by $Time(P)$. A program $P$ is {\em compute-bound} if majority of instructions are arithmetic/control instructions, 
{\em memory-bound} if majority of instructions are data movement, load/store, and memory operations, and 
{\em mixed}, if it has a balanced or near-balanced mix 
of compute-bound and memory-bound instructions. 


\begin{definition}\label{def:machine}
    A \bem{Machine} $M$ is an abstraction of a compute unit 
    represented as a tuple $M = \langle {\bf T}, {\bf Q} \rangle$, where ${\bf T}$ is the machine type and ${\bf Q}$ 
    is the machine quality and ${\bf T} \in [\text{CPU}, \text{GPU}, \text{Mixed}]$, ${\bf Q} \in [\text{Best}, \text{Worst}]$. Intuitively, for a given program $P$, if the execution times are $Time(P)_{Best}$ and $Time(P)_{Worst}$ with $Q = Best$ and $Q = Worst$, respectively, then $Time(P)_{Best} \ll Time(P)_{Worst}$.
\end{definition}



\begin{definition}\label{def:job}
    A \bem{Job} $J$ is an 
    abstraction of a program $P$ with uncertain execution time represented as a quadruple $J = \langle W, \hat{\epsilon}, \mathcal{P}, ID 
    \rangle$, where $W$ is the {\em weight} of $J$, $\hat{\epsilon}$ is a list of expected processing times (EPT) (\ie, how long it would take for $P$ to run on machine $M$) with $|\hat{\epsilon}| = N$ where $N$ is the number of machines, $\mathcal{P}$ is the nature/bounding 
    of a 
    corresponding program $P$, \ie, $\mathcal{P} \in [\text{Compute}, \text{Memory}, \text{Mixed}]$, and $ID \in \mathbb{Z}^+$ is a unique job identifier.
    We use $J.W$, $J.\hat{\epsilon}$ to denote the individual components of a job $J$. We compute \bem{weighted shortest processing time} (WSPT) of a job $J$ for $k^{th}$ machine as $T_k^J = J.W / \hat{\epsilon_k}, \hat{\epsilon_k} \in \hat{\epsilon}$~\cite{jager2023improved}, which can be used to rank jobs by priority. 
\end{definition}

\noindent \bem{Intuitive Example:} Lets consider a job $J$ that corresponds to a convolution neural network layer being scheduled to a system that consists of just two machines, a CPU and a GPU. $J.\EPT$ would have two entries, one each corresponding to the expected processing time of $J$ on either the CPU or GPU. Due to $J$ being a convolutional operation, we could reasonably accomplish this job on either the CPU or the GPU, but we may expect the GPU implementation to complete quicker (\ie, $J.\EPT_{GPU}<J.\EPT_{CPU}$). 
    
This EPT is a best guess, \bem{not a guarantee}. For any actual job invocation, variance (\ie, from data loading, shared memory usage, etc.) will cause the actual runtime to deviate from this EPT. In fact, these extra elements of variance can be included as part of the predicted EPT. This allows modeled network communication and data movement to be encoded into scheduling decisions, without having to be explicitly calculated and tracked for each individual decision. The Weight ($J.W$) of the job is an abstraction of this particular job's priority. We could encode it to account for metadata that may be known about this job. For example, Weight could correlate with the number of jobs that depend on the completion of this job (\ie, how many downstream task nodes this job has in a DAG Task Graph), prioritizing the minimization of start delays. Alternatively, weight could correspond to a job's source, to establish priority for OS/Super User processes. Either way, Weight here is a global Prioritization metric, whereas EPT is a per-machine estimate of processing time. How those values are calculated can be flexible, however, allowing the system to be tuned to specific systems and scheduling goals. It is important to note that in our work, the target workload granularity for these jobs consists of macro-level cluster workloads (such as the aforementioned neural network layer or extensive simulation batches) rather than micro-level atomic operations. Consequently, the Expected Processing Time (EPT) for such tasks spans seconds or minutes, a scale that informs our architectural priorities.




\begin{definition}\label{def:virtual_schedule}
    A \bem{Virtual Schedule} (VS) $V_i$ for machine $M_i$ is a 
    partial order for 
    the execution of a 
    set of jobs $\{K\}$, reflecting the relative WSPT value (and thus priority) of the jobs in $\{K\}$. 
    This is called a Virtual Schedule, because it contains all of the jobs that have been \bem{assigned to a machine}, but have \bem{not yet been released to the work queue}. It is an interim schedule where the ordering of jobs within it is still malleable.
    We use $Head.V_i$ to denote 
    the head of $V_i$. 
\end{definition}

\noindent \bem{Intuitive Example:} Let's reconsider the job discussed in the example of \Cref{def:job}. We established that in our example case that $J.\EPT_{GPU}<J.\EPT_{CPU}$, and so we shall assume that we assigned it to the GPU. However, while we are waiting to commit the job to the GPU for execution, another job $K$ is introduced to the system. $K$ is also assigned to the GPU, and while $K.\EPT_{GPU}=J.\EPT_{GPU}, K.W>J.W$. As such, the new job has a higher overall priority than job $J$. The Virtual Schedule allows us to adjust our planned schedule to reflect this, as the jobs have not yet been committed to a work queue.

\begin{figure}
    \begin{subfigure}[t]{0.4\textwidth}
    \centering
    \includegraphics[width=\textwidth]{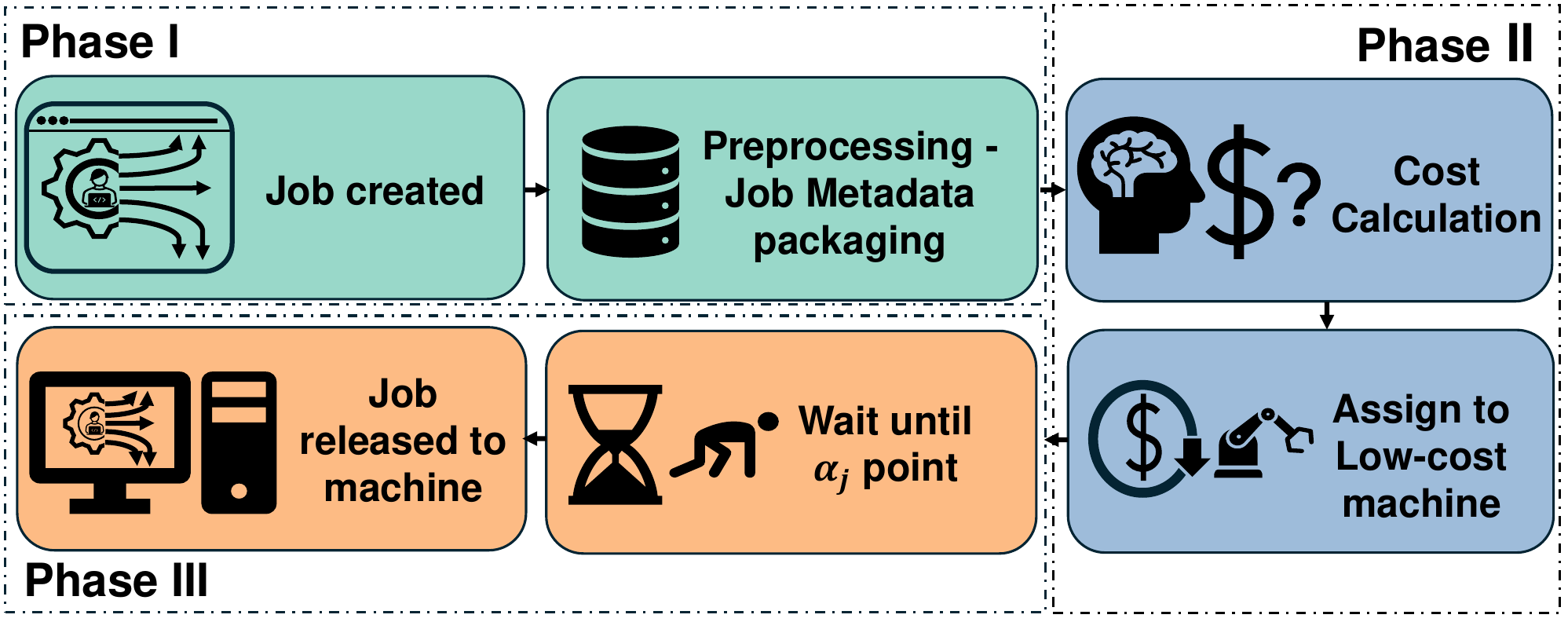}
    \caption{}
    \label{fig:alg_flow_task_perspective}
    \end{subfigure}
    \begin{subfigure}[t]{0.5\textwidth}
    \centering
    \includegraphics[width=\textwidth, trim={0 8cm 0 0}]{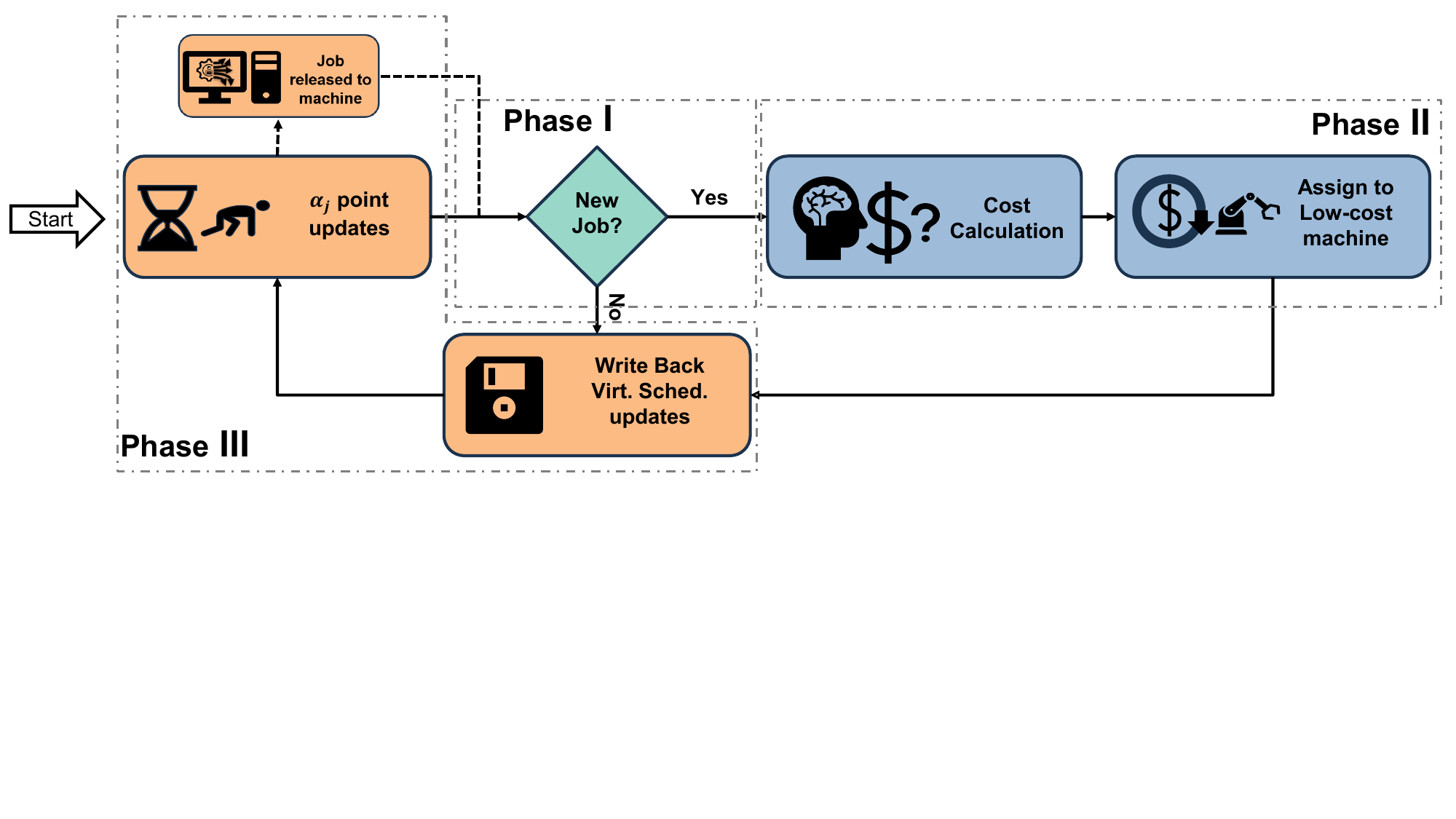}
    \caption{}
    \label{fig:STANNIC_alg_flow}
    \end{subfigure}
    \caption{
    {\bf Stochastic Online Scheduling (SOS) Algorithmic Flows}.
    \Cref{fig:alg_flow_task_perspective} shows algorithmic flow for {\em stochastic online scheduling from a task centric perspective}. 
    {\bf Phase I} prepares a job for the scheduler, {\bf Phase II} and {\bf Phase III} show the steps involved in scheduling the job. 
    \Cref{fig:STANNIC_alg_flow} revises the algorithmic flow involving 
    the same functional steps but from a persistent, virtual-schedule perspective. Due to this re-framing, the algorithmic flow is now cyclical instead of linear as in ~\Cref{fig:alg_flow_task_perspective}. For clarity, we use the same phase labeling to demonstrate 
    the functional similarity 
    of the two algorithmic flows. 
    }
    \label{fig:alg_flow}
\end{figure}

\subsection{Overview of Stochastic Online Scheduling (SOS) Algorithm}\label{sec:background_sos}

The \bem{key intuition} behind the Greedy Stochastic Online Scheduling Algorithm (SOS)~\cite{jager2023improved} is to compute the cost (in terms of expected delay) of assigning an incoming job to a machine $M$ based on the {\em currently available resources}, {\em prior assigned jobs}, and {\em job priorities}, before greedily and irrevocably assigning the job to the machine with the lowest cost for execution. This is performed in three primary phases, which we describe in \Cref{sec:task_centric}, and then reframe to the Virtual Schedule centric perspective in \Cref{sec:virt_sched}.


\subsubsection{Stochastic Scheduling Algorithmic Flow from the Task-Centric Perspective}\label{sec:task_centric}


The SOS~\cite{jager2023improved} focuses on on-the-fly intra- and inter-machine job scheduling.~\Cref{fig:alg_flow_task_perspective} shows an overview of 
the SOS. 
The key objective of the SOS is to {\em minimize} the {\em weighted sum of the expected completion time} over a set of jobs in a greedy way. 
We discuss three phases of the SOS algorithm below. 

\smallskip



\noindent {\bf Phase I (Preprocessing jobs)}: Sources produce new jobs either in burst mode or sequentially, however, the SOS algorithm considers a sequential job arrival during 
processing. The assumption of sequential job arrival allows the scheduler to tackle the uncertainty and/or stochasticity of jobs' arrival. The preprocessing steps append additional info to an arriving job, \eg, EPTs for a job ($J.\hat\ept$) for a set of target machines leveraging prior execution data or metadata obtained from the producers.  
Once a job is fully processed, it is released to the 
scheduler. 


\smallskip

\noindent {\bf Phase II (Machine Assignment)}: 
The SOS computes the cost of assigning job $J$ to machine $M_i$ based on the expected delay of starting $J$ and 
any other jobs already assigned to $M_i$ (\ie, the jobs currently in $V_i$). The cost calculation consists of two components -- $cost^H$ (delay on the new job from the set of earlier 
jobs ($\sigma^{H}$) with higher or equal priority in the $V_i$, (\cf,~\Cref{sec:CostCalMath} for more details) and $cost^L$ (delay on set of earlier 
jobs ($\sigma^{L}$) with lower priority in the $V_i$). Both costs are cumulative 
across their respective 
sets ($\sigma^{H}$ and $\sigma^{L}$) of 
assigned jobs. Internal priority is the relative WSPT value of $J$ with respect to (w.r.t.) the other jobs assigned to $M_i$, which may appear in the Virtual Schedule $V_i$ ahead of, behind, or in-between the other jobs. We 
explain $J$'s relative position in $V_i$ w.r.t. other jobs in~\Cref{sec:CostCalMath}. Once the cost for each of the $N$ machines has been calculated, the machine with the lowest cost is greedily and irrevocably chosen as the {\em assigned} machine for $J$.



\smallskip

\noindent {\bf Phase III (Job Scheduling)}: 
SOS also schedules all jobs assigned to a specific machine. 
Due to the stochastic and online nature of SOS, it is unknown 
when or if new jobs will be assigned. 
However, to ensure that future jobs that are shorter or more important can take precedent over previously assigned jobs, 
J{\"a}ger introduced the $\alpha_J$ WSPT policy, which tracks jobs in a Virtual Schedule $V_i$ and adjusts ordering as new jobs come in~\cite{jager2023improved}. The $V_i$ then releases the job at its head ($Head.V_i$) 
when the wait time of $Head.V_i \geq (\alpha_J \times J.\ept_i), \alpha_J\in(0,1]$. Once $J$ has been released from the Virtual Schedule, it is released to the end of the assigned 
machine's actual job  
queue, and is considered fully scheduled for execution. 

\smallskip

\noindent \bem{Shortcomings of the Task-Centric Perspective of the SOS Algorithm}: Although presenting the algorithmic flow from the task-centric perspective is useful in understanding the overarching flow of the greedy SOS, 
it does not inherently translate well into a dedicated hardware scheduling architecture; Though functionally, jobs arriving and then leaving scheduled to a machine is the primary goal of a scheduler, {\em this model does not consider the data flow of the persistent internal states} which inform the majority of the schedulers cost decisions. While we did this for our previous work in $\hercules$ ~\cite{HERCULES}, it leads to poor memory usage and 
routing congestion leading to a degraded overall performance (\cf,~\Cref{sec:bottlenecks}). 

\subsubsection{Re-Framing Stochastic Scheduling Algorithmic Flow from the Virtual Schedules Perspective}\label{sec:virt_sched}



For our new model, which is shown in ~\Cref{fig:STANNIC_alg_flow}, we instead consider the virtual schedules perspective as the scheduling operation is performed. This re-framing shifts the focus away from the transient elements of the algorithm (namely the scheduled jobs) and instead shifts it onto the persistent elements, the intermediary virtual schedules, enabling 
much more efficient utilization of memory elements and reducing routing congestion due to enhanced 
spatial/temporal locality. In this re-framed version, we still maintain the notions of the three phases of scheduling. However, instead of the flow being a pipeline where jobs come in and the assignment of that job is returned, this new flow projects 
the algorithm as a cyclical one, in which possible sources of alterations (\eg, a new job $J$ being assigned to machine $M$) 
to the virtual schedule 
are stepped through and considered, before finally being written back to establish the state of the virtual schedules for the next cycle. 

In adapting this perspective for the algorithmic flow, we place a higher emphasis on the tracking and coherency of the persistent values that are necessary for every cycle of operation, including cycles in which a new job does not arrive. Using this model as the foundation for a new $\mu$architecture design, we optimize the memory management of associated 
computations, thereby greatly reducing the spatial and temporal management overheads and enhancing scheduler performance.

\begin{wrapfigure}[22]{r}{2.5in}
    \centering
    \includegraphics[scale=0.2]{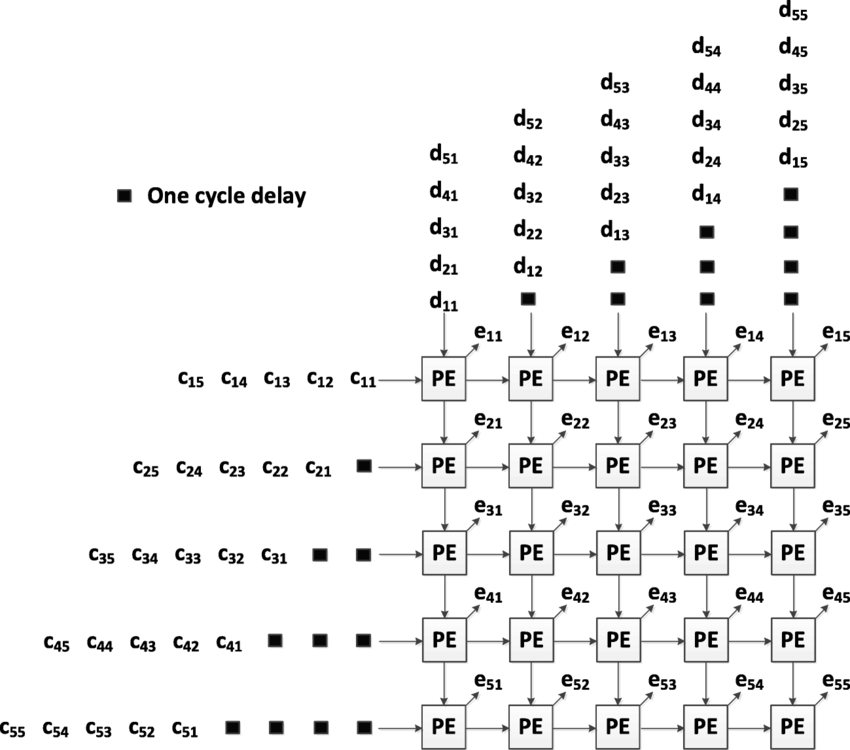}
    \caption{{\bf Systolic Flow Example.} A $5 \times 5$ systolic array for matrix multiplication~\cite{Systolic_Ex}. Each PE is responsible for accumulating the value of its corresponding index in the output matrix $e$. Each PE cascades row data from input matrix $c$ to their right neighbor and column data from input matrix $d$ to their downward neighbor. \copyright Springer Nature. 
    }
    \label{fig:syst_arr_example}
\end{wrapfigure}

\subsection{Systolic Architectures}\label{sec:Systolic_Bkgr}

Systolic Architecture is a parallel computational architecture realized with multiple, identical, and simple 
PEs 
that rhythmically process and pass data mimicking the regularity of blood flow in heart \cite{Intro_to_Systolics}. 
These PEs are interconnected with their immediate 
neighbors, and as such are particularly useful for the acceleration of tasks with high spatial data interdependence. Each PE can perform local processes, and then ``pump'' their data along to other PEs as necessary for the completion of a given task (\cf,~\Cref{fig:syst_arr_example}). 
To prototype the Virtual Schedule perspective of the SOS algorithm, we infuse systolic architecture in $\stannic$. 

Systolic architectures have been successfully used in several, diverse acceleration applications. Google's Tensor Processing Units (TPUs) utilize a systolic architecture for acceleration of matrix multiplication in neural networks \cite{TPU}. Another neural network example is Eyeriss \cite{Eyeriss}, which specifically focuses on energy efficiency by maximizing spatial memory reuse during convolution. Yu et al.,  used a systolic architecture for DNA pattern matching~\cite{DNASystolic}, while \cite{HoughTransformer, VisionTransformer} both use a systolic architecture for efficient pattern recognition and vision transformers 
aiming for real-time execution~\cite{HoughTransformer} and  
power efficiency~\cite{VisionTransformer}. Additionally, recent design automation  tools such as SuSy \cite{SuSy} and Allo \cite{allo2024chen} provide foundational 
support to generate 
optimized systolic designs. In each of these works, a spatial data dependency is leveraged to achieve acceleration.


As discussed in \Cref{sec:virt_sched}, the SOS 
maintains ordered $V_i$'s over time in successive iterations
while the scheduler is active, and in {\bf Phase II} of the algorithm, 
the relative ordering of incoming job $J$ effects the cost contributions of all other jobs $K$ in $V_i$. As such, for \stannic, we propose a systolic architecture that exploits this spatial relationship to optimize scheduling time. 

\section{Computational Mathematics of SOS 
in Continuous and Discrete Time}\label{sec:sos_overview}


In this section, we first elaborate 
the cost computation for machine assignment for continuous time (\Cref{sec:CostCalMath}). 
Next, we explain the enhancements and approximations to extend the cost computation to discrete time, making it amenable for hardware implementation (\Cref{sec:discretization}).



\begin{align}
    \iota _K (t_J) = 1 - \dfrac{1}{K.\EPT_i} \cdot \overbracket{\int_{0}^{t_J} F_{K}^{V_i}(s)\, ds}^{\Omega},~K\in V_i \label{eqn:int_virt_over_expexted} \\ 
    F_{K}^{V_i}(s) = \begin{cases}  
                        1, & \mbox{if}~J \in Head.V_i~\mbox{at time}~s \\
                        0, & \mbox{otherwise}
                     \end{cases}    \nonumber
\end{align}


\subsection{Cost Computation in Continuous Time}\label{sec:CostCalMath}

The notion of {\bf Virtual Work} ($VW$) is crucial for computing the cost of scheduling a job in a machine. Intuitively, $VW$ {\em captures the amount of time a job} $K$ has spent at the head of the Virtual Schedule ($Head.V_i$). The $\Omega$ in~\Cref{eqn:int_virt_over_expexted} captures the amount 
of Virtual Work of job $K$ completed at time $t_J$ (\ie, when job $J$ is created) 
and $\iota_K(t_J)$ represents the remaining fraction of Virtual Work 
of $K$. 
The $VW$ 
is directly related to the $\alpha_J$ release point of that assigned job, as $\alpha_J$ sets the percentage threshold of completed Virtual Work at which the job is released (\ie, 
Phase III in \Cref{fig:alg_flow_task_perspective}).

The cost of scheduling job $J$ in machine $M_i$, $cost(J \rightarrow M_i)$, denoted as $cost$ for brevity, is computed as follows~\cite{jager2023improved}.


\begin{align}
\begin{split}
    cost =& \overbracket{(J.W) \cdot \left( J.\EPT_i +
    \sum_{K \in V_i,~T_i^K \geq T_i^J} \iota_K(t_J) \cdot K.\EPT_i\right)}^{cost^{H}} 
    + \overbracket{\sum_{K \in V_i,~T_i^K < T_i^J}
    K.W \cdot \iota_K(t_J) \cdot J.\EPT_i}^{cost^L}
    \label{eqn:cost_equation}
\end{split}
\end{align}



The $cost$ in~\Cref{eqn:cost_equation} has two parts -- $cost^H$ and $cost^L$. $cost^H$ captures 
the set of jobs ($\sigma^{H}$) in $V_i$ whose WSPT ratio is higher than or equal to the WSPT of $J$ and $cost^L$ captures the set of jobs ($\sigma^{L}$) in $V_i$ whose WSPT ratio is lower than or equal to the WSPT of $J$. $\sigma^{H}$ would delay the start of the job $J$ as they have the higher WSPT priority whereas $\sigma^{L}$ will be delayed by $J$ as they have lower WSPT priority. This splitup of jobs in two sets is crucial 
to the performance of the cost calculation 
and non-trivial as the two sets of jobs affect the cost computation differently. Note that both $cost^H$ and $cost^L$ include the term \(\iota_K(t_J)\). 
Intuitively, w.r.t. 
cost calculation, 
\(\iota_K(t_J)\) correlates to the percentage of job $K$'s wait time is left before being released from $V_i$, by tracking how much of $K.\hat\ept_i$ it has waited. 
Inclusion of \(\iota_K(t_J)\) reduces  
delay incurred by the previously assigned job $K$ 
onto the new job $J$ (or vice versa), by this ratio. This is due to $K$ being closer to release from $V_i$, and thus incurring a reduced 
delay. 

\subsection{Cost Computation in Discrete Time}\label{sec:discretization}

A key necessity to port the SOS algorithms 
to digital hardware (\eg, FPGA) is to discretize certain parameters such as time. This modification leads to a considerable reduction in the cost computation complexity 
resulting in a simpler yet high-efficiency hardware 
design with reduced logic 
footprint. 





Quantizing time allows to rewrite the integration ($\Omega$) in~\Cref{eqn:int_virt_over_expexted} as  $n_K(t_J) = \sum_{0}^{t_J} F_{K}^{V_i}(t_J)$, where $n_K(t_J)$ represents the 
number of cycles a job $K$ has performed Virtual Work in $V_i$. We track and update $n_K(t_J)$ in every clock cycle due to its importance in cost calculation and $\alpha_J$ release point determination. Such 
update forgoes detailed job tracking (\eg, when a job was added in the $V_i$), lengthy summations to compute $n_K(t_J)$, and complex reconstruction of $V_i$ every time a new job is added to it in favor of a singular lookup. Substituting $n_K(t_J)$ for the integration in 
~\Cref{eqn:int_virt_over_expexted}, the remaining fraction of virtual work simplifies as follows.  

\begin{align}
    \hat\iota _K (t_J)= 1 - \dfrac{n_K(t_J)}{K.\EPT_i},\ K \in V_i
    \label{egn:n_virt_over_expected}
\end{align}

Substituting $\hat\iota _K (t_J)$ in~\Cref{eqn:cost_equation}, 
$cost^H$ and $cost^L$ simplify as follow.

\begin{align}  
    cost^H = (J.W) \cdot \ (J.\EPT_i + \underbracket{\sum_{K \in V_i,\ T^K_i \geq T^J_i} \overbracket{(K.\EPT_i - n_K(t_J)}^{sum^H}}_{sum^{HI}}))
    \label{eqn:costh_discretized} \\
    cost^L = J.\EPT_i \cdot \ \underbracket{\sum_{K \in V_i,\ T^K_i < T^J_i} \overbracket{\left(K.W - n_K(t_J) \frac{K.W}{K.\EPT_i}\right)}^{sum^L}}_{sum^{LO}} \label{eqn:costl_discretized}
\end{align}




\noindent \bem{Remark}: Although we are subtracting terms in $sum^H$ and $sum^L$, we do not 
risk of having a previously assigned job $K$ contributing a negative cost to a potential job's calculation. For either $sum^H$ or $sum^L$  
to reduce to 0, $n_K(t_J)$ would have to equal $K.\EPT_i$. However, with the 
$\alpha_J$ release policy, $K$ will be released from $V_i$ either at or before this point, as the job releases when $n_K(t_J)\geq \alpha_J \cdot K.\EPT_i,\text{and }\alpha_J\in (0,1]$.

\subsection{Additional Design-Based Optimizations}\label{sec:cost_optimization}


We optimize~\Cref{egn:n_virt_over_expected,eqn:costh_discretized,eqn:costl_discretized} further to optimize the computation for efficient hardware implementation.

\begin{enumerate}
    \item {\bf Reductions in Division Operations}: We store $T^K_i = K.W/K.\hat{\epsilon_i}$ to reuse it to compute $cost^L$ and when comparing $T^J_i\text{ to }T^K_i$ to sort $K$ in $cost^H$ and $cost^L$ for cost calculation. 
    Additionally, the earliest possible time to calculate $T^K_i$ is when job $K$ is first created and $cost(K \rightarrow M_i)$ is calculated. 
    When $J$ is assigned to $V_i$, we store $T^J_i$ until $J$ is released from $V_i$. When combined, these optimizations save numerous computationally costly division operations. 


    \item {\bf Incremental Update for Virtual Work}: In addition to $n_K(t_J)$, the $sum^H$ and $sum^L$ 
    solely rely on the attributes of the job $K \in V_i$. Note $n_K(t_J)$ is essentially a cycle count since $K$ has started its Virtual Work, requiring frequent updates. A key observation is that all other attributes of $K$ (\eg, $\hat{\epsilon_i}$) are constants when $n_K(t_J)$ is updated. Consequently, we {\em initialize} $sum^H$ to its maximum value of $K.\hat{\epsilon_i}$ and {\em decrement} it by {\bf 1} in every cycle $K$ is virtually worked on. For $sum^L$, we initialize it to its maximum value of $K.W$ and decrement it by $T^K_i$ (note $T^K_i = K.W/K.\EPT_i$ is the WSPT of $K$ in machine $M_i$). These set of optimizations save considerable amount of lengthy summations and divisions contained in $sum^H$ and $sum^L$ of~\Cref{eqn:costh_discretized,eqn:costl_discretized} 
    making it faster and amenable for hardware implementation. It is worth noting that updating of $sum^H$ and $sum^L$ happens 
    in parallel with the 
    $\alpha_J$ release checks, {\em overlapping} 
    the processing time of these updates, and preventing the need for explicit evaluation across each job $K$ when $cost(J\rightarrow M_i)$ is computed.
\end{enumerate}

\medskip

\noindent \bem{Remark.} These are the 
optimizations utilized in both $\hercules$ anmd $\stannic$ 
$\mu$architectures. \stannic\ further leverages the systolic organization and spatial ordering of jobs $K$ within each $V_i$ to precalculate $sum^{HI}$ (\cf,~\Cref{eqn:costh_discretized}) and $sum^{LO}$ (\cf,~\Cref{eqn:costl_discretized}) for all possible separations of the high and low WSPT sets. The exact systolic mechanics that enable and maintain these pre-calculations will be discussed 
in ~\Cref{sec:Systolic_SOSA}.

\section{$\hercules$: Task-centric Hardware Implementation of SOS Algorithm
}\label{sec:asic_design}



In this section, we discuss in detail the $\mu$architectural components of $\hercules$, the task-centric perspective (\cf,~\Cref{sec:task_centric}) hardware implementation of SOS algorithm. 


\begin{figure}
    \centering
    \includegraphics[scale=0.3]{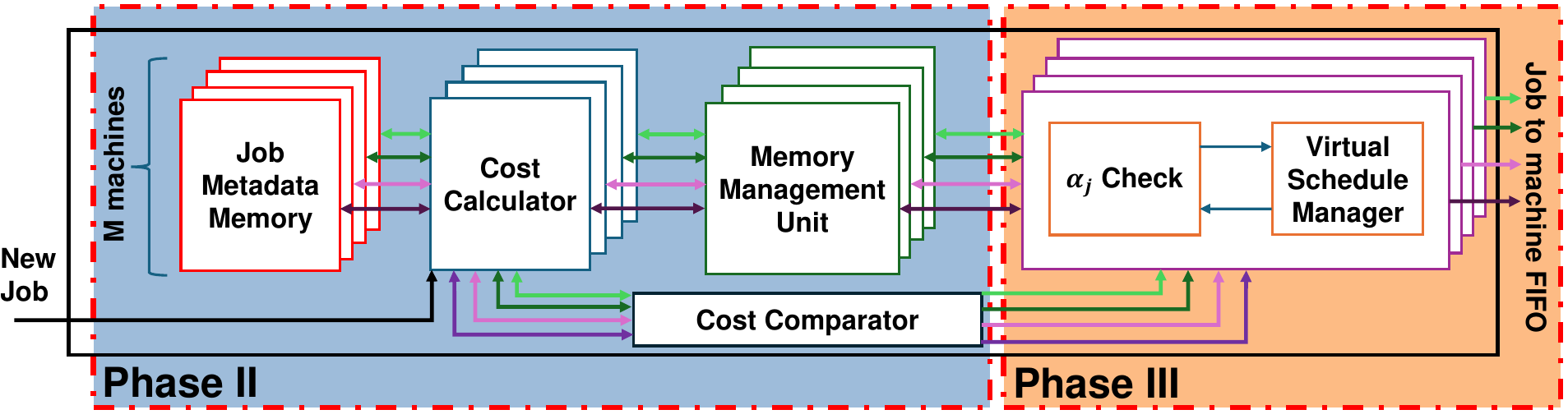}
    \caption{{\bf Top-level block diagram of the $\hercules$ scheduler}. Phase II and III are the phases shown in ~\Cref{fig:alg_flow_task_perspective}.}
    \label{fig:top_level_block}
\end{figure}

\subsection{\hercules\ $\mu$architecture}\label{sec:micro_arch}

\Cref{fig:top_level_block} shows the $\mu$architectural 
design of the scheduler in \hercules. The scheduler 
implements Phase II and III of 
~\Cref{fig:alg_flow_task_perspective} 
to identify the machine with the lowest compute cost and release the job to the machine at the 
designated $\alpha_J$ point. To compute the $cost$ 
and track job progress in machine $M_i$, 
the following attributes for all jobs in the virtual schedule $V_i$ needs to 
be retained -- 
(i) $J.W$, (ii) $J.\hat{\epsilon_i}$ 
(iii) WSPT ratio ($T_i^J$), 
and (iv) $\alpha_J$ point until each job is released for execution. After a job is released, it no longer contributes to the cost calculation, and its metadata can be safely discarded by the scheduler. The Virtual Schedule must be updated in two events -- (1) when a job is released for execution 
and (2) when a new job is scheduled. 
To perform these updates, the scheduler must track the job at the head of the Virtual Schedule ($Head.V_i$). 
{\bf Job Metadata Memory} stores the job metadata and $sum^H$ and $sum^L$; the {\bf Cost Calculator} interacts with Job Metadata Memory and computes $cost(J\rightarrow M_i)$
; the {\bf Cost Comparator} identifies the minimum-cost machine for job $J$;  the {\bf Memory Management Unit} acts as a gatekeeper to this metadata to ensure consistent and efficient read/write access; the {$\bm \alpha_J$} {\bf Check} module determines whether a job has reached its $\alpha_J$ scheduling threshold and is eligible for execution; and the {\bf Virtual Schedule Manager}  maintains the ordering of the jobs in the Virtual Schedule. In the next few subsections, we detail each $\mu$architectural block.

\begin{figure}
    \centering
    \includegraphics[scale=0.4]{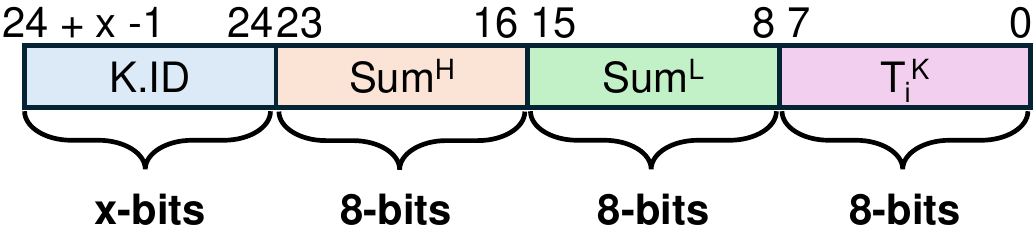}
    \caption{{\bf Job Metadata Memory register implementation}. {$x$}: Configurable based on the maximum number of jobs across all machines computed as $\ceil{log_2 (M \times N)}$. {\bf M}: Number of machines. {\bf N}: Max. number of jobs in $V_i$ of machine $M_i$. This leads to a total register width of $x+24$ bits.}
    \label{fig:metadata_register}
\end{figure}

\subsubsection{Job Metadata Memory (JMM)}\label{sec:jmm}

The JMM is implemented as an $M \times N$ register array, where $M$ is the number of machines and $N$ is the maximum number of jobs that can 
reside in the $V_i$. 
A key insight is that each job's metadata must be accessed in every cycle for cost updates and scheduling decisions. A 
RAM-based implementation would impose limitations on simultaneous read and write via limited number of memory ports and would add considerable access latency, thereby severely degrading scheduler performance. {\em To avoid the performance bottleneck}, we use a fully register-based implementation of JMM as shown in~\Cref{fig:metadata_register}. Each register is $24 + x$ bits wide (\cf, \Cref{fig:metadata_register}), where $x = \ceil{log_2 (M \times N)}$. Each job attribute is 8 bits wide. We discuss the rationale for selecting 8-bit wide attributes in~\Cref{subsec:quantization}. 



\begin{figure}
    \begin{subfigure}[b]{0.45\textwidth}
        \centering
        \includegraphics[width=\textwidth]{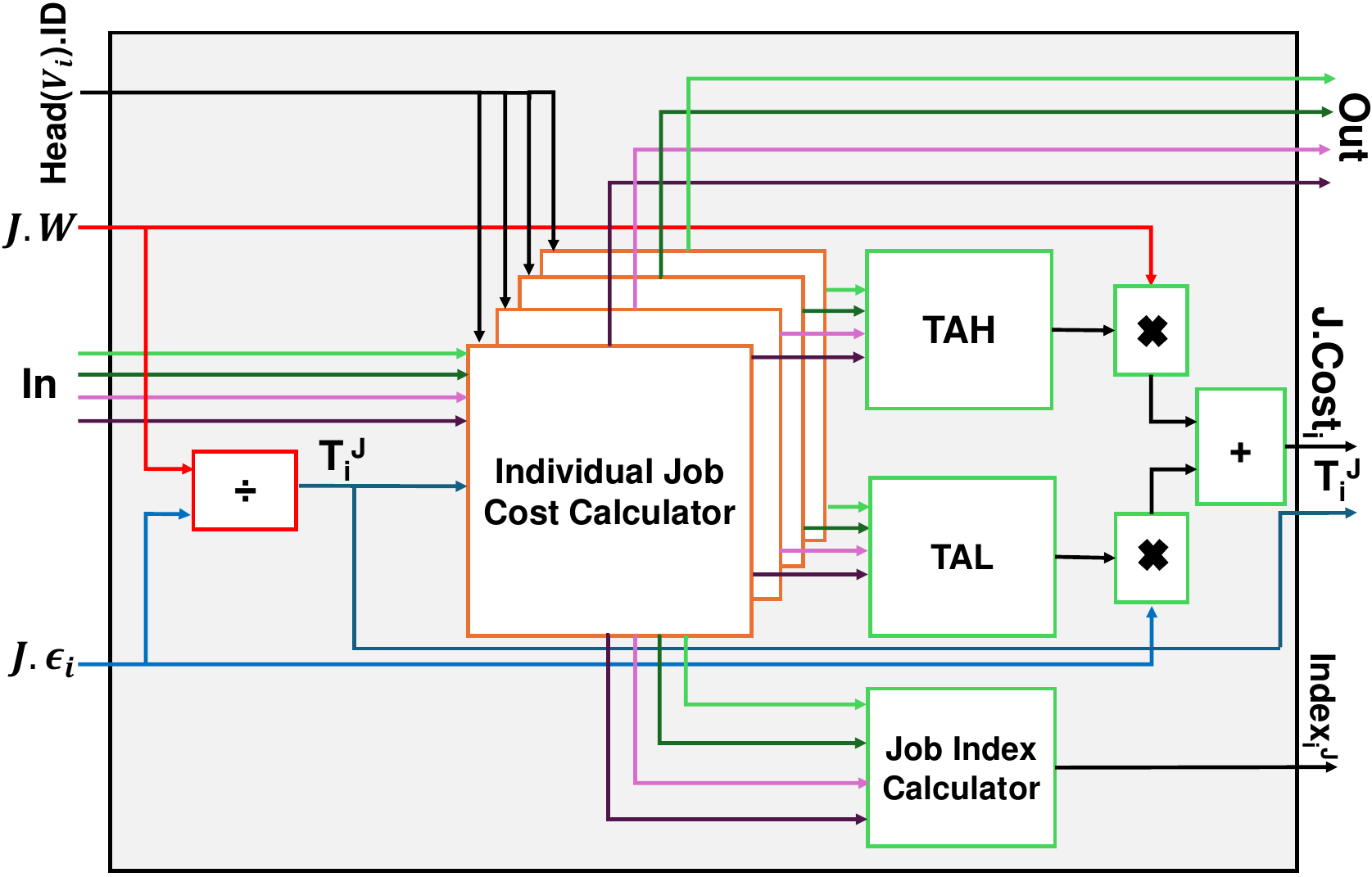}
        \captionsetup{justification=centering}
        \caption{\label{fig:cost_calculator}}
   \end{subfigure}
   \begin{subfigure}[b]{0.45\textwidth}
        \centering
        \includegraphics[width=\textwidth]{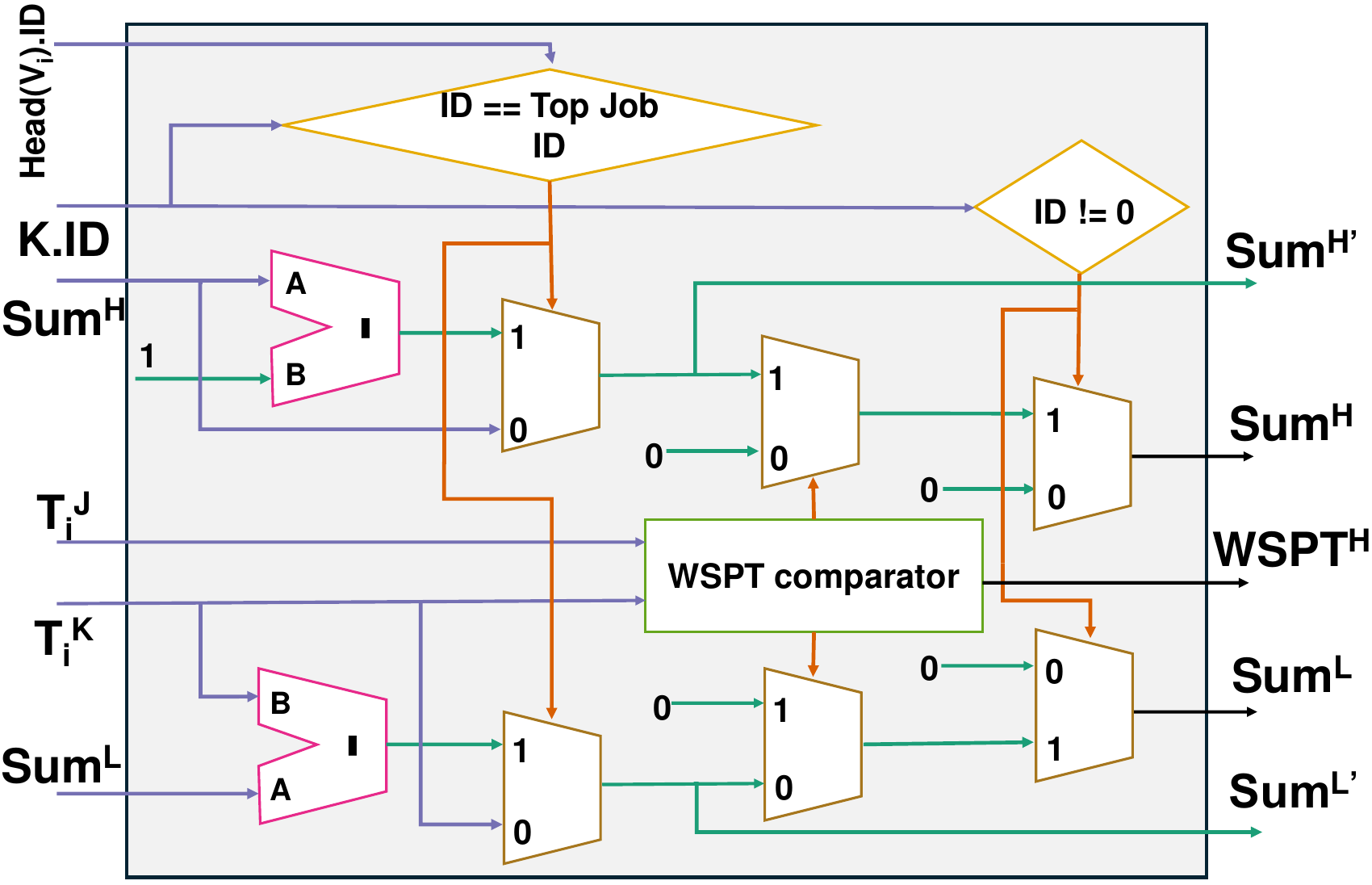}
        \captionsetup{justification=centering}
        \caption{\label{fig:individual_job_cost_calculator}}
   \end{subfigure}   
   \begin{subfigure}[c]{0.45\textwidth}
        \centering
        \includegraphics[width=\textwidth]{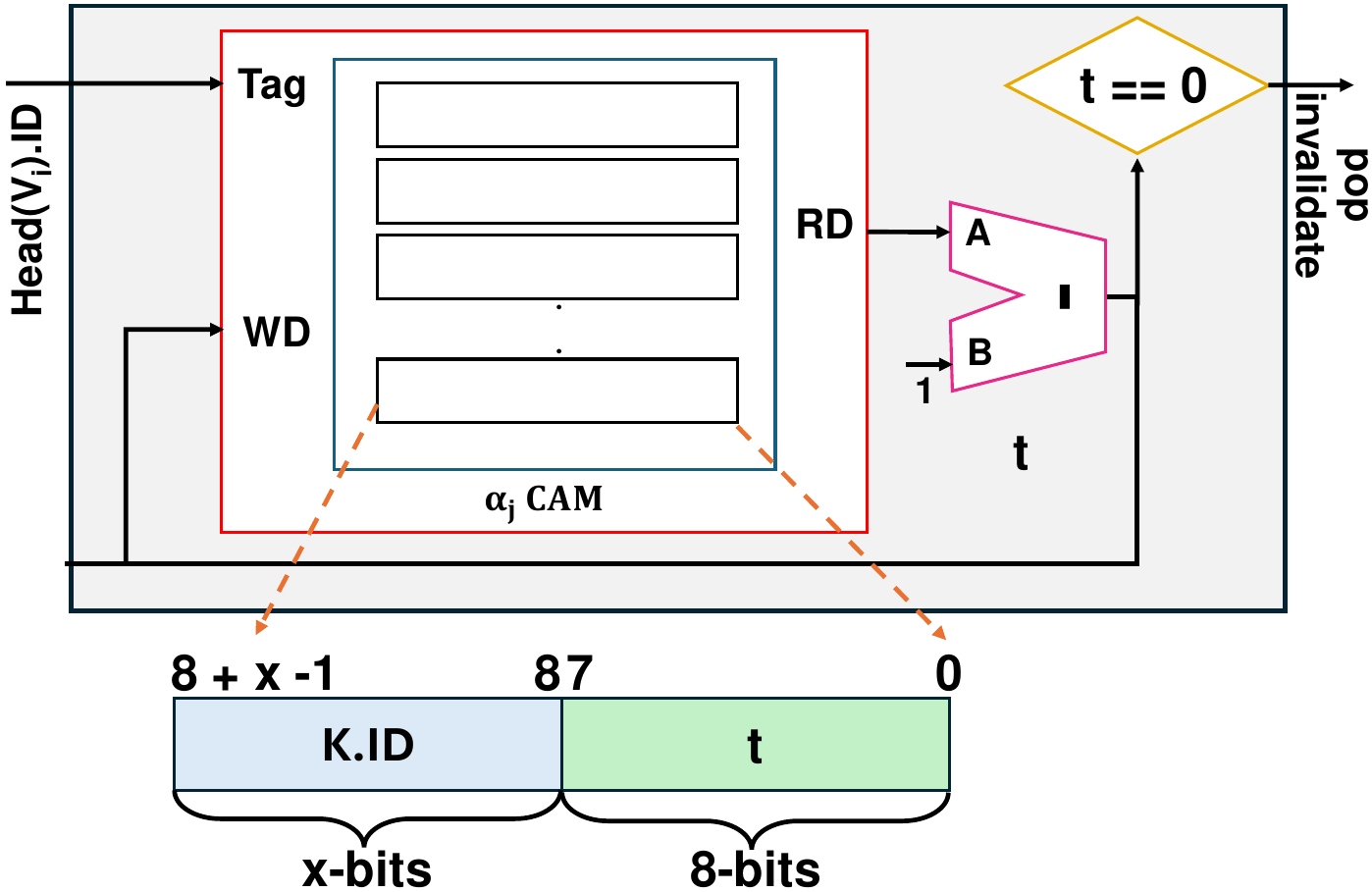}
        \captionsetup{justification=centering}
        \caption{\label{fig:alpha_j_check}}
   \end{subfigure}
   \begin{subfigure}[c]{0.45\textwidth}
       \centering
       \includegraphics[width=\textwidth]{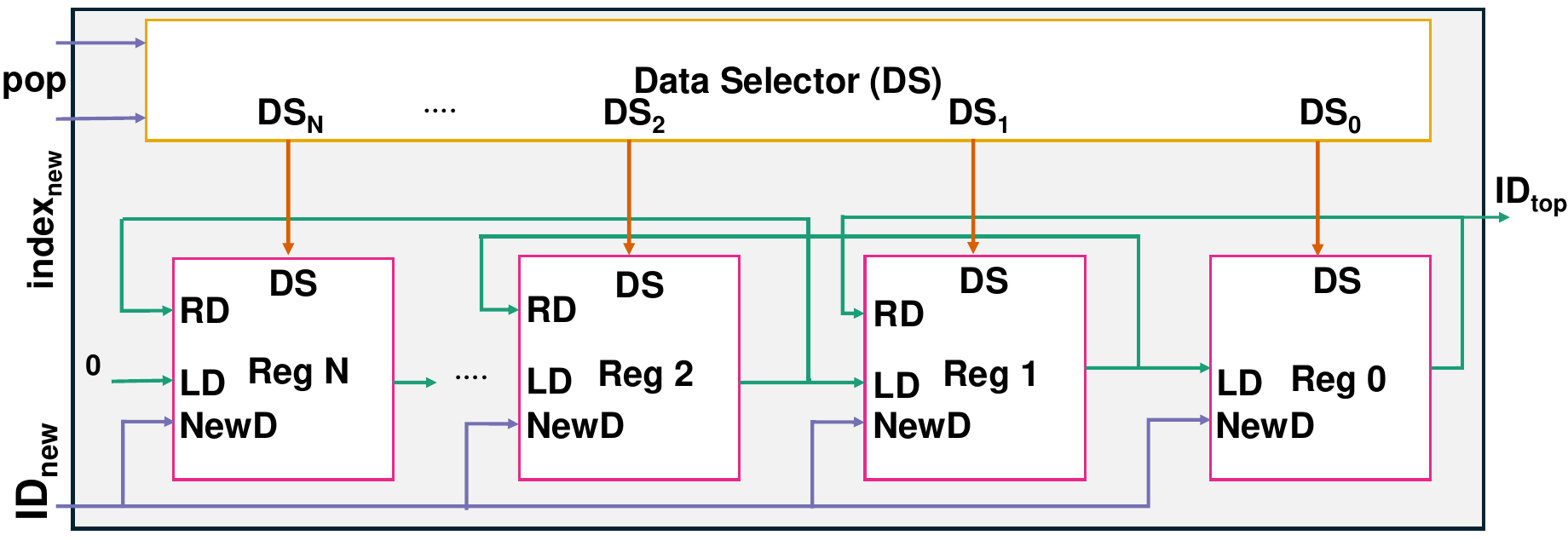}        
            \caption{ 
            \label{fig:virtual_fifo}}
   \end{subfigure}
   \caption{{\bf $\mu$architectural components of $\hercules$}. (a): {\bf Cost Calculator}. {\bf TAH}: Tree Adder to compute $cost^H$. {\bf TAL}: Tree Adder to compute $cost^L$. {\bf N}: \# of jobs in each machine. {\bf In}: \{K.ID, $\bm{sum^H}$, $\bm{sum^L}$, $\bm{T_i^K}$\} $\times N$. {\bf Out}: \{$\bm{sum^H}$, $\bm{sum^L}$\} $\times N$.~(b): {\bf Individual Job Cost Calculator}.~(c): {\bf $\alpha_J$ check module}. {\bf CAM}: Content Addressable Memory. $K.ID$ is used as the tag for content matching and data retrieval.~(d) {\bf Virtual Schedule Manager}
   }
\end{figure}

\subsubsection{Cost Calculator (CC)}\label{sec:arch_cost_calculator}

~\Cref{fig:cost_calculator} shows the $\mu$architecture of the CC to compute scheduling cost. 
The inputs to the CC include the metadata of 
jobs currently scheduled in the machine and the weight ($J.W$) and EPT ($J.\hat{\epsilon}_i$) of the new job. 
The outputs of the CC are -- (1) the updated values of $sum^H$ and $sum^L$ for all jobs in 
$M_i$, (2) the cost of assigning the new job, (3) the WSPT ratio of the new job, and (4) its index in the $V_i$ based on WSPT ratio comparison. Additionally, the CC updates job-specific costs, $sum^H$ and $sum^L$, and the index of the new job in the $V_i$ 
based on the WSPT comparison. The $sum^H$ and $sum^L$ are stored in the JMM for future computations, and 
the cost 
and job index are forwarded to the Cost Comparator. Each machine is equipped with 
a 
CC to concurrently compute 
cost 
for a new job across all machines within a single cycle. 

A key observation from~\Cref{sec:discretization} is that the $sum^H$ and $sum^L$ can be computed in parallel. 
To exploit this parallelism, the CC includes up to $N$ instances of {\em Individual Job Cost Calculator} (\cf,~\Cref{arch:individual_cost}). We choose {\em Tree Adders to minimize computation latency by enabling single-cycle summation}. Each Tree Adder consists of $N - 1$ adders arranged in $\log_2 N$ stages. 
Although an accumulator-based design would reduce area, it would require multiple cycles per computation, thus degrading scheduler performance. The Tree Adder provides an optimal trade-off between the scheduler performance and hardware cost. We use two Tree Adders per CC, one for $sum^H$ (TAH) and another for $sum^L$ (TAL). We multiply output of TAH by the weight of new job to compute $cost^H$ and output of TAL by the expected processing time of new job to compute $cost^L$.  The \textbf{Job Index Calculator} acts as a {\tt popcount}~\cite{popcount} to compute the number of 1's in its input. 



\subsubsection{Individual Job Cost Calculator (IJCC)}\label{arch:individual_cost}

~\Cref{fig:individual_job_cost_calculator} shows the $\mu$architecture of the IJCC. Each job in $V_i$ contributes either to the $cost^H$ or the $cost^L$ based on its WSPT classification relative to the new job (\cf,~\Cref{sec:CostCalMath}). However, the IJCC computes both $cost^H$ and $cost^L$ and masks out irrelevant cost term as needed. Specifically, the $cost^H$ output is zero ({\bf 0}) if \bem{either} 
the new job has an invalid ID (\ie, no job is present), \bem{or} 
the WSPT of the job under consideration is lower than that of the new job. Similarly, $cost^L$ is set to zero ({\bf 0}) if the WSPT of the job under consideration is greater than that of the new job. Incorporating this decision logic within the IJCC eliminates the need for additional job-specific condition checks from CC and propagation of job attributes to other scheduler components, reducing routing congestion and resource utilization while improving scheduler performance.
Additionally,  IJCC computes $sum^H$ and $sum^L$. 
The job at the head of $V_i$ performs $VW$, requiring modification of its attributes. To enforce this, each job's ID is compared with the ID of the job at $Head.V_i$. If the IDs match, the updated values are written back to the JMM. Otherwise, the original values are preserved. The output of the WSPT comparator is {\bf 1} when $T^K_i \ge T^J_i$ and is forwarded to the Job Index Calculator. 


\subsubsection{Memory Management Unit (MMU)}\label{arch:mmu}

The primary function of the MMU is to manage access to the JMM. MMU acts as a bridge between Phase II and Phase III of the scheduler aiding each scheduler component to access necessary job metadata information quickly. A dedicated MMU helps to write the new job metadata quickly at a free JMM location instead of time-consuming 
search. 
MMU maintains two 
data structures -- (1) a lookup table (LUT) that maps each Job ID to its metadata address and (2) a FIFO of free memory addresses. The LUT is used during metadata invalidation. 
A Job's metadata is discarded 
upon receiving an \texttt{invalidate} signal from the $\alpha_J$ Check and its address is queued in the FIFO for future use. 
When a new job is scheduled, the CC requests a free metadata address from the MMU. The MMU responds by popping an available address from the FIFO and returning it to the CC. 

\subsubsection{Cost Comparator (CR)}\label{sec:cost_comparator}

The CR compares the costs 
across machines and sends the new job's ID and its index in $V_i$ 
to $\alpha_J$ check module. The CR also informs the CC of the machine selected, which 
in turn interacts with the MMU to find the next available memory address and pass this information on to JMM to store the new job's metadata. 





\subsubsection{$\alpha_J$ Check (AC)}\label{sec:alphaj_check}

\Cref{fig:alpha_j_check} shows the architecture of the $\alpha_J$ check module. AC tracks the amount of time (calculated as $t = \alpha_J \cdot \hat{\epsilon_i}$) a job $J$ spends at $Head.V_i$ 
and decrements it by one ({\bf 1}) every clock cycle. Once the counter reaches zero ({\bf 0}), the job is popped from the Virtual Schedule Manager (VSM) and sent to the designated machine. The AC consists of a Content Addressable Memory (CAM) of size $N$ with job IDs as the tag and 
$t$ as the content. 
When a job is popped from the CAM, the corresponding entry is invalidated in the MMU and the job is also popped from VSM. Intuitively, using a CAM enables to dynamically reorder the jobs as per the WSPT values with minimal computational overhead. Note, a new job $J$ may replace the job at  $Head.V_i$ if $J$'s WSPT is higher than that of the head job, requiring a job reordering. 


\begin{figure*}
    \centering
    \begin{subfigure}[c]{0.4\textwidth}
      \resizebox{\textwidth}{!}{
        \begin{tabular}{|c|c|c|c|c|c|}
            \hline
            \textbf{Precision} & \textbf{Weight } & $\bm{\alpha_J}$ & $\bm{\epsilon}$ & \textbf{WSPT} & \textbf{Cost}\\
            \hline\hline
            {\tt INT4} & 4-bit  & 4-bit  & 4-bit  & 4-bit  & 8-bit   \\
            \cellcolor{green!50}{\tt INT8} & \cellcolor{green!50}{8-bit}  & \cellcolor{green!50}{8-bit}  & \cellcolor{green!50}{8-bit}  & \cellcolor{green!50}{8-bit}  & \cellcolor{green!50}{16-bit}  \\
            {\tt Mixed} & 4-bit  & 4-bit  & 8-bit  & 8-bit  & 16-bit  \\ 
            {\tt FP16} & 16-bit  & 16-bit  & 16-bit  & 16-bit  & 16-bit  \\
            {\tt FP32} & 32-bit  & 32-bit  & 32-bit  & 32-bit  & 32-bit  \\
            \hline
        \end{tabular}
        \vspace{8mm}
        }
    \caption{}
    \label{tab:precision_explanation}
    \end{subfigure}
    \hspace{6mm}
    \begin{subfigure}[c]{0.4\textwidth}
        \centering
        \includegraphics[width=\textwidth]{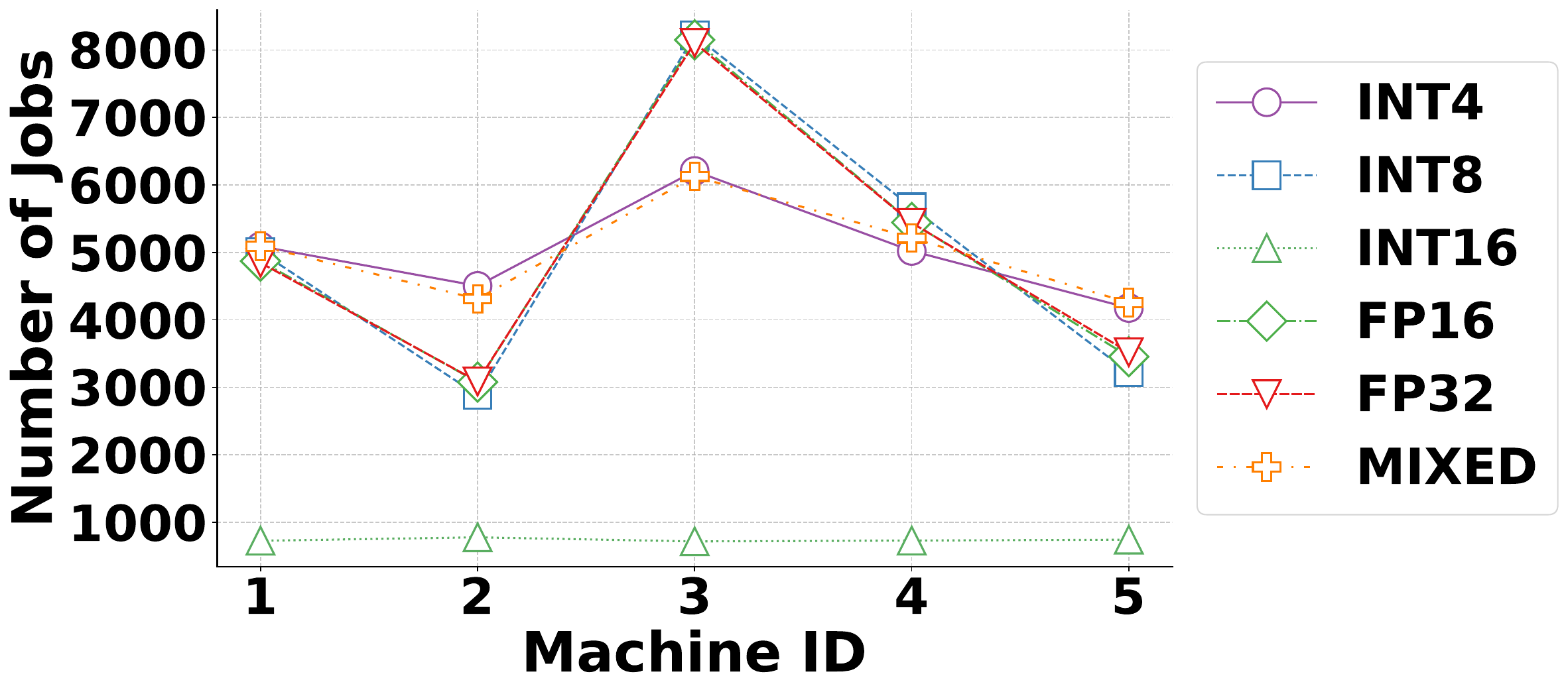}
    \caption{}
    \label{fig:job_distribution}
    \end{subfigure}
    \begin{subfigure}[t]{0.4\textwidth}
        \centering
        \includegraphics[scale=0.2]{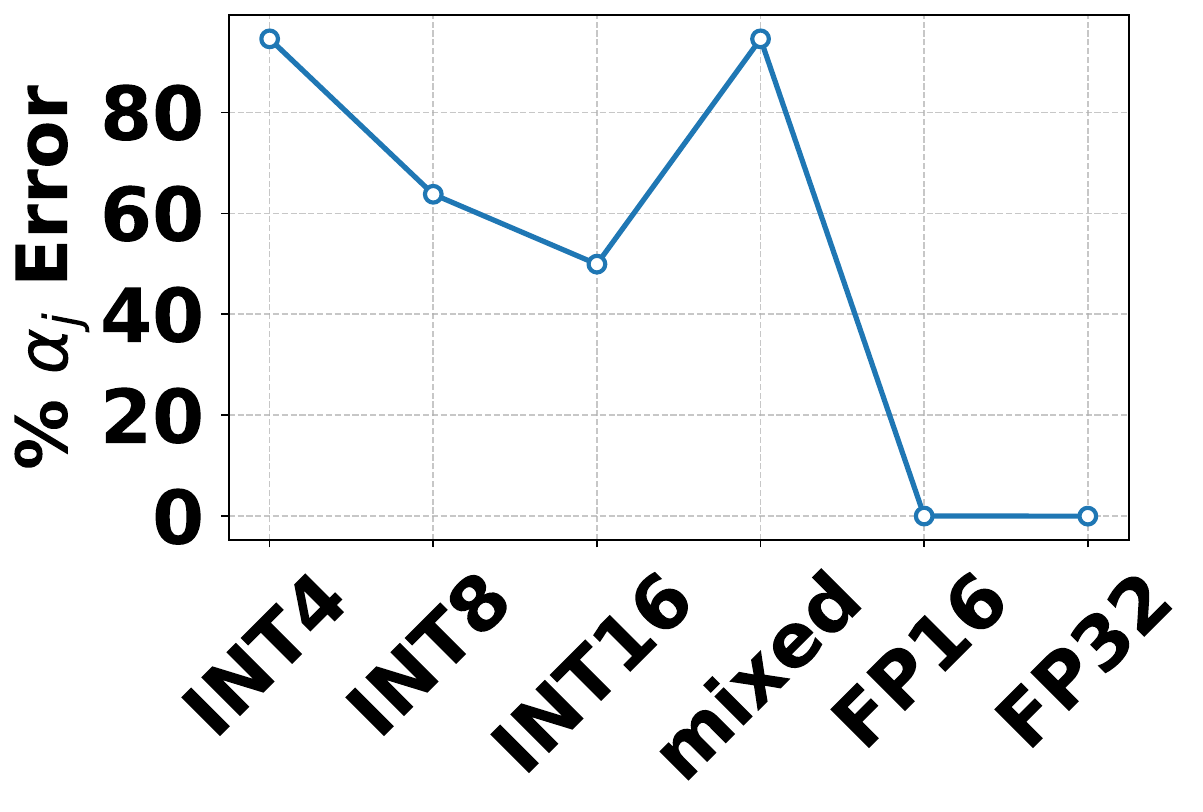}
    \caption{}
    \label{fig:alpha_j_error}
    \end{subfigure}   
    \hspace{6mm}
    \begin{subfigure}[t]{0.4\textwidth}
        \centering
        \includegraphics[scale=0.2]{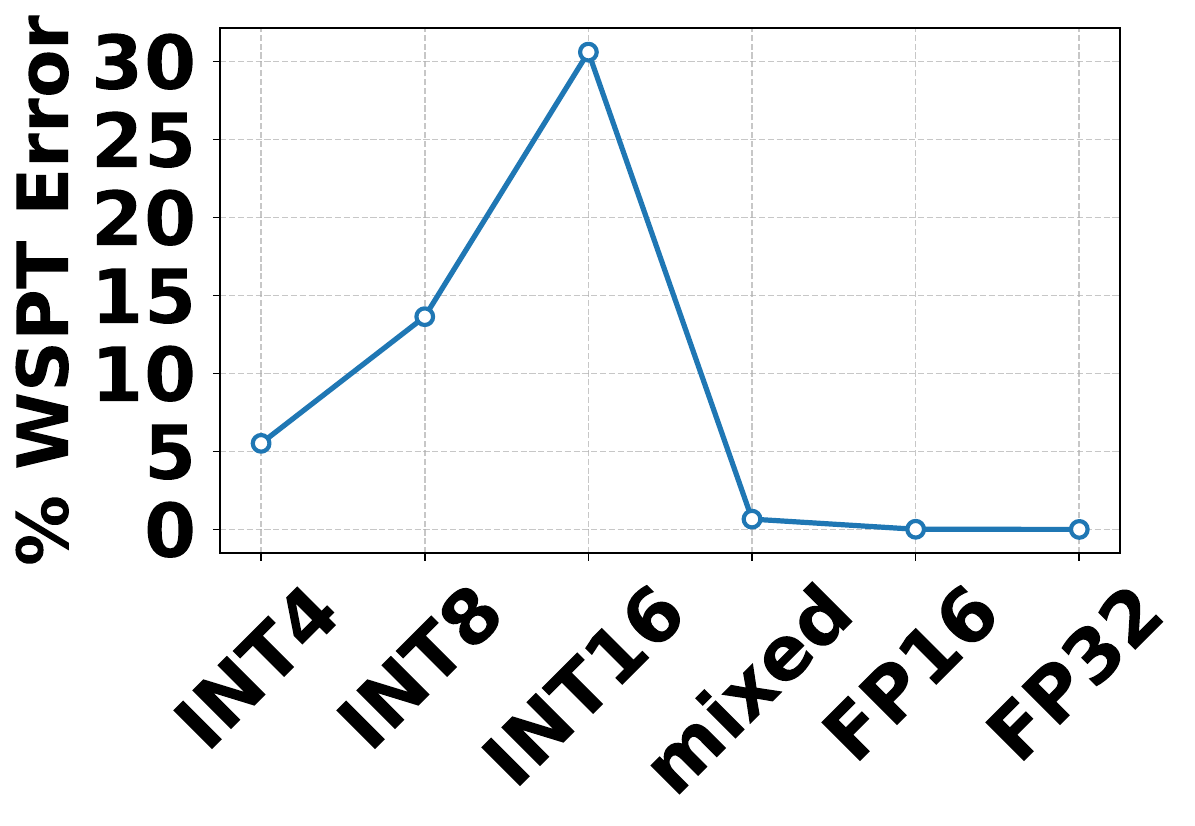}
    \caption{}
    \label{fig:wspt_error}
    \end{subfigure}   
    \caption{(\subref{tab:precision_explanation}): {\bf Various quantization techniques applied to each job attribute to evaluate effect of quantization on final schedule computation}. 
    Green highlights the most suitable quantization. 
    ~(\subref{fig:job_distribution}): {\bf Scheduled job distribution in each machine}.~(\subref{fig:alpha_j_error}): {\bf \% Error in $\alpha_J$}.~(\subref{fig:wspt_error}): {\bf \% Error in WSPT}. 
    }
    \label{fig:quantization_results}
\end{figure*}



\subsubsection{Virtual Schedule Manager (VSM)}\label{sec:vsm}

\Cref{fig:virtual_fifo} shows the 
architecture of the VSM which maintains the ordered list of jobs scheduled on a given machine. On receiving a {\tt pop} 
from AC, VSM releases the head job to the designated machine. 
We use a configurable shift-register structure, where each register stores the Job ID ($J.ID$) of one scheduled job and supports left shifts, right shifts, and partial shifts, enabling dynamic reordering based on job's arrival and departure. The VSM updates either (a) when a job is released to the machine (departure) or (b) is assigned to the machine (arrival). Note, arrival and departure may occur at the same time. The maximum capacity of VSM is $N$. The job at index $k$ is referred to as $J_k, k \in [0, N - 1]$ and $J_0$ represents the job at $Head.V_i$. 
When a job is released, all remaining jobs are right-shifted to preserve job ordering such that $J_{k - 1} \leftarrow J_k$. When a new job is scheduled, it can be inserted at any index $p \in [0, N - 1]$. To accommodate the new job at position $p$, jobs from $J_p$ to $J_{N - 2}$ are shifted left by one position (\ie, $J_{p + 1} \leftarrow J_p$), while entries before $p$ remain unchanged. The register at index $p$ is then updated with the new Job ID. A full left shift occurs when $p = 0$ (\ie, when WSPT of $J_0$ is lower than the WSPT of new job), and a partial left shift is performed when $p > 0$.








To implement this behavior, each register 
is connected to a \textit{Data Selector} (DS) which chooses one of four inputs -- the job ID from the left or right neighbor, the new job, or the current value (no change). The DS receives the job Index from CC and the \texttt{pop} signal from AC. Based on these inputs, DS generates a control signal for each register to perform the appropriate update. The ID of the job at $Head.V_i$ 
is shared with both AC and CC to coordinate $\alpha_J$ point tracking and cost computation.





\subsection{Quantization
Selection Rationale}\label{subsec:quantization}


The SOS algorithm contains complex mathematical operations that can degrade scheduler performance (\eg, increased dynamic power at a higher job arrival rate) if done at full floating-point precision (\ie, at {\tt FP64}). To ensure that the scheduler exhibits high performance (\ie, scheduling job in near real time) while minimizing area and power consumption, 
the SOS algorithm requires us to operate with reduced numerical precision with modest reduction in scheduling accuracy. 
To identify an appropriate numerical precision, we evaluate SOS algorithm using various numerical precision detailed in~\Cref{tab:precision_explanation}. 

We choose five different machine configurations and varying workload (\cf, \noindent \bem{Workload generation} in~\Cref{sec:exp_setup} for details) to empirically identify the suitable numerical precision. We set the minimum job weight to one ({\bf 1}), and the minimum expected processing time to 10. We choose SOS algorithm's performance at {\tt FP32} as the baseline for identifying the suitable numerical precision for the scheduler.~\Cref{fig:job_distribution} shows that {\tt INT8} quantization closely replicates the {\tt FP32} job distribution. Additionally, we analyzed errors in key job attributes. According to~\Cref{eqn:costh_discretized,eqn:costl_discretized}, cost accuracy is only influenced by the jobs currently scheduled on a machine. 
Hence, errors in a Job's WSPT and $\alpha_J$ can significantly impact scheduling decisions. ~\Cref{fig:wspt_error,fig:alpha_j_error} 
present the 
error in WSPT and $\alpha_J$, respectively. {\tt INT8} exhibits the second-highest WSPT error, while {\tt INT4} and {\tt Mixed}-precision approaches show lower WSPT errors. However, {\tt INT8} demonstrates lower $\alpha_J$ error than {\tt INT4} and {\tt Mixed} quantization. Consequently, the latter schemes 
release jobs for execution earlier than intended, resulting in erroneous cost calculation and increased cost error. 
\bem{These observations form the basis for choosing {\tt INT8} as our preferred precision level}.

\section{Systemic Bottlenecks in \hercules\ $\mu$Architecture}\label{sec:bottlenecks}

We will now re-explore 
design and implementation of \hercules\ and 
correlate different design decisions to various performance bottlenecks. 
Our primary bottleneck concerns relate to the maximum system size \hercules\ could support, which was capped at 10 machines. Similarly, while \hercules\ vastly outperforms a software SOS implementation, individual scheduling took 466 cycles (\cf, \Cref{sec:arch_comp_analysis}). While iterations can overlap, this is still a significant delay to an individual job being assigned to a machine.
In doing this, we will 
identify elements of operation and instantiation that 
need to be enhanced and/or augmented to create an improved scheduling accelerator architecture.

\smallskip

\noindent {\bf Memory Interface}: 
The memory interface of $\hercules$, used to read in job metadata for scheduling and writing out resultant schedules, operates by sending over batches of $\mathcal{X}$ number of jobs at a time, and then returning the information for those 
$\mathcal{X}$ jobs when all of them are completed. This was useful for timing of releases on the scheduler's end, and allowed us to track individual jobs closely throughout their processing. However, this batching method 
imposes 
severe overheads on the scheduling operation --  while waiting for new jobs, we are waiting for the host to read and organize $\mathcal{X}$ jobs at a time, delaying their initial arrival. This method also imposes a hardware overhead in the form of requiring the FPGA to write and track a table $\mathcal{X}$ entries long, where any arbitrary machine needs to be able to write popped jobs to any arbitrary entry in that table, which then needs to be sent over in one go. This imposes both a resource utilization overhead as 
we need to track extra information across cycles,  
but also affects scalability, as each new machine in the configuration increases the number of components that need to be routed to this table.

\smallskip

\noindent {\bf Iterative Cost Comparator}: In \hercules\ the calculated costs for each machine are compared in an iterative, sequential manner. This imposes some control overhead, but more importantly also imposes direct iteration latency costs that correspond to the scaling of 
the configuration, as this comparison operation is computationally bounded by $O(\mathcal{N})$, $\mathcal{N}$: Number of machines.

\smallskip

\noindent {\bf Redundant Circuitry in Cost Calculator}: In \hercules, the cost calculation has been vastly simplified over the theoretical basis, making it much more amenable to hardware acceleration. However, in the design of $\hercules$, our goal was to streamline the overall pipeline of assigning a given task to a machine. As such we used CAM for the job metadata that was fully sent over to the cost calculator. The cost calculator performed every possible calculation for each job. The memory management unit then had to select which computations were actively being kept, depending on the ordering of the jobs, whether they were receiving virtual work, and whether they were still valid within the schedule. As such, the Cost Calculators in \hercules\ were filled with redundant circuitry and needed to wait on multiple components to achieve coherency before the final cost was accurate. This imposes severe latency and resource cost penalties, contributing to \hercules's 466 cycle iteration. 

\smallskip

\noindent {\bf Decentralized Memory Management}: In \hercules, to facilitate the job assignment, 
the responsibility of tracking distinct attributes of the Virtual Schedules for each machine was split into three distinct components. The JMM tracks the associated attributes of any given job that are required for the cost calculations. The VSM tracks and maintains the WSPT ordering of jobs within a given Virtual Schedule, inserting and removing jobs as needed. Lastly the MMU enforces coherency between the other two elements, ensuring Virtual Schedule Manager knows when changes to the schedule are required, and so the Job Metadata Memory knows which job values need to be updated or marked invalid based on updates in the Virtual Schedule. The \bem{time required 
for these discrete components to achieve coherency imposes a delay on iteration speed and and the intense 
intercommunication 
between each of these elements imposes significant routing congestion}. By requiring each of these components to be capable of full intercommunication about arbitrarily ordered data, each component requires dense interconnectivity, leading to routing for designs tracking more than 10 machines to fail using the \hercules\ architecture. 
Consequently, we identify this decentralized memory management system as the 
crucial bottleneck on system scalability in $\hercules$.
\section{$\stannic$: Virtual Schedule-Centric Hardware Implementation of SOS 
}\label{sec:Systolic_SOSA}

In this section, we present a 
systolic-array-based SOSA 
called \stannic\ to alleviate the limitations of \hercules\ as detailed in~\Cref{sec:bottlenecks}. We discuss the primary subcomponents, and establish their functional parity with components within the \hercules\ SOSA. In doing so, we establish a 
high-level functional 
similarity of the two architectures, while also clarifying how the difference in organization (and core model) results in a centralized design with improved scheduling performance and hardware resource usage. 

\subsection{Systolic 
Architecture for $\stannic$}\label{sec:systolic_arch}

\begin{figure}
    \centering
    \includegraphics[scale=0.4
    ]{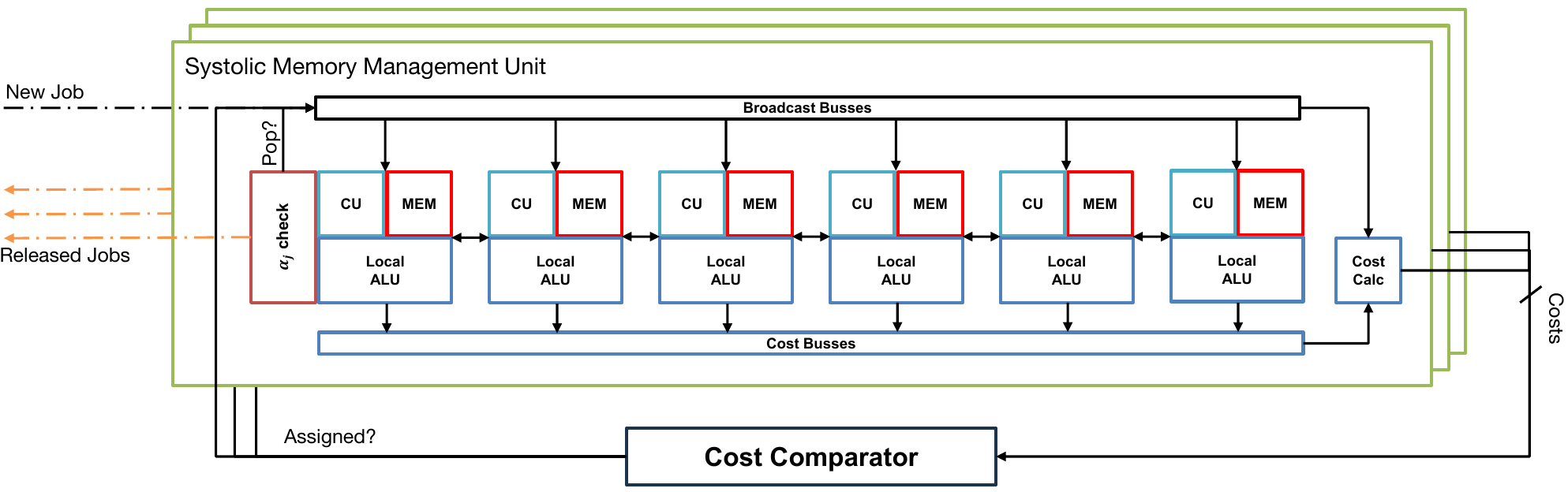}
    \caption{
    {\bf The Systolic SOS $\mu$architecture, as used in $\stannic$.} This architecture places maintaining the order of the Virtual Schedules as the primary function of scheduling. To do so, it uses one dimensional systolic arrays that track each individual job, and leverages the ordered nature of the virtual schedules to enhance the performance of other functionalities.
    }
    \label{fig:STANNIC_block_diagram}
\end{figure}

In this section, we describe the 
$\mu$architecture of $\stannic$ in detail. 
\Cref{fig:STANNIC_block_diagram} shows the entire 
block diagram of the $\stannic$ $\mu$architecture. 

\subsubsection{Systolic Memory Management Unit (SMMU)}\label{sec:smmu_arch}
\stannic\ reorganizes and unifies each machine's individual scheduler functionality within a set of Systolic Memory Management Units (SMMUs). A 
SMMU is dedicated to each machine in the corresponding system we are scheduling for. The SMMU primarily consists of a one-dimensional systolic array, which tracks the corresponding machine $M_i$'s virtual schedule $V_i$ across each of it's \bem{
PEs}. 
Each PE represents an index in the $V_i$, and is responsible for both moving and updating cost metadata for job $K$ while it is in that index. Each SMMU also contains a \bem{Cost Calculator} that performs necessary computations for the final cost calculation 
when a new job is considered. Lastly, the SMMU contains two busses, one to broadcast metadata (\bem{Broadcast Bus}) of a new incoming job $J$ to each PE for local WSPT comparison, and the other to send the cost sub-components to the Cost Calculator (\bem{Cost Bus}). 

\begin{wraptable}[14]{r}{0.4\textwidth}
    \centering
    \resizebox{0.4\textwidth}{!}{
    \begin{tabular}{|c|c|}
        \hline
        \textbf{HERCULES} & \textbf{STANNIC}\\
        \hline
        Memory Management Unit & Systolic Memory Management Unit\\
	   $\alpha_j$ check & $\alpha_j$ check (head only)\\
	   Virtual Schedule Manager & Control Unit\\
	   Cost Calculator & Local ALUs, Cost Calc\\
	   Job Metadata Memory & MEM\\
	   Cost Comparator & Cost Comparator\\
        \hline
    \end{tabular}
    }
    \caption{{\bf Mapping between the $\mu$architectural components of $\hercules$ and $\stannic$ 
    }. This correspondence establishes similarities between 
    the high-level abstract responsibilities of individual components, but the actual methodologies implemented to perform those responsibilities differ across the two $\mu$architectures.}
    \label{tab:Module_parity}
\end{wraptable}

\subsubsection{Processing Element (PE)}\label{sec:PE_arch}

Each PE contains the logic and memory necessary to independently track a single job $K$ and it's associated cost attributes. \bem{Memory (MEM)} stores job metadata, such as $\alpha_J$ point, $T^K_i$, $K.ID$, $n_K(t_C)$ (where $t_C$ is the current cycle), as well as local pre-calculated $sum^{HI}_K$ and $sum^{LO}_K$ values as mentioned in \Cref{sec:cost_optimization,sec:discretization}. Across iterations, these precalculations are updated in the \bem{Local Arithmetic Logic Unit (Local ALU)}, with the exact mathematics detailed 
in \Cref{sec:Systolic_Operation}. Lastly, the \bem{Control Unit (CU)} collects global and local signals from the broadcast bus and neighboring PEs, respectively. These signals are decoded to determine local data movements and arithmetic updates as required to maintain the WSPT ordering of the overarching $V_i$.

The \bem{Head PE}, which corresponds to $Head.V_i$, differs from the rest of the PEs in that it includes the \bem{$\alpha_J$ check} module and has a modified CU. The $\alpha_J$ check module checks whether $n_{Head.V_i}(t_C) \geq (\alpha_J \times Head.V_i.\EPT_i)$, and thus should be released from the schedule. When this happens, it sends a {\tt pop} signal across the SMMU's broadcast bus, alerting all other PEs of the coming change. The CU is modified to account for the fact that the Head PE has no left neighbor, and thus requires different control logic to ensure proper job insertions. This is not necessary for the tail, as job insertions can only occur into $V_i$'s which have space, and thus have an invalid job in their tail PE.

\subsubsection{Cost Comparator (CC)}
The CC is the only module that is separate from the SMMU. It is a singular shared module that receives all of the individual machine costs calculated by the SMMUs. This is necessary to complete the inter-machine scheduling phase (\bem{Phase II}) of the SOS algorithm. This is done with an iterative comparator, like in \hercules.

\subsubsection{Architectural Similarities Between $\hercules$ and $\stannic$}

The \stannic\ architecture achieves the same funtional requirements as \hercules\ by consolidating the decentralized components into the the centralized SMMU with it's systolic array $V_i$. We show a component mapping between the two architectures in \Cref{tab:Module_parity}.  



This architectural shift moves the responsibility of memory management and WSPT ordering maintenance from three discrete, intercommunicating components to the local decision making of PEs in a Centralized Systolic Array structure. This design choice leads to significant reduction in routing congestion and iteration latency due to communication overheads. In~\Cref{sec:exp_results}, we will show that the \stannic\ architecture results in an average $7.5\times$ iteration speedup and a $14\times$ increase in maximum routeable system configuration size as compared to $\hercules$.

\subsection{Systolic SOS Operation} \label{sec:Systolic_Operation}


\begin{figure*}
    \centering
     \begin{subfigure}[b]{0.4\textwidth}
        \centering
        \includegraphics[scale=0.2 
        ]{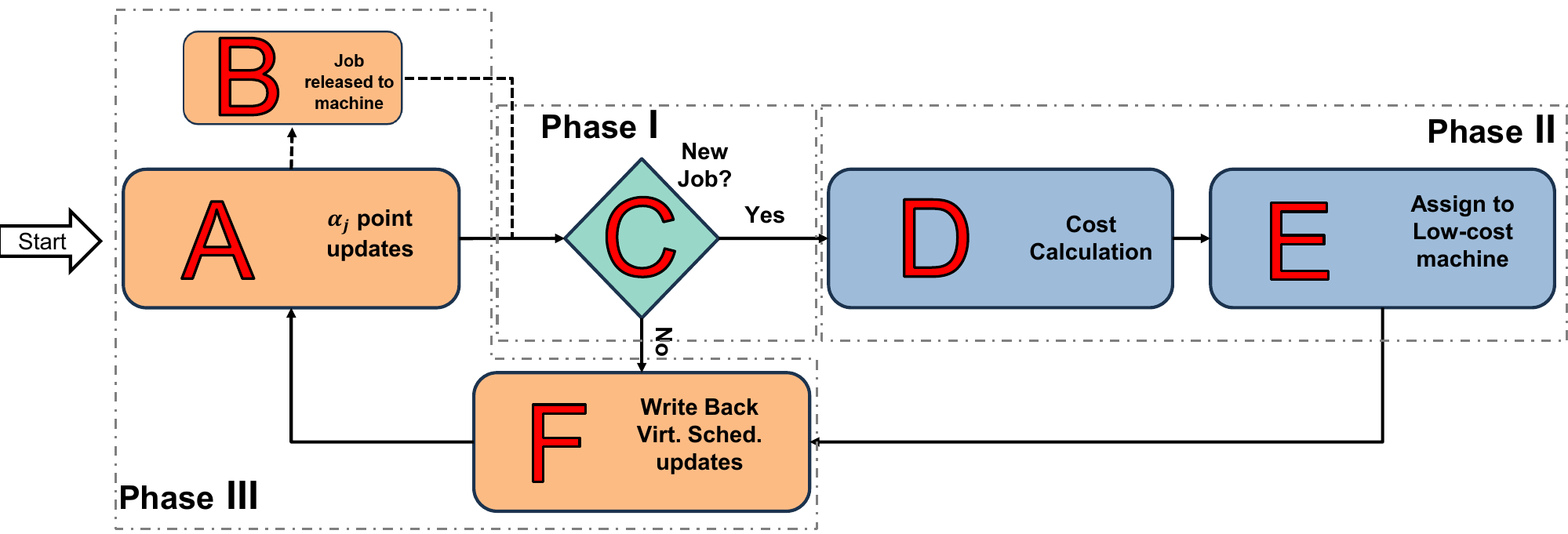}
    \caption{}
    \label{fig:annotated_flow}
    \end{subfigure}
    \hfill
    \begin{subfigure}[b]{0.4\textwidth}
      \resizebox{\textwidth}{!}{
        \begin{tabular}{|c|c|}
            \hline
            \textbf{Iteration} & \textbf{Path}\\
            \hline\hline
            Standard & $A \rightarrow C \rightarrow F$\\
            Pop & $A \rightarrow B \rightarrow C \rightarrow F$\\
            Insert & $A \rightarrow C \rightarrow D \rightarrow E \rightarrow F$\\
            Pop and Insert & $A \rightarrow B \rightarrow C \rightarrow D \rightarrow E \rightarrow F$\\
            \hline
        \end{tabular}
        \vspace{8mm}
        }
    \caption{}
    \label{tab:flow_loops}
    \end{subfigure}
    \caption{(\subref{fig:annotated_flow}): {\bf Annotated Algorithmic Flow} derived from~\Cref{fig:STANNIC_alg_flow} with additional annotations.
    ~(\subref{tab:flow_loops}): {\bf Iteration Mapping} showing 
    the four paths through (\subref{fig:annotated_flow}), classifying each as a specific type of iteration.
    }
    \label{fig:annotated_loops}
\end{figure*}

Using our Virtual-Schedule based perspective on the SOS algorithm (\cf, \Cref{fig:STANNIC_alg_flow}), we can see that the algorithm has four looping paths (\Cref{tab:flow_loops}), but each returns us back to the start point of the scheduling iteration . The operational efficiency of \stannic\ relies on maintaining an \textit{ordered} $V_i$ in our systolic array as a fundamental loop invariant. 
The systolic array enables parallelism by allowing each PE to use this ordering and local data to accelerate cost calculation and maintain state coherence with minimal routing overhead. By \Cref{def:virtual_schedule}, a $V_i$'s ordering is done by WSPT ratio ranking. We will expand upon this to define 
\textit{Properly Ordered Systolic Virtual Schedule}.

\begin{definition}\label{def:ordered_vi}
    \bem{Properly Ordered Systolic Virtual Schedule:} A Virtual Schedule $V_i$, tracked by a systolic array, is properly ordered if:
    \begin{itemize}
        \item The \bem{Head PE} (PE$_0$) is either empty or contains the job $K, K \in V_i$ with the highest WSPT ratio $T^K_i$ of all jobs in the virtual schedule.
        \item For any valid job $K$ in PE$_i$, the neighbor to it's immediate ``right'' (PE$_{i+1}$) contains either a job $K_R$ with $T^{K_R}_i\leq T^K_i$, or an invalid job.
        \item There are no invalid jobs (bubbles) between two PEs that contain valid jobs.
    \end{itemize}
\end{definition}

\subsubsection{Systolic Cost Calculation}\label{sec:syst_cc}

\begin{figure}
    \centering
    \includegraphics[scale=0.45
    ]{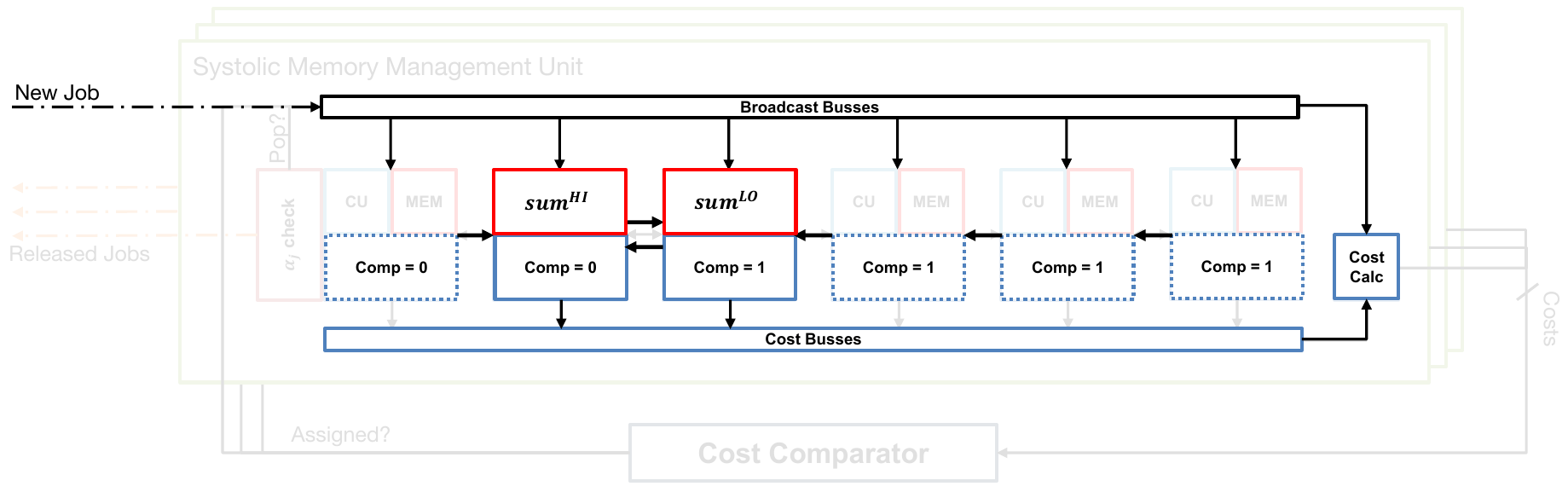}
    \caption{{\bf An illustrative example of cost calculation within the systolic cirtual schedule architecture}. Each PE is capable of performing a local WSPT comparison, self identifying which set they are in. By then looking at the direct neighboring PEs, local elements can self identify as threshold elements, and thus contribute a memoized summation of all lower/higher costs as needed, removing the summation iteration latency and improving overall speed.}
    \label{fig:Systolic_CC}
\end{figure}

$\stannic$ 
accelerates cost calculation by leveraging the ordered 
systolic $V_i$ to precalculate and memorize the summation terms $sum^{HI}$ and $sum^{LO}$ (\cf, \Cref{eqn:costh_discretized,eqn:costl_discretized}). \Cref{fig:Systolic_CC} illustrates an example of how the systolic $V_i$ contributes to cost calculation.

As detailed in \Cref{sec:sos_overview}, the cost of assigning a new job $J$ onto machine $M_i$ is calculated using two discretized terms, $cost^H$ and $cost^L$. Individual jobs $K\in V_i$ contribute to one of these two terms based on the WSPT of job $K$ to the new incoming job $J$. As such, upon the arrival $J$, it's WSPT $T^J_i$ is broadcast to all PEs. Each PE, tracking job $K$ with WSPT $T^K_i$, performs a local comparison $\mathcal{C}$:

\begin{align}
    \mathcal{C} = \begin{cases}  
                        0, & \mbox{if}~T^K_i \geq T^J_i \\
                        1, & \mbox{otherwise}
                  \end{cases} 
    \label{eqn:cost_compari_sys_pe}
\end{align}

A comparison value of $\mathcal{C} = 0$ indicates that job $K$ contributes to $sum^{HI}$, whereas a value of $\mathcal{C} = 1$ indicates that job $K$ contributes to $sum^{LO}$ for the cost calculation of job $J$.
Note that $\mathcal{C} = 1$ even when a PE is not tracking a valid job. As such, assuming the invariant of a properly ordered systolic $V_i$ as defined in \Cref{def:ordered_vi}, reading the comparison values $\mathcal{C}$ from the Head PE to the Tail PE results in a string of zeros followed by a string of ones. 

This means the \textit{comparison threshold} between the two cost sets for $sum^{HI}$ and $sum^{LO}$ can be found between two PEs: the last PE with $\mathcal{C} = 0$, and the first PE with $\mathcal{C} = 1$. After each PE has calculated its own comparison value $\mathcal{C}$, it can then check the comparison values of its neighboring PEs, and self-identify as part of the threshold. We notate a PE's neighbors comparison values as $\mathcal{C}_R$ and $\mathcal{C}_L$ for the right and left neighbor, respectively. As noted in \Cref{sec:PE_arch}, each PE contains two local memorized values: $sum^{HI}_K$ and $sum^{LO}_K$. $sum^{HI}_K$ 
computes if the PE's tracked job $K$ were the last element in the higher priority set, whereas $sum^{LO}_K$ 
computes if the PE's tracked job $K$ were the first element in the lower priority set. Initial calculation and maintenance of $sum^{HI}_K$ and $sum^{LO}_K$ is discussed in \Cref{sec:invariant}. During cost calculation, 
instead of iteratively summing over the individual contributions in the systolic array, the two PEs at the threshold can self-identify and volunteer their memorized values to the Cost Calculator. The PE with local $\mathcal{C} = 0$ contributes its $sum^{HI}_K$ for the calculation of $cost^H$, while the PE with local $\mathcal{C} = 1$ contributes its $sum^{LO}_K$ for the calculation of $cost^L$. In doing this, we entirely remove the latency associated with summation across the depth of $V_i$, converting the summation into a single-cycle lookup operation.

\subsubsection{Maintaining Virtual Schedule Ordering}\label{sec:invariant}

To accurately 
perform cost calculation, we need to maintain the proper ordering of the virtual schedule $V_i$ as a loop invariant across 
iterations of the algorithmic flow (\cf, \Cref{fig:annotated_loops}). Maintaining this loop invariant can be done with local, and thus parallel, PE decisions 
and updates by individual PEs in response to global signals. There are four categories of iteration through the algorithmic flow, each of which requires different actions in order to maintain the loop invariant.

\begin{figure}
    \centering
    \includegraphics[scale=0.45
    ]{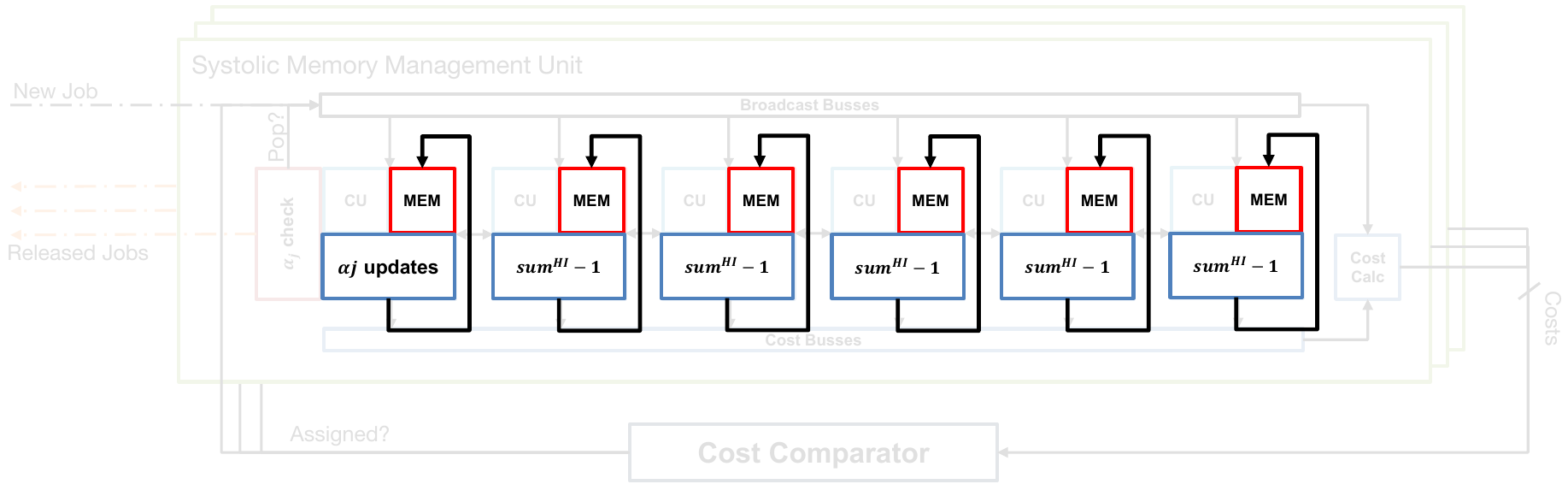}
    \caption{{\bf Standard Iteration updates and writeback paths.} Local memoized cost updates are written back directly to local memory {\bf $\alpha_j$ updates:} At the head, it performs the same incremental updates as described in part two of~\Cref{sec:cost_optimization}. Each other job only needs to perform the updates to $sum^{HI}$ as the job in the head PE is by definition in the set of jobs with a higher WSPT than themselves.} 
    \label{fig:Standard_itr}
\end{figure}

\begin{enumerate}[leftmargin=14pt, itemsep=3pt]
    \item \bem{Standard Iteration.} A standard iteration occurs when no new job has been assigned to machine $M_i$, and the job at $Head.V_i$ has not yet reached it's $\alpha_J$ release point. In a standard iteration, schedule ordering remains stationary but the memorized cost values across the systolic array need updating to reflect the accrual of Virtual Work for $Head.V_i$. The operation of a standard iteration proceeds in parallel across the array, and is illustrated in \Cref{fig:Standard_itr}. There are two PE actions -- (a) \bem{Head PE Action:} The Local ALU updates both $sum^{HI}_K$ and $sum^{LO}_K$ by decrementing them. $sum^{HI}_K$ is decremented by $1$, and $sum^{LO}_K$ is decremented by $T^K_i$ representing one cycle of completed Virtual Work as described in \Cref{sec:cost_optimization}; and (b) \bem{Subsequent PE Action:} Every other PE that is currently tracking a valid job has a job $K$ that by definition has a lower or equal WSPT than the job at $Head.V_i$. As such, their local $sum^{HI}_K$ represents a summation that includes $sum^{HI}_{Head.V_i}$. The local ALUs in the other PEs must therefore also decrement the local $sum^{HI}_K$ by one. $sum^{LO}_K$ remains unchanged. Each PE's CU selects it's own Local ALU as the source for memory write-back. This operation maintains the WSPT order while ensuring that all memorized cost values accurately reflect the virtual work completed by the highest priority job.

\begin{figure}
    \centering
    \includegraphics[scale=0.45
    ]{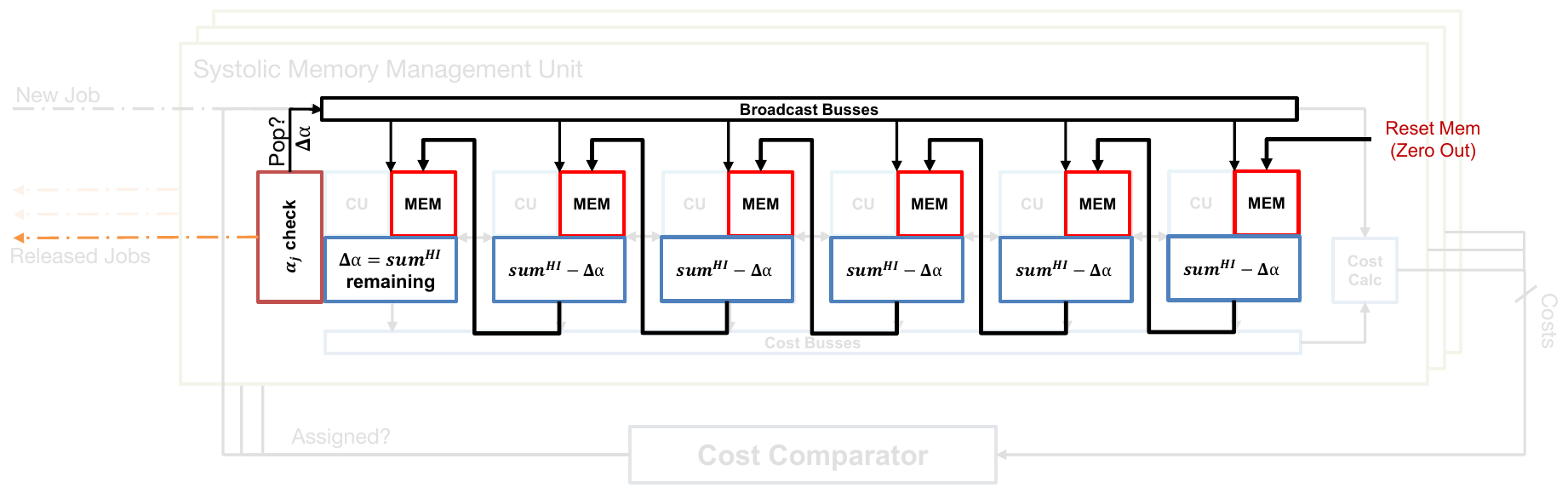}
    \caption{{\bf POP Iteration updates and writeback paths.} The head PE recognizes that its current job has reached its $\alpha_j$ point, and as such notifies the relevant machine queue to finalize scheduling. The rest of the jobs are still in order, but simply clearing head would insert a ``bubble'' which would break proper ordering definition. Instead, each PE calculates the new $sum^{HI}$ value of their given job to account for the head job leaving, and then the CU selects the ALU of the right neighbor as the source of the local memory writeback. The tail ``inserts'' an invalid job at the end.}
    \label{fig:POP_itr}
\end{figure}


    \item \bem{POP Iteration.} A POP iteration occurs when the job in the Head PE has reached its $\alpha_J$ release point, requiring that job to be released to the machine's work queue, removing it from the $V_i$. The goal is to remove the job from the Head PE, and shift all subsequent jobs to the left in order to preserve the WSPT ordering and maintain the no-bubbles element of our invariant. The process of doing this is illustrated in \Cref{fig:POP_itr}. POP iteration has three different types of iterations -- (a) \bem{Job Release and Broadcast}: This iteration starts with the $\alpha_J$ check module identifying that the $Head.V_i$ job has reached its release point. It releases the job's $ID$ to the work queue, and simultaneously broadcasts both a {\tt pop} flag and the remaining value $\Delta\alpha=sum^{HI}_{Head.V_i}$. This value represents the remaining contribution of the current job at $Head.V_i$ to all the other jobs $K\in V_i$'s $sum^{HI}_K$; (b) \bem{Cost Updates}: Each PE performs the subtraction $sum^{HI}_K - \Delta\alpha$ in their Local ALU. This subtraction is necessary as $Head.V_i$ is now being removed from the set and should not be considered for any subsequent cost calculations; and (c) \bem{Synchronous Left-Shift:} Lastly, in the write-back stage of this iteration, each PE synchronously writes back the data from its right neighbors ALU (\ie, $PE_i \leftarrow PE_{i+1}$). The tail's right neighbor ALU inputs are hardwired to zero, in essence resetting the memory and inserting an invalid job at the end of the virtual schedule. This maintains proper ordering while simultaneously removing the released job from the schedule and any memorized cost components.

\begin{figure}
    \centering
    \includegraphics[scale=0.45
    ]{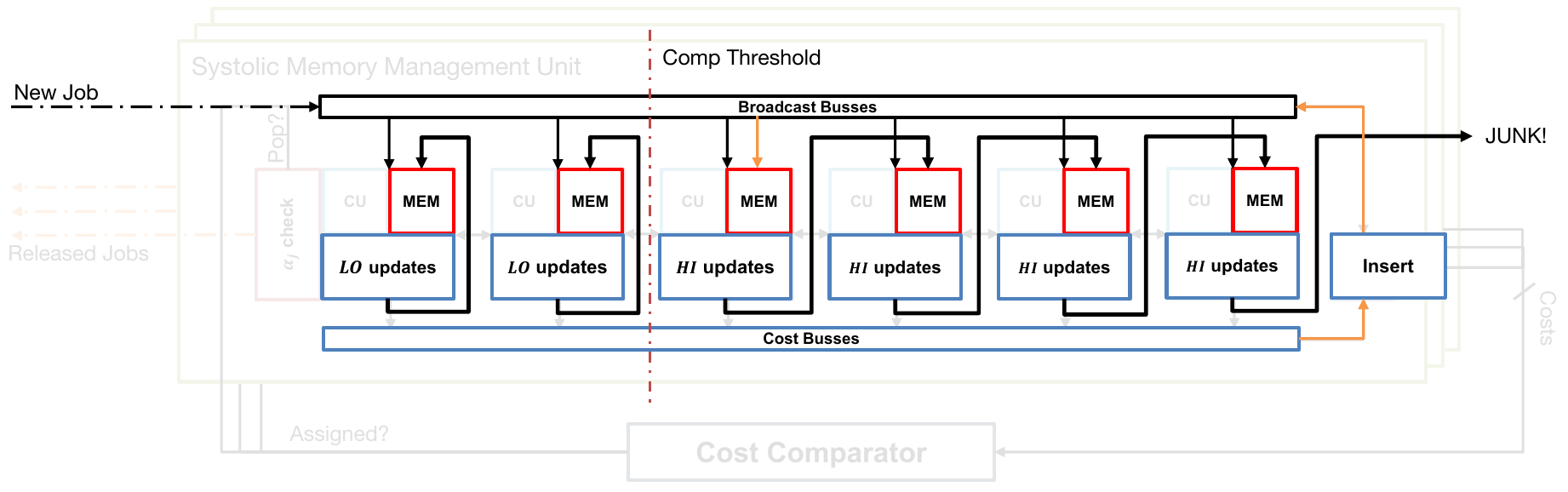}
    \caption{{\bf Insert Iteration updates and writeback paths.} Job insertion implies that cost calculation comparisons have been performed in this cycle for the current jobs already. These local comparisons can be reused to recognize which of two categories a given job is in. {\bf HI set:} All jobs with a higher WSPT than the new job. They perform the standard $\alpha_j$ updates, but additionally add the low cost of the new job to their $sum^{LO}$ as it will be inserted below them. {\bf LO set:} All jobs with a lower WSPT instead account for the new job in their $sum^{HI}$ total. Additionally, to make space for the new job each PE in this set writes the new data from their left neighbor, except the PE at the threshold, which instead writes the data associated with the new job, whose initial $sum^{HI}$ and $sum^{LO}$ values are calculated in the cost calculator.}
    \label{fig:INS_itr}
\end{figure}
\begin{table}
    \centering
    \resizebox{\textwidth}{!}{
    \begin{tabular}{|c|c|c|c|}
        \hline
        \textbf{PE Set} & \textbf{Comparison Value $C$} & \textbf{Reordering Action} & \textbf{Cost Updates} \\
        \hline\hline
        $HI$ Set & $C=0$ & Stationary & $sum^{LO}_K+J.W$\\
        $LO$ Set & $C=1$ & Synchronous Right-Shift & $sum^{HI}_K+J.\EPT_i$\\
        New Job $J$ & $C=1, C_L=0$ & Stores New Job & Loads Initial $sum^{HI}_J$ and $sum^{LO}_J$*\\
        \hline
    \end{tabular}
    }
    \caption{{\bf Insert Iteration Behavior for different sets of PEs.} Each PE can identify which set they are a part of using $C$ and $C_L$. *The initial memoized costs for $J$ are calculated in the overall cost calculator. The PE that will store job $J$ still needs to calculate $sum^{HI}_K+J.\EPT_i$, so that the next PE can store this updated value during the right-shift.}
    \label{tab:insert_itr}
\end{table}

    \item \bem{Insert Iteration.} An Insert iteration occurs when the machine corresponding with a particular Virtual Schedule $V_i$ has been selected as the minimum-cost machine for a new job $J$. At this point, the $V_i$ must reorder itself to insert $J$ in the correct WSPT ordering position. Additionally, new memorized costs need to be calculated for $J$ and all other jobs need to update their local cost to account for the new job now being present in either their $sum^{HI}_K$ or $sum^{LO}_K$ set. An illustrated example can be seen in \Cref{fig:INS_itr}, while \Cref{tab:insert_itr} shows the three distinct sets of behaviors. The selection of machine $M_i$ as the lowest cost machine implies that a cost calculation has been performed earlier in this iteration (\cf,~\Cref{sec:syst_cc}). As such, each PE still has a local comparison value ($\mathcal{C}$) to self-identify as part of the high or low WSPT set, relative to the new job $J$. Furthermore, by checking the $\mathcal{C}$ value of a neighbor, the threshold between these two sets can be locally self-identified by PEs as well. We can use this information to inform a specific PEs behavior, both in terms of cost update calculation and reordering. Insertion iteration comprises three distinct computation -- (a) \bem{Cost Update Calculation}: All jobs that have a higher WSPT than $J$ (\ie, the $sum^{HI}$ set) 
    need to account for the new job inserted below them, and so need to update $sum^{LO}_K$. All jobs in the $LO$ set instead need to calculate an updated value of $sum^{LO}_K$. The initial memorized values for job $J$ are calculated in the overall cost calculator. The cost calculator has $sum^{HI}$, $sum^{LO}$, $J.W$ and $J.\EPT_i$ already when performing the cost calculation, and so can calculate $sum^{HI}_J = sum^{HI} + J.\EPT_i$ and  $sum^{LO}_J = sum^{LO} + J.W_i$ simultaneously. For each of these calculations, they happen on top of the $\alpha_J$ cost updates, as detailed in the Standard Iteration; (b) \bem{Reordering}: To maintain proper ordering, $J$ must be inserted after all the jobs in the $HI$ set, and right before all the jobs in the $LO$ set, with each of the jobs in both sets maintaining their current orders. As such, PEs with $\mathcal{C} = 0$ store the values from their own ALU during write back, and most of the PEs with $\mathcal{C} = 1$ store the information from their left neighbor. An 
    exception to this is the PE with $\mathcal{C} = 1$ and $\mathcal{C}_L = 0$, which is the PE that was the low side of the comparison threshold. This PE corresponds to the index at which $J$ needs to be inserted, and as such, it loads all of it's new data from the broadcast bus; and (c) \bem{Edge Cases -- Inserting at Head PE}: When the incoming job $J$ has a higher WSPT than all existing jobs, it must be inserted into the Head PE. Since the Head PE has no left neighbor, the Head PE's CU has different control logic that recognizes if $\mathcal{C} = 1$, the Head PE is the insertion point. This also covers inserting $J$ into an otherwise empty $V_i$. \bem{Full $V_i$} if we were to try an insert a job into a full systolic array (\ie, a systolic array with a valid job in each of the PEs), the job in the tail PE would be lost, as no PE will store it during the writeback cycle. As such, full $V_i$s can not be assigned new jobs. 

\begin{figure}
    \centering
    \includegraphics[scale=0.45
    ]{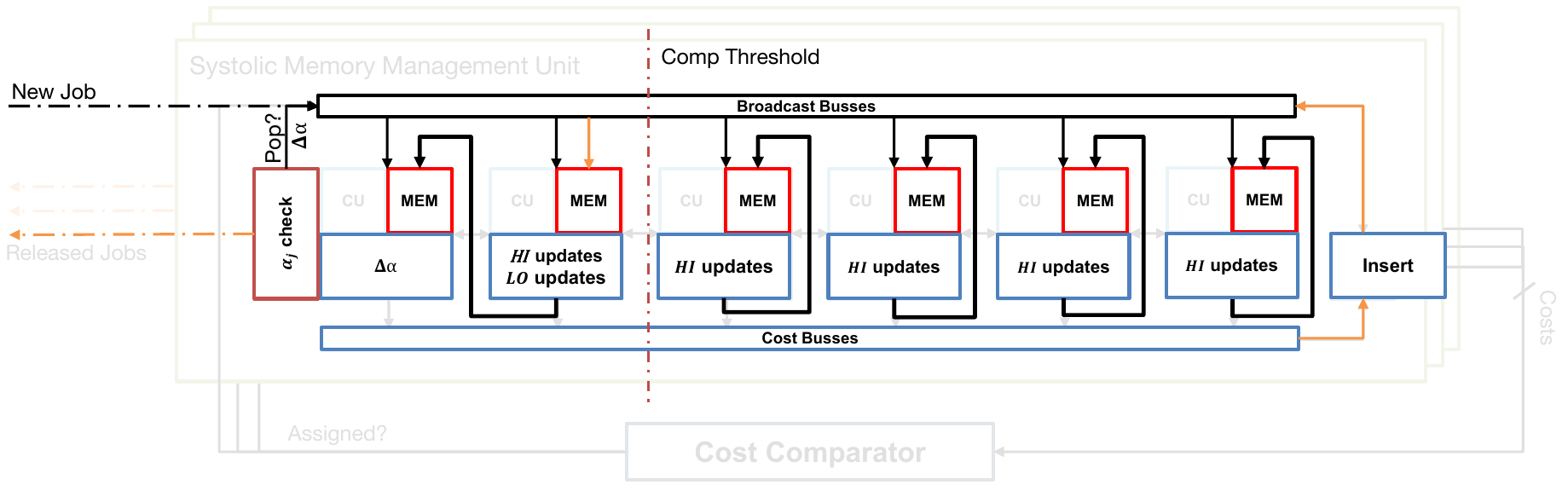}
    \caption{{\bf POP and Insert Iteration updates and writeback paths.} In the case that a POP and insert occur in the same iteration, the two schedule reordering operations are overlapped, causing the high set to shift left and the low set to remain stationary due to effectively left and right shifting simultaneously. The new job is inserted on the left of the threshold to accoun for the POP's left shift. Relevant cost updates account for not only the addition of the new job, but also the subtraction of the head job's remaining cost $\Delta\alpha$.}
    \label{fig:POP_INS_itr}
\end{figure}
\begin{table}
    \centering
    \resizebox{\textwidth}{!}{
    \begin{tabular}{|c|c|c|c|}
        \hline
        \textbf{PE Set} & \textbf{Comparison Value $C$} & \textbf{Reordering Action} & \textbf{Cost Updates} \\
        \hline\hline
        $HI$ Set & $C=0$ & Left-Shift & $sum^{LO}_K+J.W, sum^{HI}_K-\Delta\alpha$\\
        $LO$ Set & $C=1$ & Synchronous Right-Shift & $sum^{HI}_K+(J.\EPT_i-\Delta\alpha$)\\
        New Job $J$ & $C=0, C_R=1$ & Stores New Job & Loads Initial $sum^{HI}_J$ and $sum^{LO}_J$*\\
        \hline
    \end{tabular}
    }
    \caption{{\bf POP and Insert Iteration Behavior for different sets of PEs.} Each PE can identify which set they are a part of using $C$ and $C_R$. *The initial memoized costs for $J$ are calculated in the overall cost calculator. The PE that will store job $J$ still needs to calculate $sum^{LO}_K+J.W, sum^{HI}_K-\Delta\alpha$, so that the previous PE can store this updated value during the left-shift.}
    \label{tab:pop_insert_itr}
\end{table}

    \item \bem{POP and Insert Iteration.}
    Finally, a POP and Insert Iteration happens when the job ad $Head.V_i$ reaches it's $\alpha_J$ release point in the same iteration in which machine $M_i$ is selected as the lowest-cost machine for an incoming job $J$. This iteration combines the functional requirements of the individual POP iteration and Insert iteration. Memorized cost updates need to account both for the $\Delta\alpha$ of the departing job, as well as the $J.W$ or $J.\EPT_i$ of the incoming job (depending on relative WSPT). Lastly the array must account for the left-shift and insertion in a single update. An illustrated example can be seen in \Cref{fig:POP_INS_itr}, with the three sets of behaviors being described in \Cref{tab:pop_insert_itr}. (a) \bem{Cost Updates}: The $HI$ set needs to calculate updates for both of its memorized values, as the job at $Head.V_i$ is being removed from its $sum^{HI}_K$ summation, and $J$ is being added to its $sum^{LO}_K$ summation. For $LO$ set PEs, their $sum^{HI}_K$ needs to remove $Head.V_i$ while adding $J$ into its summation. For the initialization of $sum^{HI}_J$ and $sum^{LO}_J$, the only difference compared to the Insert Iteration is that $sum^{HI}_J$ must now also account for $Head.V_i$ leaving. As such $sum^{HI}_J = sum^{HI} + (J.\EPT_i - \Delta\alpha)$; (b) \bem{Overlapped Reordering}: Rather than performing the two movements separately, they can be composed into a singular transformation. For a POP iteration, every PE shifts left. In an insert operation, the $HI$ set remains stationary, while the $LO$ set right shifts to create a space for job insertion. Combining these two results in a net left-shift for the $HI$ set, and the $LO$ set remaining stationary. To properly account for the new job being inserted and then shifted to the left, we can simply insert it one index to the left compared to a regular Insertion Iteration. This is equivalent to inserting it at the $HI$ end of the comparison threshold (\ie, the PE with $\mathcal{C} = 0$, $\mathcal{C}_R = 1$); and (c) \bem{Edge Case -- $J$ Has Highest WSPT}: In the case that $J$ has the highest WSPT, the Head PE would have a $\mathcal{C} = 1$, and there would be no PE with $\mathcal{C} = 0$ in the entirety of the $V_i$. However, in such a scenario, job $J$ would have to be inserted into the Head PE anyway to maintain proper WSPT ordering. As such, when the Head PE recognizes a POP, it sets $\mathcal{C} = 0$. This way the CU can check if $\mathcal{C}_R = 1$ and correctly self-identify as the insertion point PE. 

\end{enumerate}

\section{Experimental Setup}\label{sec:exp_setup}


In this section, we outline the experimental setups for the evaluation of our scheduling architectures. We set up two broad categories for these evaluations. The first of these categories (outlined in ~\Cref{sec:exp_setup_inter}) is the output schedule evaluation, which we use to compare the produced schedules of our accelerated algorithm to other preexisting methods. The second catagory (outlined in section ~\Cref{sec:exp_setup_intra}) is the setup for direct architectural performance comparisons between \hercules\ and \stannic.


\subsection{Output Schedule Evaluation}\label{sec:exp_setup_inter}

\noindent \bem{Target machine configurations}: We have used five (5) 
machine configurations -- {\bf M1}: \bangle{CPU, Best}, {\bf M2}: \bangle{CPU, Worst}, {\bf M3}: \bangle{Mixed, Best}, {\bf M4}: \bangle{GPU, Best}, and {\bf M5}: \bangle{GPU, Worst}.

\smallskip

\noindent \bem{Workload generation}: We have developed an 
in-house workload generator (WG) to emulate job dispatch in heterogeneous systems with varied job distributions, reflecting real-world scenarios such as CPU-heavy/GPU-heavy bursts. The WG has multiple configurable parameters -- (a) \bem{Job Composition} (JC) captures the fraction of compute intensive, memory intensive, and mixed jobs, summing to {\bf 1.00}; (b) \bem{Machine Composition} (MC) captures numbers of CPU/GPU/Mixed machines; (c) \bem{Burst Factor} (BF) captures maximum number of jobs that may be released in a single clock tick; (d) \bem{Burst Type} (BT) captures job arrival patterns. For {\em random}, jobs are released at randomly selected ticks and for {\em uniform}, a BF amount of jobs are released every tick; (e) \bem{Idle Time} (IT) captures number of ticks inserted after a specified number of jobs are released; and (f) \bem{Idle Interval} (II) capture maximum number of jobs released before inserting an idle period. BF and BT model the \bem{uncertainty} in job arrivals in realistic scenarios whereas IT and II imitate time spans where \bem{new jobs are not created until ongoing jobs are completed}. 

\smallskip

\noindent \bem{Baseline schedulers}: We compare 
performance of $\hercules$ 
against four baseline scheduling algorithms -- Round Robin (RR)~\cite{silberschatz2012operating}, Greedy~\cite{dong2015greedy}, Work Stealing Round Robin (WSRR), and Work Stealing Greedy (WSG)~\cite{taskflow2022huangtcad}. 

\smallskip

\noindent \bem{Metrics for comparison}: We use four metrics for comparisons. 
\bem{\fairness} 
measures if low-performing machines are not starved. 
\bem{\lbalance} measures equality of job distribution 
across machines 
and  
is computed as the Coefficient of Variation (CV) in the number of jobs assigned to a machine across scheduling intervals. Lower CV 
indicates 
better load balancing. 
\bem{\latency} captures the average delay between job creation 
and its scheduling time. 
Because the macro-level tasks scheduled by SOSA operate on a scale of seconds to possibly hours, the nanosecond-scale hardware execution time required to assign them constitutes an effective $0\%$ of the total wall time. Therefore, this metric focuses strictly on queue delay rather than computational scheduling overhead. Lower latency reflects faster scheduling and results in higher system throughput.

\smallskip

\noindent \bem{Hardware for SOS scheduler}: We have used an AMD Alveo U55C~\cite{u55c} as our target FPGA 
to implement the SOS scheduler. We used Allo/HeteroCL~\cite{allo2024chen, heterocl2019lai} programming language to design and implement \hercules\ using their {\bf Vitis HLS} backends. \stannic\ was implemented directly in {\bf C++ Vitis HLS} to allow finetuning of the micro architectural behaviors. To facilitate host-device communication, the scheduler is managed as an OpenCL kernel via Xilinx XRT, utilizing standard AXI4 Memory Map interfaces for efficient PCIe transactions. The operating frequency of the scheduler is 371.47 MHz. 

\subsection{Architectural Comparisons}\label{sec:exp_setup_intra}

For our architectural comparison experimentatin, we introduce four more metrics.

\smallskip

\noindent {\bf Iteration Latency}: The integer number of clock cycles it takes for the accelerator to complete one iteration of the scheduling process. In a strictly online context where job arrivals and system dynamics change rapidly, this metric represents the hardware's active decision-making resolution. Minimizing iteration latency ensures the scheduler can maintain high throughput and avoid becoming a bottleneck during massive burst arrivals, allowing us to effectively quantify operational speed.


\smallskip

\noindent {\bf Resource Utilization}: This is the number of a discrete computational elements present on the FPGA that are required to physically implement our design. This makes it a good metric for evaluating relative size, and therefore relative resource efficiency of two designs.

\smallskip

\noindent {\bf Maximum Rout-able Configuration Size}: This number represents the largest number of discrete machines a design can track while still producing a rout-able design. This is our metric defines a designs scalability.

\smallskip

\noindent {\bf Power Profile}: This is the average measured power usage in watts of the FPGA while the scheduler is running. This metric will help us further quantify design efficiency, as physical space is not the only resource of consequence.

\subsubsection{Data Gathering Methodology}

For qualitative comparison, Iteration Latency and Resource Utilization are gathered by referring to the Xilinx Vitis and Vivado toolchains {\tt csynth} \cite{amdWebsite} 
reports after synthesis and implementation runs were completed on the two architectures using different system configurations. Maximum Rout-able Configuration size is measured by incrementally increasing the number of machines in the configuration by ten. The largest configuration in which the design is properly synthesized is presented as the final result. 

Power Draw is measured using {\tt xbtop}, a utility available from Xilinx to gather real time performance data on a connected FPGA. To obtain measurements, the design binary under test is first loaded onto the FPGA, and then the design is used to schedule jobs in a workload suite using the same workload generation as described in the experimental setup in \Cref{sec:exp_setup_inter}. The Power Draw measurements could then be taken during runtime.

For the sake of direct comparison of Iteration Latency, Resource Utilization, and Power Draw, these metrics are gathered for equivalent system configurations for both \hercules\ and \stannic. The four configurations (\bem{C1}-\bem{C4}) used for comparison are $5\times10,\ 5\times20,\ 10\times10,\ \text{and } 10\times20$, where $m\times d$ denotes a system with $m$ machines and a per machine $V_i$ depth of $d$ jobs. For each of our metrics, we also consider the average across all configurations for both designs.

\section{Experimental Results}\label{sec:exp_results}

In this section, we present the results of a series of experiments using the metrics, configurations, and hardware setups described in ~\Cref{sec:exp_setup}. The first two experiments (\Cref{sec:exp_1,sec:exp_2}) demonstrate the inherent scheduling capabilities of the SOS algorithm, and the benefits of accelerating this scheduling algorithm with hardware. 

~\Cref{sec:arch_comp_analysis} compares the performance and resource utilization metrics of \hercules\ and \stannic, empirically quantifying the performance improvements gained from the optimizations outlined in \Cref{sec:Systolic_Operation}. This section presents out numeric inter-architectural comparison that corroborates the improvement claims made throughout the rest of the paper.

~\Cref{sec:hetero_workload} compares the efficiency of the resulting schedules from SOSA with alternative baseline schedulers, demonstrating the benefit of using the SOS algorithm while scheduling in an online, stochastic, and heterogeneous context. Due to the two architectures implementing the same scheduling algorithm, the resulting schedules from both \hercules\ and \stannic\ are identical. As such, for the schedule analysis experiments of \Cref{sec:exp_1,sec:hetero_workload}, we present one scheduling result for the SOS algorithm.



\subsection{Effectiveness of SOSA On Varying Workloads}\label{sec:exp_1}

\begin{figure}
    \centering
    \begin{subfigure}[b]{0.4\columnwidth}
        \centering
        \includegraphics[width=\textwidth, center]{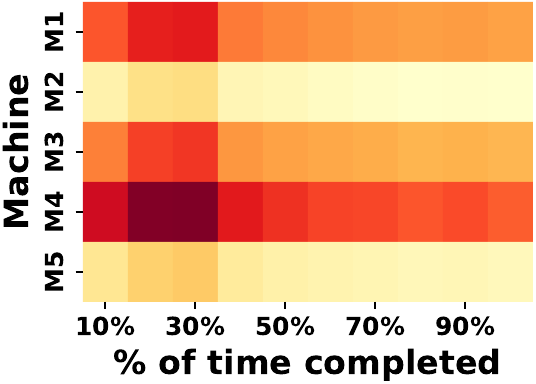}
    \caption{}
    \label{fig:algo_fairness}
    \end{subfigure}
    \hspace{5mm}
    \begin{subfigure}[b]{0.4\columnwidth}
        \centering
        \includegraphics[width=\textwidth, center]{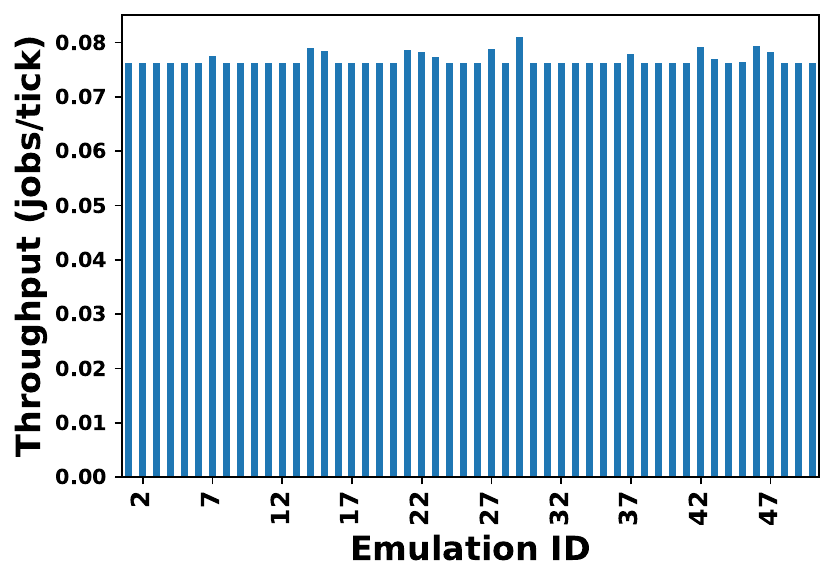}
    \caption{}
    \label{fig:alg_throughput}
    \end{subfigure}   
    \caption{(\subref{fig:algo_fairness}): {\bf Average machine utilization across emulations}. The darker the color, the more jobs are assigned to the machine.~(\subref{fig:alg_throughput}): {\bf Scheduler throughput across emulations}.}
    \label{fig:enter-label}
\end{figure}

\begin{figure}
    \centering
    \begin{subfigure}[c]{0.35\columnwidth}
        \centering
        \includegraphics[width=\textwidth, center]{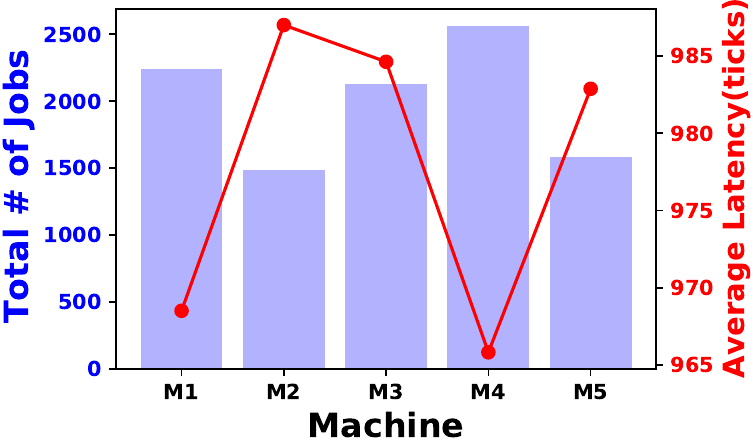}
    \caption{}
    \label{fig:job_latency_distribution}
    \end{subfigure}
    \hspace{5mm}
    \begin{subfigure}[c]{0.55\columnwidth}
    \resizebox{\textwidth}{!}{
    \begin{tabular}{|c|c||c|c|c||c|c|c|}
        \multicolumn{2}{c}{}&\multicolumn{3}{c}{\hercules}&\multicolumn{3}{c}{\stannic}\\
        \hline
        \textbf{C} & \textbf{ST} & \textbf{HT} & \textbf{SU} & \textbf{FPC} & \textbf{HT} & \textbf{SU} & \textbf{FPC}\\      
        {} & {\bf (secs)} & {\bf (secs)} & {} & {Watts} & {\bf (secs)} & {} & {Watts}\\
        \hline\hline
        {\bf C1} & 58.8 & 0.09 & 653$\times$ & 20.83 & 0.04 & {1469$\times$} & 20.94 \\
        {\bf C2} & 43.26 & 0.10 & 433$\times$ & 21.11 & 0.05 & {865$\times$} & 20.91\\       
        {\bf C3} & 98.41 & 0.09 & 1093$\times$ & 20.91 & 0.05 & {1968$\times$} & 20.72\\
        {\bf C4} & 95.40 & 0.09 & 1060$\times$ & 21.39 & 0.05 & {1908$\times$} & 20.91\\
        \hline
    \end{tabular}
    }
    \caption{}
    \label{tab:speedup}
    \end{subfigure}   
    \vspace{-2mm}
    \caption{(\subref{fig:job_latency_distribution}): {\bf Jobs and average latency per machine}.~(\subref{tab:speedup}): {\bf SOSA vs. software implementation}. {\bf M}: No. of machines. {\bf JP}: Jobs/machine. {\bf ST}: Software execution time. {\bf HT}: Hardware execution time. {\bf SU}: Speedup. {\bf FPC}: Power consumption. Here we see that \stannic\ offers nearly double the speedup of \hercules\ with similar power draw.}
    \label{fig:latency_power}
\end{figure}

In this experiment, we explore the effectiveness of SOSA in terms of {\em fairness} and {\em load balancing}. Toward that, we have generated 50 different workloads by varying the workload parameters (\cf,~\Cref{sec:exp_setup}) using a Monte-Carlo simulation and then use our hardware based schedulers to schedule jobs on {\bf M1}-{\bf M5} for all 50 workloads. In~\Cref{fig:algo_fairness}, we show the average number of jobs assigned to each machine over all 50 workloads at different fraction of time points during their run. We observe that machines 
{\bf M1}, {\bf M3}, and {\bf M4} consistently  exhibit high utilization as they are best performing machines. However, despite their higher capability, the scheduler intelligently identifies when these machines reach their scheduling capacity and assign jobs dynamically to the remaining two low-performing machines, \ie, {\bf M3} and {\bf M5}, preventing them from starving. {\em Due to such intelligent scheduling, the throughput (measured in terms of jobs scheduled per clock tick) of the scheduler almost remains constant across all the 50 workloads as shown in}~\Cref{fig:alg_throughput}. This observation indicates that SOSA is highly capable of load balancing while maintaining high throughput and utilization of the available heterogeneous computing resources. Additionally,~\Cref{fig:job_latency_distribution} shows that {\bf M1}, {\bf M3}, and {\bf M4} exhibit lowest average latency as expected since they are high performance machines. This experiment shows that \bem{SOSA is robust, adaptable to varying workload without compromising performance all the while ensuring high utilization of the available resources making it a prime candidate  for scheduling in 
systems containing a plethora of heterogeneous computing resources}.


\subsection{Speedup Compared to Software Implementation}\label{sec:exp_2}

We compare the execution time of the 
SOSA with a single-threaded {\bf C} implementation (SOSC) of the SOS on an Intel Xeon\textsuperscript\textregistered~W5-3433 processor running at 4.00 GHz with 512GB RAM. We consider up to 10 machines of varying capabilities (such as {\bf M1}, {\bf M2}, etc.), up to 20 jobs per $V_i$, and 10,000 jobs to schedule as shown in~\Cref{tab:speedup}. 
The SOSC took {\em up to 98.41 seconds} to schedule, whereas \hercules\ and \stannic\ took only {\em up to 0.10 and 0.05 seconds} to schedule all the jobs, achieving a {\em speedup of up to 1060$\times$ and 1968$\times$} respectively. Both of these architectures achieve this while consuming only {\em up to 21 Watts of power}. 
This experiment shows that \bem{a dedicated hardware accelerator-based scheduler can efficiently and effectively schedule jobs on-the-fly within an acceptable power envelope}.


\begin{figure}
    \centering
    \includegraphics[scale=0.25, trim={3cm 1cm 0 0}]{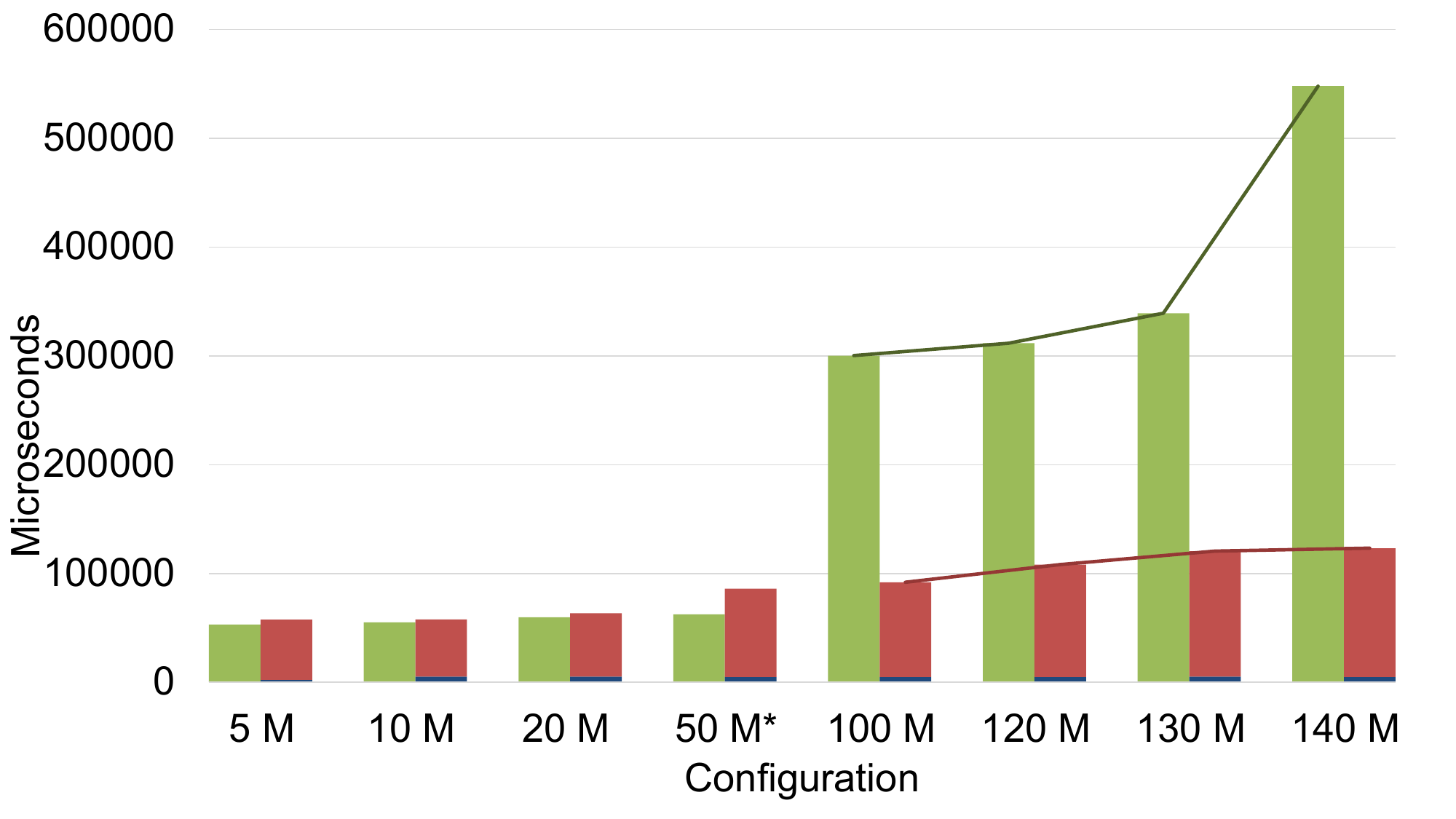}
    \caption{
    {\bf Latency of AVX SIMD Implementation (Green) vs. \stannic\ (Red) across multiple   system configuration sizes}. All configurations use a $V_i$ depth of ten jobs. The part of \stannic's runtime that consists of PCIe communication overhead is marked in dark blue. While the AVX Implementation of the SOS algorithm marginally outperforms \stannic\ in small system configurations, it faces drastic drops in performance as the system scales, causing the intermediary $V_i$ data to misalign with the AVX Vector Bounds. As such, we see that \stannic\ drastically outperforms even AVX enabled schedulers at scale. 
    *Due to heuristic boundaries triggering at different system configurations, Vitis HLS was not capable of routing certain mid-range configurations of \stannic\ with the most aggressively optimized pragmas, leading to performance loss.
    }
    \label{fig:AVX_V_STANNIC}
\end{figure}

We also compare the execution time of \stannic\ versus an AVX SIMD accelerated implementation of the SOS algorithm. The AVX version was run on the same system as the single-threaded {\bf C} implementation previously mentioned. ~\Cref{fig:AVX_V_STANNIC} shows these experimental results. We notice that at small configuration sizes, AVX-enabled software indeed outperforms STANNIC, though only marginally when compared to a well-optimized version of the design (i.e., machine configurations 5, 10, and 20). However, as the machine configuration scales, the AVX SIMD implementation encounters  significant impacts to performance. Specifically, this occurs when all the required machine data misaligns with AVX vector memory boundaries. At these boundaries, the AVX implementation incurs significant overhead, requiring: i) supplementary instructions to execute identical procedures, ii) complex and unaligned memory accesses, as values must be compared across discrete vector registers, and iii) a progressively inflating memory footprint as additional machine state vectors are introduced.

In contrast, \stannic\ runtime scales linearly. While the hardware footprint on the FPGA increases to accommodate the tracking of more virtual schedule data, the host memory usage remains low, as during runtime, host memory is only needed to communicate job data/scheduling decisions over PCIe. We also see that the PCIe communication overhead is on average 4789 Microseconds per 10,000 jobs across all tested configuration sizes. As such, the PCIe communcation overhead is negligible in the overall scheduling latency cost of our SOSA implementations. As such, at larger system configurations, \stannic\ significantly outperforms the AVX-enabled software implementation of the SOS algorithm. This demonstrates that \stannic ’s { \bf remains a highly viable architecture for small-scale systems, particularly in memory-constrained environments, while exhibiting vastly superior scalability compared to the AVX alternative as system configurations expand.}

\subsection{Architectural Comparison Evaluation} \label{sec:arch_comp_analysis}

\begin{figure*}
    \begin{subfigure}[b]{0.45\columnwidth}
        \centering
        \includegraphics[width=.9\textwidth]{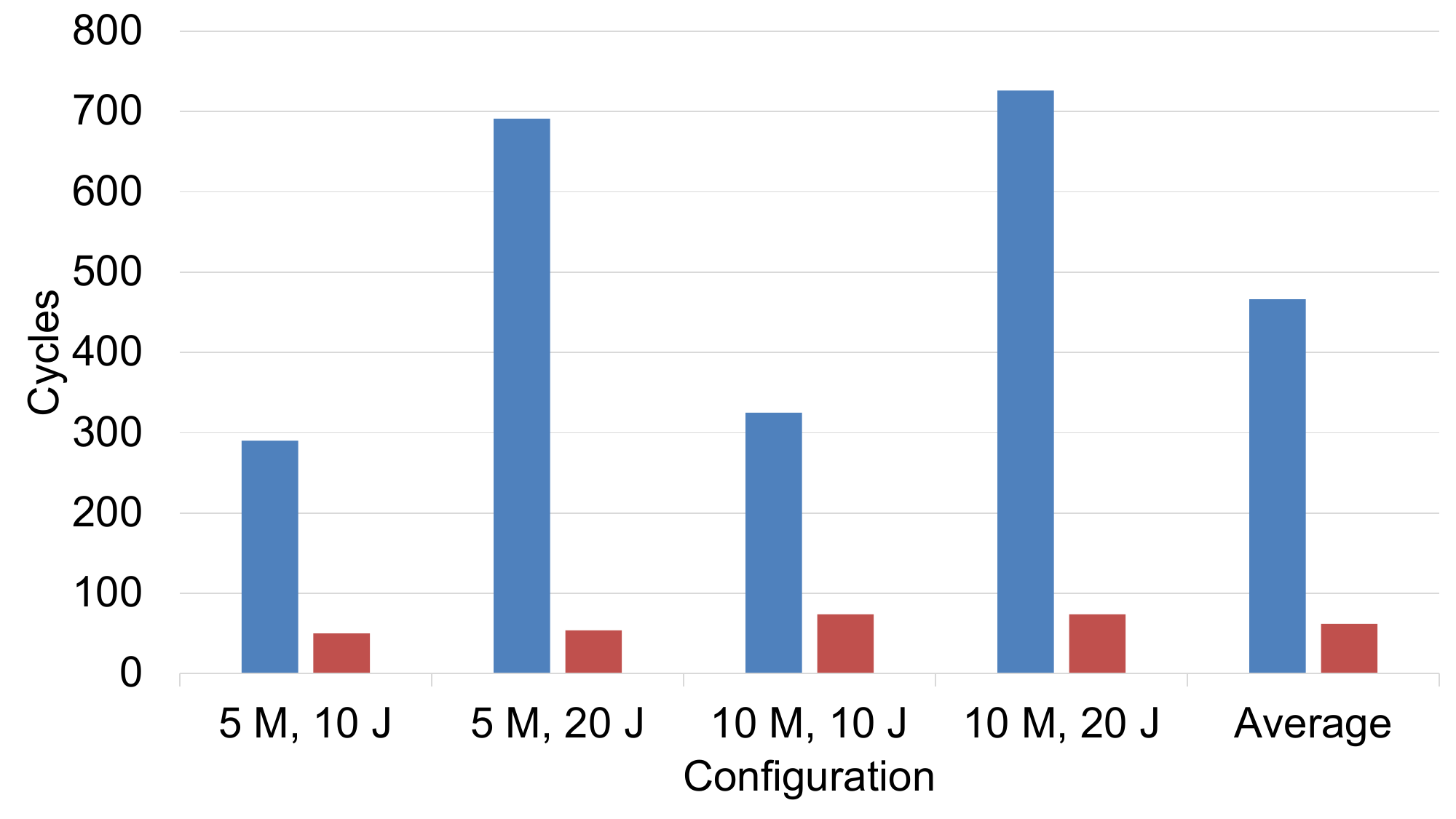}
        \captionsetup{justification=centering}
        \caption{\label{fig:Itr_latency}}
   \end{subfigure}
   \begin{subfigure}[b]{0.45\columnwidth}
        \centering
        \includegraphics[width=.9\textwidth]{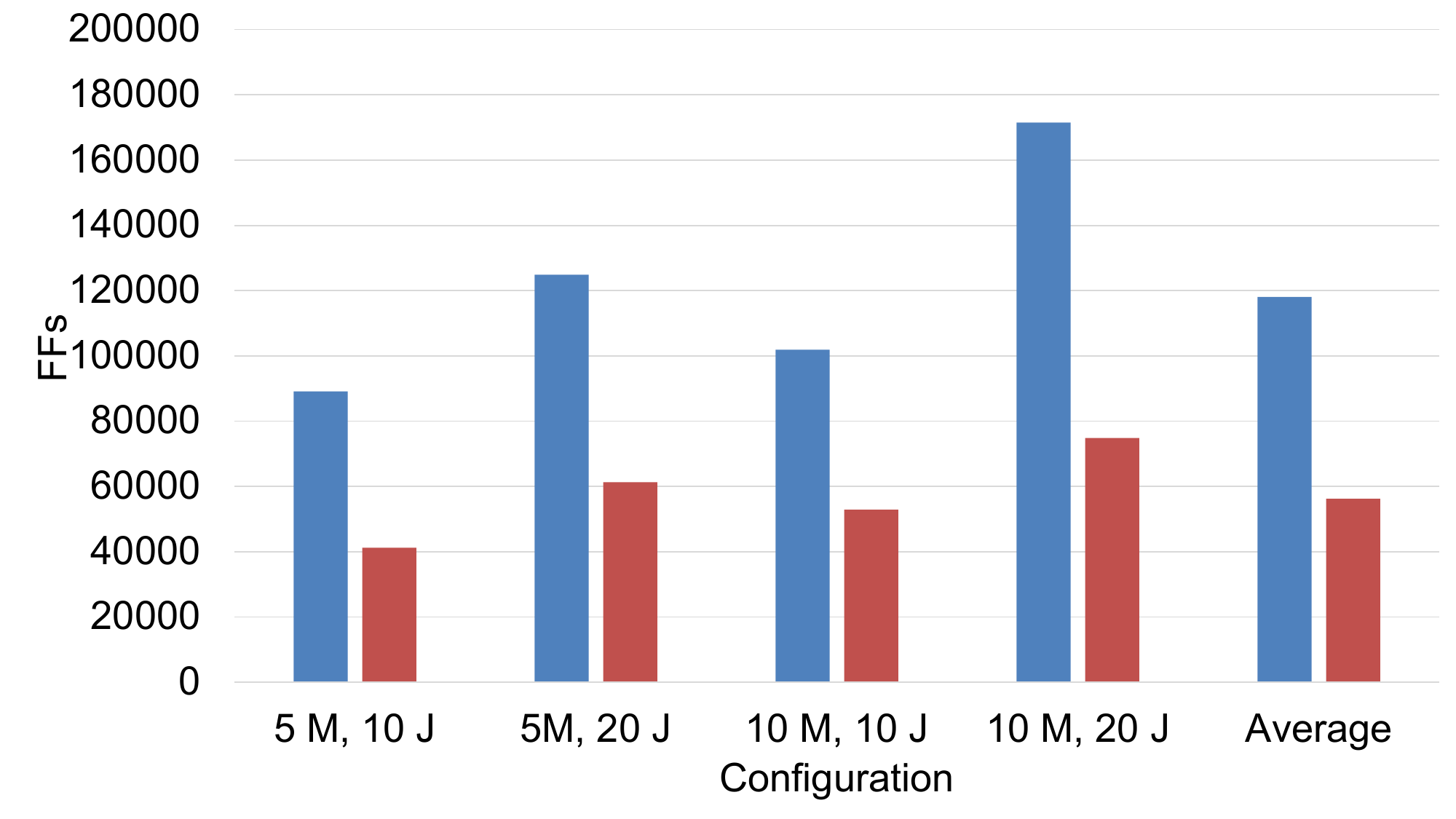}
        \captionsetup{justification=centering}
        \caption{\label{fig:FF_Util}}
   \end{subfigure}   
   \begin{subfigure}[b]{0.45\columnwidth}
        \centering
        \includegraphics[width=.9\textwidth]{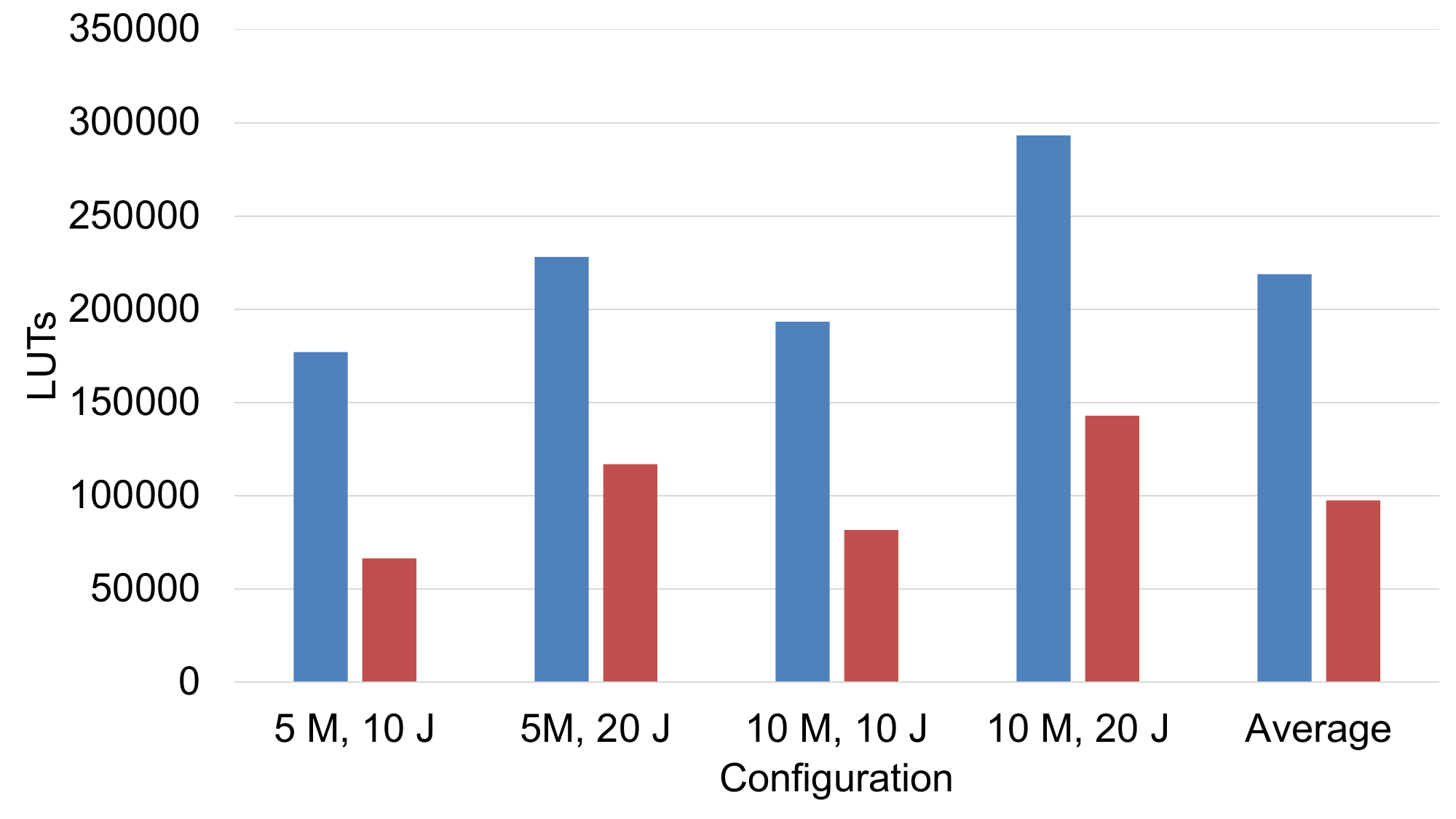}
        \captionsetup{justification=centering}
        \caption{\label{fig:LUT_Util}}
   \end{subfigure} 
   \begin{subfigure}[b]{0.45\columnwidth}
        \centering
        \resizebox{\textwidth}{!}{
        \raisebox{.5\height}{
        \begin{tabular}{|c|c|c|}
            \hline
            & \hercules & \stannic\\
            \hline\hline
            Maximum Routed Configuration & 10 Machines & 140 Machines\\
            Avg. Iteration Latency & 466 Cycles & 62 Cycles\\
            Avg. LUT Utilization & 218,762 LUTs & 97,607 LUTs\\
            Avg. FF Utilization & 118,086 FFs & 56,284 FFs\\
            \hline
        \end{tabular}
        }}
        \vspace{.5cm}
        \captionsetup{justification=centering}
        \caption{\label{tab:Avg_Comp}}
   \end{subfigure}     
   \caption{\bem{Quantitative Comparison Results} for \hercules\ (blue) and \stannic\ (red). All results are from synthesis runs using Xilinx tools. Average values from \Cref{fig:Itr_latency,fig:FF_Util,fig:LUT_Util} reported in \Cref{tab:Avg_Comp}.}
    \label{fig:fpga_resource_utilization}
\end{figure*}

\subsubsection{Iteration Latency}
\Cref{fig:Itr_latency} shows the recorded Iteration Latencies of the four comparison configurations, as well as their average. The average for \hercules\ was 466 cycles, while \stannic 's was 62 cycles. This means that \textit{on average the \stannic\ architecture has a 7.5 times reduction in latency} compared to \hercules . It is also important to notice that where the latency of \hercules\ significantly increases with the increased depth of the Virtual Schedules, \stannic 's latency is negligibly impacted. Additionally, in \hercules, the latency increase for each machine added (with the same Virtual Schedule depth) was $\approx7$ cycles per machine, whereas for \stannic\ the latency cost of a new machine was only $\approx5$ cycles per machine. This shows that not only does \stannic reduce iteration latency across the tested configurations, its latency also increases slower with scaling when compared to \hercules .

\subsubsection{Resource Efficiency}
\Cref{fig:LUT_Util} shows the recorded Look Up Table Utilization, and \Cref{fig:FF_Util} shows the recorded Flip Flop Utilization of the four comparison configurations, as well as their averages. Across all configurations in both designs, the LUT usage was higher than the FF usage, but the exact utilization in both designs corresponded directly with the configuration size the recording represents. This makes sense, as this scaling in configuration will directly result in the need for more corresponding hardware to perform the virtual schedule tracking/cost calculation processes for each machine. On average across the tested configurations, \hercules\ used 218,762 LUTs and 118,086 FFs, while \stannic\ used 97,607 LUTs and 56,284 FFs. This correlates to a 2.24 times reduction and 2.1 times reduction in LUT and FF utilization, respectively. As such we cans see that \stannic\ uses less than half of the hardware resources that \hercules\ would require for an equivalent configuration.

\subsubsection{Maximum Routing Configuration and Power Usage}

Using the progressive configuration scaling until routing failure as described in the \Cref{sec:exp_setup_intra}, \hercules\ was capable of fully routing at a configuration size of 10 machines. \stannic\ was capable of routing for a configuration of 140 machines. These results are recorded in 
\Cref{tab:Avg_Comp}, and shows a 14 times increase in scalability between the two designs. This scalability is in part due to the speedups and resource efficiency described in the last two subsections, but is primarily due to the reduced inter-connectivity routing needs that stem from the local systolic operation of schedule maintenance.

The recorded power utilization for {\bf all} configurations across both designs can be found in \Cref{tab:speedup}, which shows a consistent power usage of $\approx20.5W$. Note that while not presented in these figures, this power draw holds for the 140 Machine configuration implementation of \stannic . As such, the FPGA's power draw was also measured in an idle state with no bitstream loaded, and that measured negligibly lower. This shows that both \hercules\ and \stannic\ are both power efficient designs, barely bringing the Alveo U55C above its idle power draw. 

\subsubsection{Summary of Comparison Results}

In quantitatively comparing the two designs, we can see that \stannic\ has on average a 7.5 times reduction in iteration latency and utilizes less than half the hardware resources of \hercules\ for equivalent system configurations. \stannic\ is also capable of scaling to schedule for systems that are 14 times larger than \hercules\ was capable of. \stannic\ achieves all of this while still maintaining the low power utilization of \hercules .

\subsection{Comparison of Hardware-Accelerated SOSA and Baseline Scheduling Algorithms}\label{sec:hetero_workload}

In these experiments, we compare SOSA against four baseline algorithms under varied workloads 
in terms of average latency and total number of jobs assigned to each 
machine {\bf M1} - {\bf M5}.

\smallskip

\noindent {\bf \ding{172}  Performance under evenly distributed workload}: 
We generate an evenly distributed workload consisting of 35\% memory-intensive jobs, 35\% compute-intensive jobs, and 30\% mixed-type jobs. From~\Cref{fig:001_sos_lnj,fig:001_rr_lnj,fig:001_greedy_lnj,fig:001_wsrr_lnj,fig:001_wsgreedy_lnj}, 
we observe that SOSA demonstrates superior performance in terms of fairness and load balancing targeting heterogeneous systems. 
However, SOSA exhibits slightly higher latency compared to other baseline methods as 
SOSA schedules 
by controlling the job ordering through WSPT ratio. 
Use of WSPT ratio helps in scheduling jobs with higher WSPT earlier, whereas lower WSPT ratio will have a higher latency. 

\smallskip

\noindent {\bf \ding{173} Performance under memory-skewed workload}: In this experiment, we generate memory-skewed workload consisting of 
70\% memory-intensive jobs, 10\% compute-intensive jobs, and 20\% mixed-type jobs. From~\Cref{fig:002_sos_lnj,fig:002_rr_lnj,fig:002_greedy_lnj,fig:002_wsrr_lnj,fig:002_wsgreedy_lnj}, 
we observe that SOSA outperforms all other schedulers in terms of fairness and load balancing implying that SOSA maintains its efficiency and decision-making consistency 
under significant job distribution skew. This robustness is due 
to its cost function, which solely relies on job weights and EPTs, allowing it to adapt dynamically to varying workloads without requiring explicit workload profiling. These findings validate that SOSA is equally 
effective under memory-skewed workload and can achieve high performance 
in real-world 
scenarios with significant load variations. 

\smallskip

\noindent {\bf \ding{174} Performance under compute-skewed workload}: 
We generate compute-skewed workload consisting of  
70\% compute-intensive jobs, 10\% memory-intensive jobs, and 30\% mixed jobs. 
From~\Cref{fig:003_sos_lnj,fig:003_rr_lnj,fig:003_greedy_lnj,fig:003_wsrr_lnj,fig:003_wsgreedy_lnj} 
we observe that SOSA 
adapts equally well to compute-skewed workload, 
validating the effectiveness of SOSA's scheduling. 

\smallskip

\noindent {\bf \ding{175} Performance under homogeneous workload}: 
We generate a memory-intensive job workload. 
The objective is to evaluate whether SOSA maintains consistent performance targeting heterogeneous machines under a fully homogeneous workload. From~\Cref{fig:004_sos_lnj,fig:004_rr_lnj,fig:004_greedy_lnj,fig:004_wsrr_lnj,fig:004_wsgreedy_lnj} 
we observe that 
SOSA does not outperform WSRR and WSG in terms of latency. However, 
SOSA, WSRR, and WSG assign a nearly identical number of jobs to each machine. 
FIFO-based schedulers (Greedy, WSRR, WSG) dispatch jobs in arrival order, whereas SOSA uses WSPT-based prioritization. As a result, SOSA introduces controlled delays to favor jobs with higher scheduling priority, which may increase average latency while still 
minimizing the weighted expected completion time. {\em The higher latency is not a symptom of inefficiency but a side effect of intelligent scheduling prioritization}. Furthermore, SOSA deliberately buffers jobs internally to prevent overloading machine queues -- an effect not reflected in baseline scheduling algorithms. 
Therefore, although latency may appear higher, SOSA 
optimizes performance 
under 
job homogeneity.

\smallskip

\newcommand{\imagewidthlnj}{0.18\textwidth}
\newcommand{\imagescalelnj}{\linewidth}

\begin{figure*}
   \centering
    \begin{subfigure}[b]{0.03\textwidth}
       {\rotatebox{90}{\parbox[c]{4cm}{\centering \bf \tiny Com := 35\%, Mem := 35\%, \\Mixed := 30\%, \ding{172}}}}
   \end{subfigure}
   \begin{subfigure}[b]{\imagewidthlnj}
        \centering
        \includegraphics[width=\imagescalelnj]{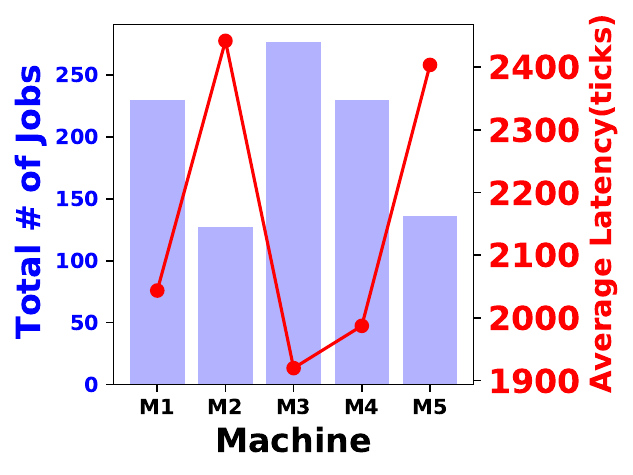}
        \captionsetup{justification=centering}
        \vspace{-5mm}
        \caption{SOS\label{fig:001_sos_lnj}}
   \end{subfigure}
   \begin{subfigure}[b]{\imagewidthlnj}
        \centering
        \includegraphics[width=\imagescalelnj]{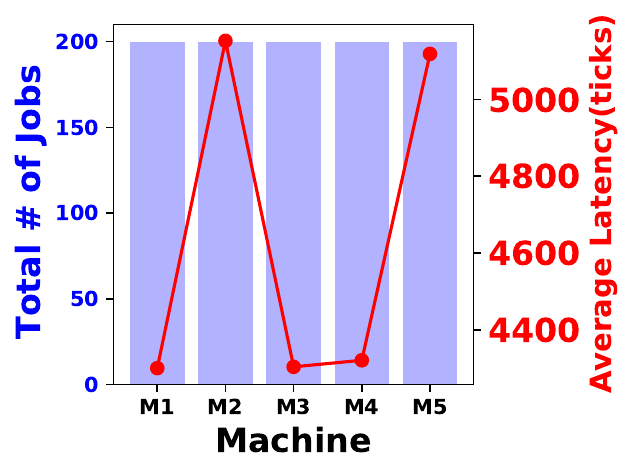}
        \captionsetup{justification=centering}
        \vspace{-5mm}
        \caption{RR\label{fig:001_rr_lnj}}
   \end{subfigure}
   \begin{subfigure}[b]{\imagewidthlnj}
        \centering
        \includegraphics[width=\imagescalelnj]{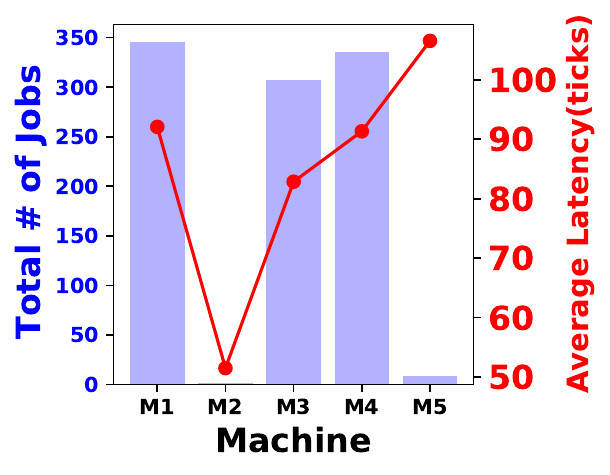}
        \captionsetup{justification=centering}
        \vspace{-5mm}
        \caption{Greedy\label{fig:001_greedy_lnj}}
   \end{subfigure}
   \begin{subfigure}[b]{\imagewidthlnj}
        \centering
        \includegraphics[width=\imagescalelnj]{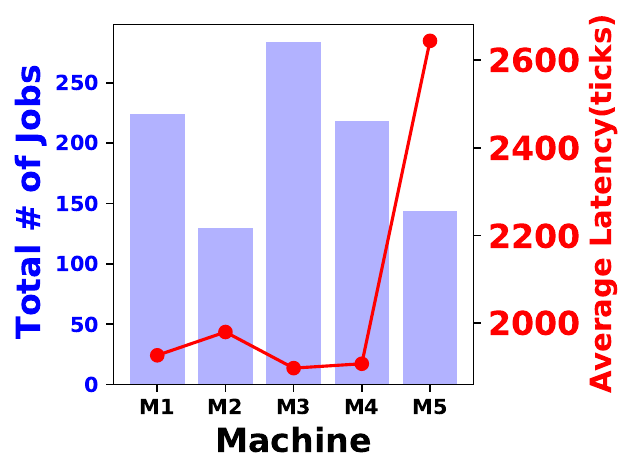}
        \captionsetup{justification=centering}
        \vspace{-5mm}
        \caption{WSRR\label{fig:001_wsrr_lnj}}
   \end{subfigure}
   \begin{subfigure}[b]{\imagewidthlnj}
        \centering
        \includegraphics[width=\imagescalelnj]{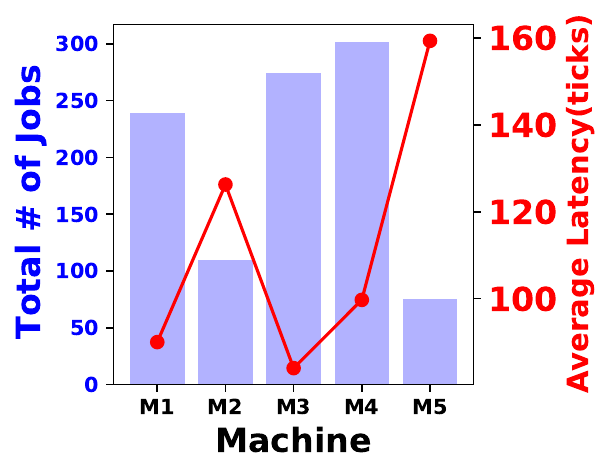}
        \captionsetup{justification=centering}
        \vspace{-5mm}
        \caption{WSG\label{fig:001_wsgreedy_lnj}}
   \end{subfigure}
   \\[-7ex]
   \begin{subfigure}[b]{0.03\textwidth}
       {\rotatebox{90}{\parbox[c]{4cm}{\centering \bf \tiny Com := 10\%, Mem := 70\%,\\Mixed := 20\%, \ding{173}}}}
   \end{subfigure}   
   \begin{subfigure}[b]{\imagewidthlnj}
        \centering
        \includegraphics[width=\imagescalelnj]{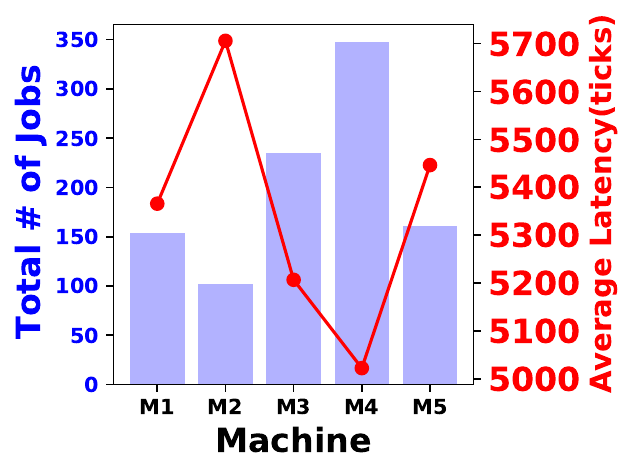}
        \captionsetup{justification=centering}
        \vspace{-5mm}
        \caption{SOS\label{fig:002_sos_lnj}}
   \end{subfigure}
   \begin{subfigure}[b]{\imagewidthlnj}
        \centering
        \includegraphics[width=\imagescalelnj]{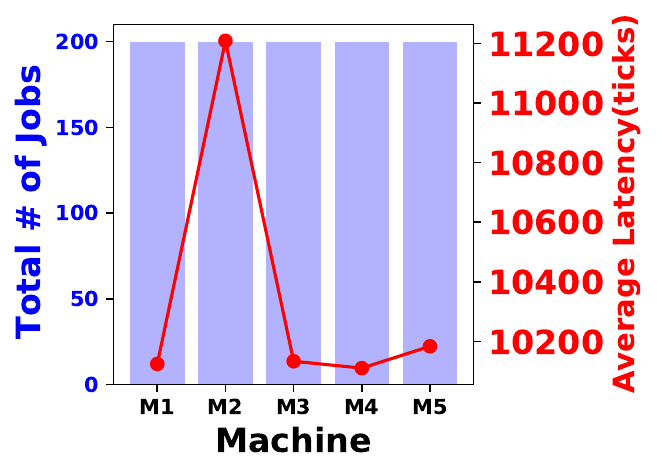}
        \captionsetup{justification=centering}
        \vspace{-5mm}
        \caption{RR\label{fig:002_rr_lnj}}
   \end{subfigure}
   \begin{subfigure}[b]{\imagewidthlnj}
        \centering
        \includegraphics[width=\imagescalelnj]{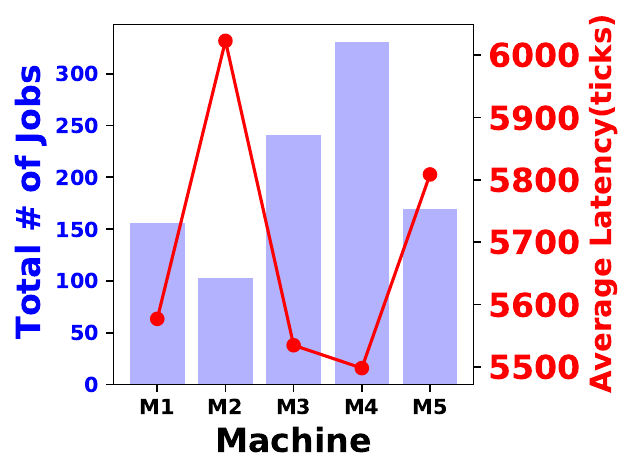}
        \captionsetup{justification=centering}
        \vspace{-5mm}
        \caption{Greedy\label{fig:002_greedy_lnj}}
   \end{subfigure}
   \begin{subfigure}[b]{\imagewidthlnj}
        \centering
        \includegraphics[width=\imagescalelnj]{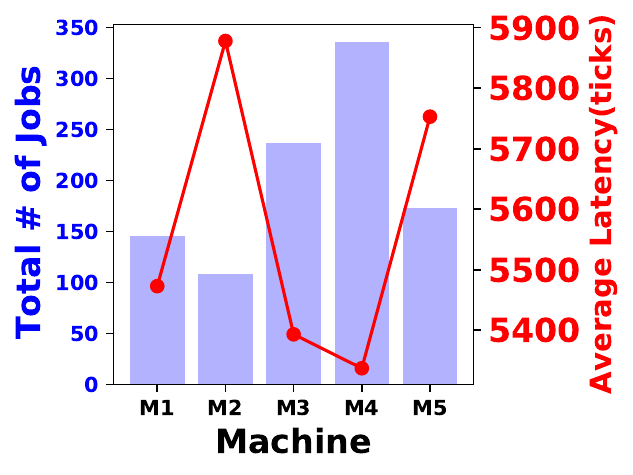}
        \captionsetup{justification=centering}
        \vspace{-5mm}
        \caption{WSRR\label{fig:002_wsrr_lnj}}
   \end{subfigure}
   \begin{subfigure}[b]{\imagewidthlnj}
        \centering
        \includegraphics[width=\imagescalelnj]{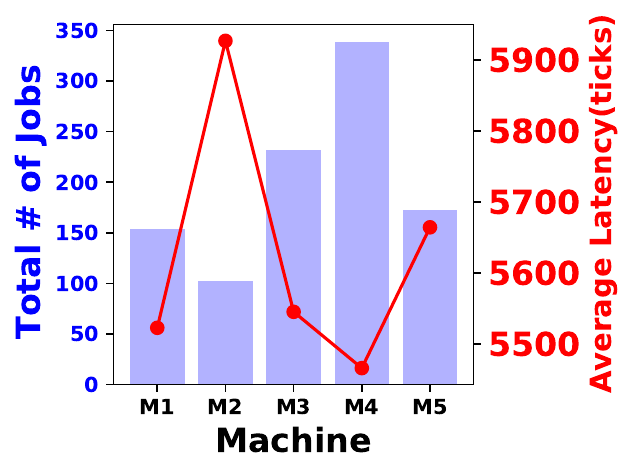}
        \captionsetup{justification=centering}
        \vspace{-5mm}
        \caption{WSG\label{fig:002_wsgreedy_lnj}}
   \end{subfigure}
   \\[-7ex]
   \begin{subfigure}[b]{0.03\textwidth}
       {\rotatebox{90}{\parbox[c]{4cm}{\centering \bf \tiny Com := 70\%, Mem := 10\%,\\Mixed := 20\%, \ding{174}}}}
   \end{subfigure}   
   \begin{subfigure}[b]{\imagewidthlnj}
        \centering
        \includegraphics[width=\imagescalelnj]{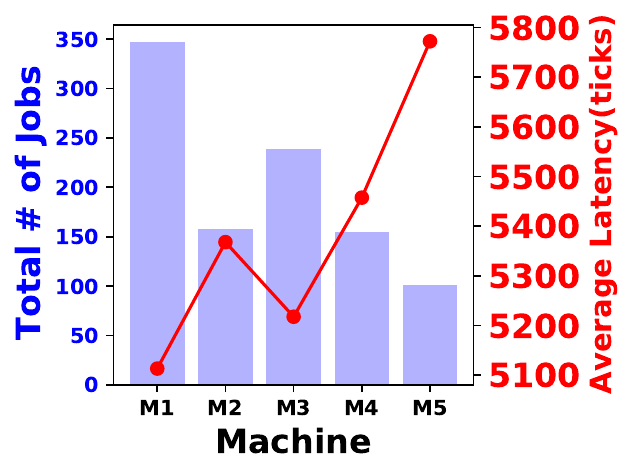}
        \captionsetup{justification=centering}
        \vspace{-5mm}
        \caption{SOS\label{fig:003_sos_lnj}}
   \end{subfigure}
   \begin{subfigure}[b]{\imagewidthlnj}
        \centering
        \includegraphics[width=\imagescalelnj]{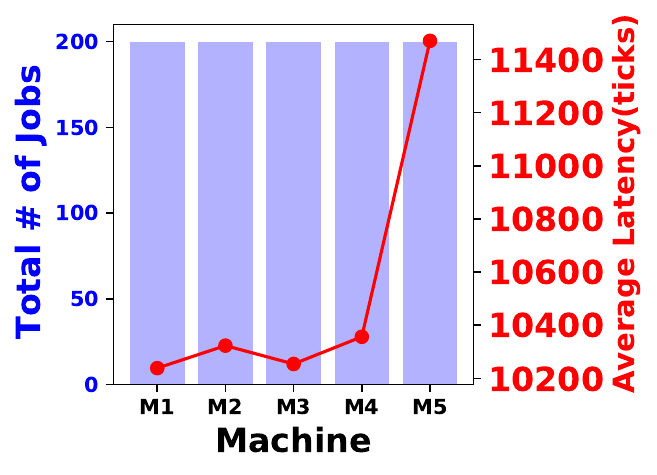}
        \captionsetup{justification=centering}
        \vspace{-5mm}
        \caption{RR\label{fig:003_rr_lnj}}
   \end{subfigure}
   \begin{subfigure}[b]{\imagewidthlnj}
        \centering
        \includegraphics[width=\imagescalelnj]{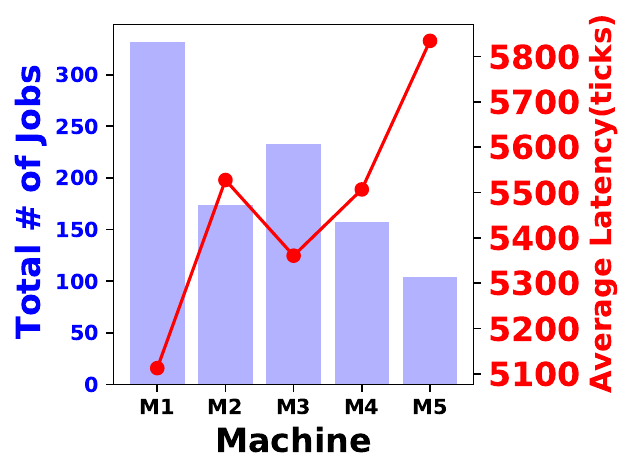}
        \captionsetup{justification=centering}
        \vspace{-5mm}
        \caption{Greedy\label{fig:003_greedy_lnj}}
   \end{subfigure}
   \begin{subfigure}[b]{\imagewidthlnj}
        \centering
        \includegraphics[width=\imagescalelnj]{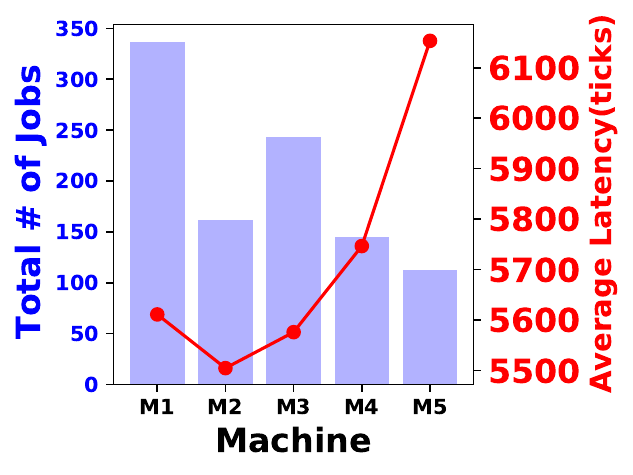}
        \captionsetup{justification=centering}
        \vspace{-5mm}
        \caption{WSRR\label{fig:003_wsrr_lnj}}
   \end{subfigure}
   \begin{subfigure}[b]{\imagewidthlnj}
        \centering
        \includegraphics[width=\imagescalelnj]{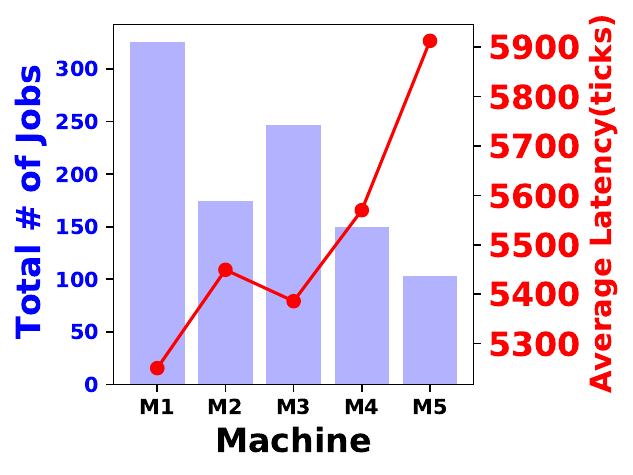}
        \captionsetup{justification=centering}
        \vspace{-5mm}
        \caption{WSG\label{fig:003_wsgreedy_lnj}}
   \end{subfigure}
   \\[-6ex]
   \begin{subfigure}[b]{0.03\textwidth}
       {\rotatebox{90}{\parbox[c]{4cm}{\centering \bf \tiny Mem := 100\%, \\ \ding{175}}}}
   \end{subfigure}   
   \begin{subfigure}[b]{\imagewidthlnj}
        \centering
        \includegraphics[width=\imagescalelnj]{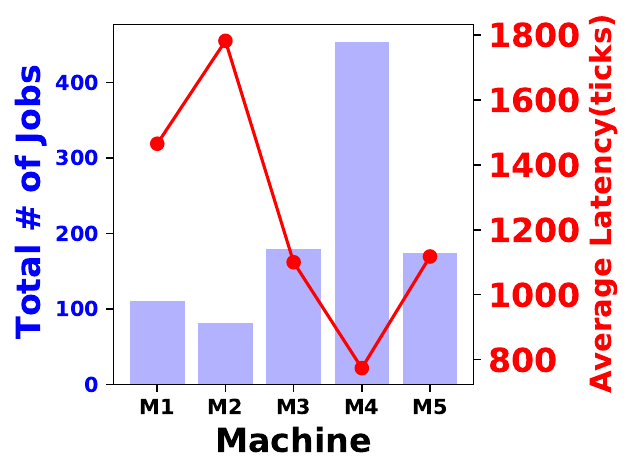}
        \captionsetup{justification=centering}
        \vspace{-5mm}
        \caption{SOS\label{fig:004_sos_lnj}}
   \end{subfigure}
   \begin{subfigure}[b]{\imagewidthlnj}
        \centering
        \includegraphics[width=\imagescalelnj]{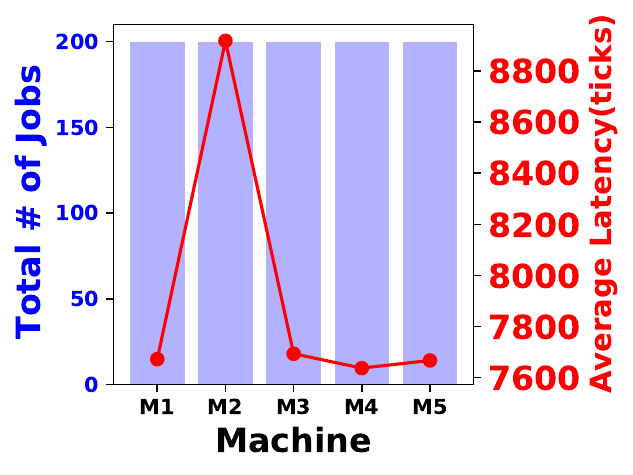}
        \captionsetup{justification=centering}
        \vspace{-5mm}
        \caption{RR\label{fig:004_rr_lnj}}
   \end{subfigure}
   \begin{subfigure}[b]{\imagewidthlnj}
        \centering
        \includegraphics[width=\imagescalelnj]{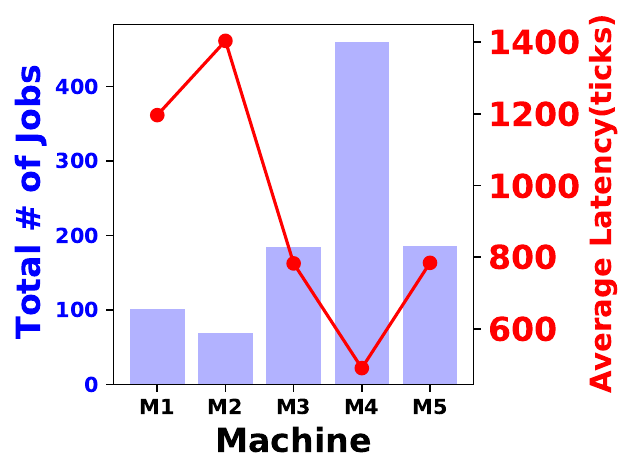}
        \captionsetup{justification=centering}
        \vspace{-5mm}
        \caption{Greedy\label{fig:004_greedy_lnj}}
   \end{subfigure}
   \begin{subfigure}[b]{\imagewidthlnj}
        \centering
        \includegraphics[width=\imagescalelnj]{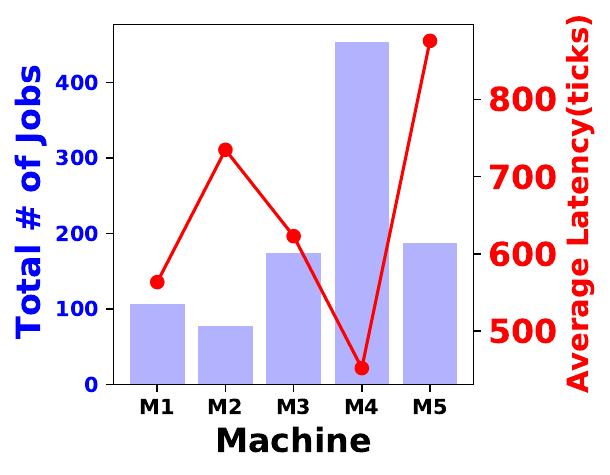}
        \captionsetup{justification=centering}
        \vspace{-5mm}
        \caption{WSRR\label{fig:004_wsrr_lnj}}
   \end{subfigure}
   \begin{subfigure}[b]{\imagewidthlnj}
        \centering
        \includegraphics[width=\imagescalelnj]{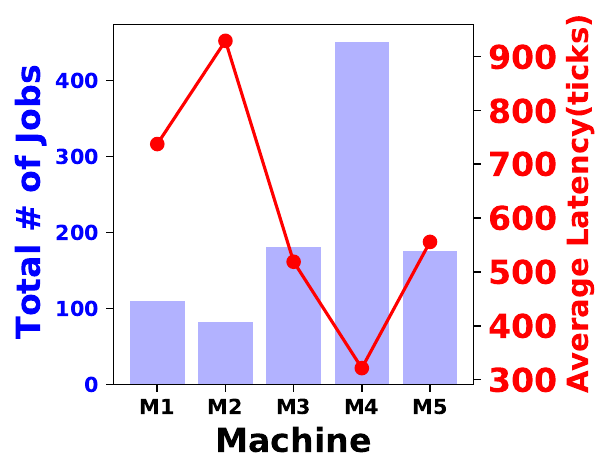}
        \captionsetup{justification=centering}
        \vspace{-5mm}
        \caption{WSG\label{fig:004_wsgreedy_lnj}}
   \end{subfigure}
   \\[-6ex]
   \begin{subfigure}[b]{0.03\textwidth}
       {\rotatebox{90}{\parbox[c]{4cm}{\centering \bf \tiny Com := 100\%, \\ \ding{176}}}}
   \end{subfigure}   
   \begin{subfigure}[b]{\imagewidthlnj}
        \centering
        \includegraphics[width=\imagescalelnj]{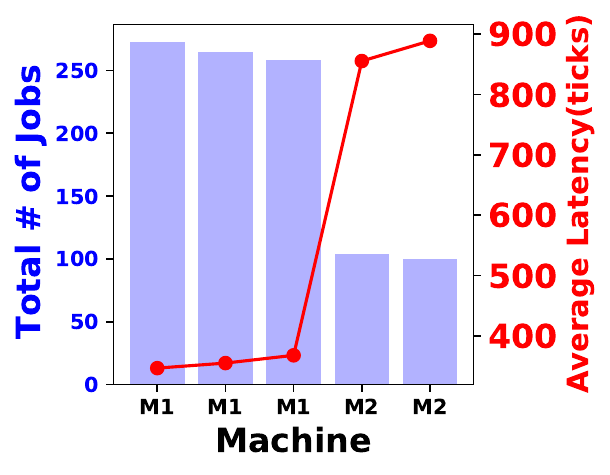}
        \captionsetup{justification=centering}
        \vspace{-5mm}
        \caption{SOS\label{fig:005_sos_lnj}}
   \end{subfigure}
   \begin{subfigure}[b]{\imagewidthlnj}
        \centering
        \includegraphics[width=\imagescalelnj]{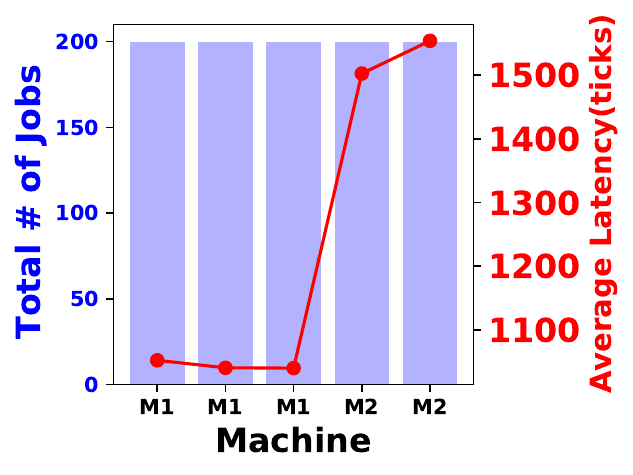}
        \captionsetup{justification=centering}
        \vspace{-5mm}
        \caption{RR\label{fig:005_rr_lnj}}
   \end{subfigure}
   \begin{subfigure}[b]{\imagewidthlnj}
        \centering
        \includegraphics[width=\imagescalelnj]{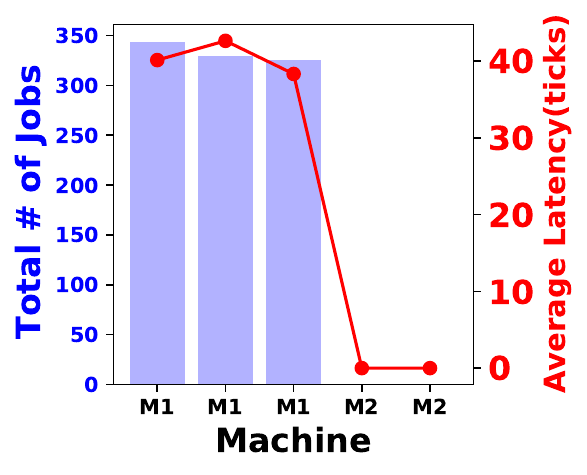}
        \captionsetup{justification=centering}
        \vspace{-5mm}
        \caption{Greedy\label{fig:005_greedy_lnj}}
   \end{subfigure}
   \begin{subfigure}[b]{\imagewidthlnj}
        \centering
        \includegraphics[width=\imagescalelnj]{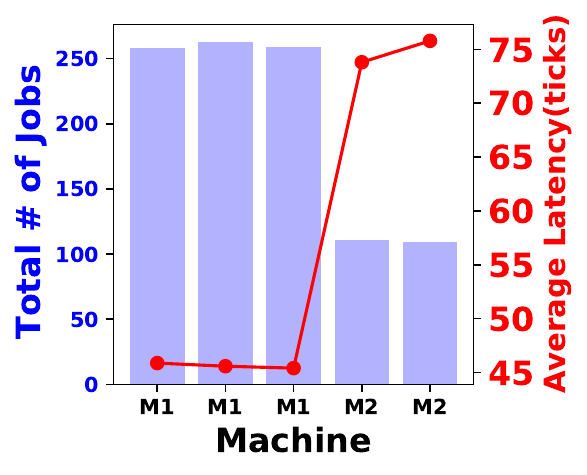}
        \captionsetup{justification=centering}
        \vspace{-5mm}
        \caption{WSRR\label{fig:005_wsrr_lnj}}
   \end{subfigure}
   \begin{subfigure}[b]{\imagewidthlnj}
        \centering
        \includegraphics[width=\imagescalelnj]{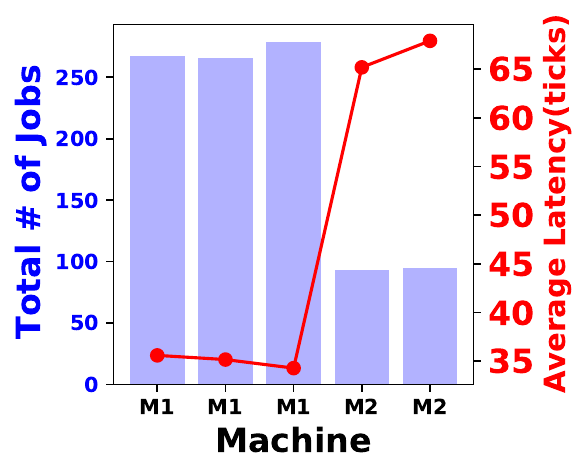}
        \captionsetup{justification=centering}
        \vspace{-5mm}
        \caption{WSG\label{fig:005_wsgreedy_lnj}}
   \end{subfigure}
   \caption{{\bf Job distribution and average latency across M1 -- M5 under varied workloads}. 
   {\bf SOS}: Stochastic Online Scheduler; {\bf RR}: Round Robin Scheduler; {\bf WSRR}: Work Stealing Round Robin Scheduler; {\bf WSG}: Work Stealing Greedy Scheduler. 
   }
   \label{fig:latency_num_jobs}
\end{figure*}

\noindent {\bf \ding{176} Performance on homogeneous machines}: 
We generate a compute-intensive job workload 
and consider only one type of target machine 
(CPU) with varying quality. Although 
SOSA is designed for heterogeneous systems, it is important to evaluate its performance for homogeneous systems, which may occur in practical deployments. From~\Cref{fig:005_sos_lnj,fig:005_rr_lnj,fig:005_greedy_lnj,fig:005_wsrr_lnj,fig:005_wsgreedy_lnj} 
we observe that 
SOSA does not outperform the 
WSRR and WSG 
in terms of latency for its WSPT-based scheduling. 
However, job distribution across machines is nearly identical for all schedulers. Despite the homogeneous nature of 
workload and machines, 
SOS maintains its scheduling principles and performs comparably to baseline schedulers. 




These experiments show that \bem{SOSA is an efficient, effective, and adaptable scheduler under varying realistic workloads targeting heterogeneous and homogeneous hardware}.

\section{Related Work}\label{sec:rel_work}

Several hardware-based schedulers have been developed for multicore systems. 
SR-PQ~\cite{kuacharoen2003configurable, schRTOSyitang2015} 
enabled configurable real-time scheduling with limited scalability. 
HRHS~\cite{hrhsdanesh2020} improved flexibility through partitioned scheduling, while TCOM~\cite{tcomnorollah2021} extended support for task dependencies. 
HD-CPS~\cite{hdcps2021} addressed communication bottlenecks using per-core queues and priority drift using centralized coordination and 
SchedTask~\cite{kallurkar2017schedtask} reduced I-cache pollution by grouping tasks with similar instruction footprints but introduced 
hardware overhead and latency. 
Heuristics-based ~\cite{schheteroserifyesil2022,exptaskschjuanfang2020,habibpour2024improved} and dynamic allocation methods~\cite{fdetsadetswan2021} improve system utilization with increased 
scheduling overhead or inconsistent convergence. Learning-based schedulers~\cite{du2019feature} adapt to workloads 
while OPADCS~\cite{liu2024uncertainty} prioritizes deadline adherence in uncertain, online scenarios. Additional software-based efforts target optimal task mapping~\cite{orr2021optimal}, system reliability~\cite{naithani2017reliability}, and energy-constrained execution~\cite{raca2022runtime}. 
\bem{However, none of these approaches address heterogeneous stochastic online scheduling, and thus, direct comparison to SOSA is not applicable.}

Of pre-existing related works, the most relevant prior work can be found in Hardware HEFT~\cite{HEFT}. In this work, the authors present a hardware-accelerated version of the Heterogeneous Earliest Finish Time scheduling algorithm. In their paper, they use their accelerator to perform scheduling decisions in a small SoC context consisting of up to 16 processing elements of two distinct types: either an ARM core or an FFT processing module. In this sense, SOSA (specifically in the \stannic\ implementation) displays the capability of addressing significantly larger systems with higher levels of heterogeneity. Additionally, while HW-HEFT is demonstrated to be a capable SoC scheduler in dealing with dynamically arriving tasks by performing real-life signal processing, it also requires knowledge of all tasks, ahead of time, making it non-applicable to full stochastic online task arrival. This makes it a great fit for smaller, individual SoCs, but in larger, shared/networked systems, these unknowns may make HW-HEFT falter.




\section{Conclusion} \label{sec:Conclusion}

We introduced two hardware-accelerated online scheduler accelerator architectures, \hercules\ and \stannic , collectively referred to as SOSA. These architectures accelerate an algorithm capable of adapting to diverse workloads targeting heterogeneous and homogeneous computing systems with acceptable 
latency, and job distribution while minimizing expected job completion times. Both SOSA architectures consume only 21 Watts 
and achieve a speedup of 1968$\times$ compared to a {\bf C}-based single thread scheduler. Such characteristics make SOSA a prime candidate for scheduling scientific and deep-learning workloads with significant variability.



The two architectures are presented chronologically in their development, allowing for iteration and optimizations to be explored in a solution-driven manner to justify the need for such a fundamental shift in architectural paradigm.
Though the differences in operation are significant, parity in the two designs' function output was established, showing that the difference in function performance has no effect on the resultant schedules effectiveness. The finer details of scheduling with this new design, and proof of the required assumptions, were also provided.

With the two architectures now fully understood, we were able to measure performance metrics of both to quantitatively compare the two to each other. In every metric, $\stannic$ outperforms $\hercules$, all while maintaining an equivalent power profile. $\stannic$ is on average $7.5$ times faster than $\hercules$, $~2.1 \times$  smaller than $\hercules$, and is capable of scaling to consider systems that are $14$ times larger than $\hercules$ could consider, all while maintaining minimal power usage. Through these qualitative comparisons, we can see that $\stannic$ is a fundamentally more efficient and capable scheduling accelerator than $\hercules$.

\smallskip

\noindent {\bf Acknowledgment}: This work was supported in part by NSF Grant \# 2433972. This work was supported in part by AMD Hardware Research Donation \# 5310-AUP-1-1Z2UEXF.


\bibliographystyle{ACM-Reference-Format}
\bibliography{bib/master, bib/reference}

\end{document}